\documentclass[twocolumn]{aastex62}

\newcommand{\thisgrb}{GRB~171010A}
\newcommand{\asat}{{\em AstroSat}}
\newcommand{\fermi}{{\em Fermi}}
\newcommand{\swift}{{\em Swift}}
\newcommand{\sw}[1]{\texttt{#1}}

\graphicspath{{./}{figures/}}

\shorttitle{Variable prompt emission polarization in  \thisgrb}
\shortauthors{Chand et al.}

\begin{document}

\title{\asat-CZTI detection of variable prompt emission polarization in  \thisgrb}
\email{$^\dagger$vikas.chand.physics@gmail.com}
\email{$^\ddagger$txc344@psu.edu}
\email{$^*$gor.oganesyan@gssi.it}

\author{Vikas Chand$^\dagger$}
\affiliation{Tata Institute of Fundamental Research,
Mumbai, India}
\author{Tanmoy Chattopadhyay$^\ddagger$}
\affil{Pennsylvania State University,
State College, PA, USA}
\author{Gor Oganesyan$^*$}
\affiliation{Gran Sasso Science Institute, Viale F. Crispi 7, I-67100, L’Aquila, Italy}
\affiliation{ INFN - Laboratori Nazionali del Gran Sasso, I-67100 L’Aquila, Italy}
\author{A. R. Rao}
\affiliation{Tata Institute of Fundamental Research,
Mumbai, India}
\author{Santosh V. Vadawale}
\affil{Physical Research Laboratory,
Ahmedabad, Gujarat, India}
\author{Dipankar Bhattacharya}
\affiliation{The Inter-University Centre for Astronomy and Astrophysics, Pune, India}
\author{V. B. Bhalerao}
\affiliation{Indian Institute of Technology, Bombay, India}
\author{Kuntal Misra}
\affiliation{Aryabhatta Research Institute of observational sciecES (ARIES), Manora Peak, Nainital, India}
\affiliation{Department of Physics, University of California, 1 Shields Ave, Davis, CA 95616-5270, USA}

\begin{abstract}

We present spectro-polarimetric analysis of \thisgrb\ using data from \asat, \fermi, and \swift, to provide insights into the physical mechanisms of the prompt radiation and the jet geometry. 

Prompt emission from \thisgrb\ was very bright (fluence $>10^{-4}$~ergs~cm$^{-2}$) and had a complex structure composed of the superimposition of several pulses. The energy spectra deviate from the typical Band function to show a low energy peak $\sim 15$~keV --- which we interpret as a  power-law with two breaks, with a synchrotron origin. Alternately, the prompt spectra can also be interpreted as Comptonized emission, or a blackbody combined with a Band function.
Time-resolved analysis confirms the presence of the low energy component, while the peak energy is found to be confined in the range of 100--200~keV.

Afterglow emission detected by \fermi-LAT is typical of an external shock model, and we constrain the initial Lorentz factor using the peak time of the emission. \swift-XRT measurements of the afterglow show an indication for a jet break, allowing us to constrain the jet opening angle to  $>$ 6$\degr$.

Detection of a large number of Compton scattered events by \asat-CZTI provides an opportunity to study hard X-ray polarization of the prompt emission. We find that the burst has high, time-variable polarization, with the emission {\bf have higher polarization} at energies above the peak energy. 

We discuss all observations in the context of GRB models and polarization arising due to {\bf due to physical or geometric effects:} synchrotron emission from multiple shocks with ordered or random magnetic fields, Poynting flux dominated jet undergoing abrupt magnetic dissipation, sub-photospheric dissipation, a jet consisting of fragmented fireballs, and the Comptonization model.
\end{abstract}


\section{Introduction}
The emission mechanisms of gamma ray bursts (GRBs) and the radiation processes involved in producing 
the complex structure during the prompt phase have eluded a complete understanding
in spite of the fact that these energetic cosmological 
events     have  been  studied for more than five decades. Based on the duration of the observed prompt emission, GRBs are broadly
classified into two families named as long and short GRBs with a demarcation at 2 s. The long GRBs are found to be associated with with type Ic supernovae pointing at their massive star origin. To extract energy efficiently to power a GRB such a collapse results in a black hole or a rapidly spinning and highly magnetized neutron star \citep{Woosley:2006, Cano:2017}. The recent groundbreaking
discovery of gravitational waves and its electromagnetic counterpart object (a short GRB) resolved the problem of identifying progenitors of short GRBs at least for the case of one joint GW/GRB detection \citep{Abbott:2016PhRvL}.
The commonly accepted scenario for the production of the emerging radiation 
is based on the premise of   relativistic jets  being  launched from the central engine. When the jet  
pierces through an ambient medium which is either a constant density
interstellar medium (ISM) or a wind-like medium whose density varies with distance, 
external shocks are formed and the generated
radiation has contributions from both the forward and the 
reverse shocks. The resultant  emission constitutes the widely observed afterglows in GRBs \citep{Rees:1992, Meszaros:1993, Meszaros:1997, Akerlof:1999, Meszaros:1999}. 
A plateau phase 
in the X-ray emission is, however, thought to be associated with a long term central engine activity or the stratification in Lorentz factor of the ejecta could also give extended energy injection \citep{Dai_lu:1998, Zhang:2006ApJ}. 

The prompt emission can arise due to various mechanisms, the leading candidates among them are: (i) increase of the jet Lorentz factor with time, leading to the collision of inner layers with the outer ones, generating internal shocks which produce non-thermal 
synchrotron emission as the electrons gyrate in 
the existing magnetic field \citep{Narayan:1992, Rees:1994}. A variant of internal shock model is the Internal-Collision-induced MAgnetic Reconnection and Turbulence
(ICMART) model where abrupt dissipaton occurs through magnetic reconnections in a Poynting flux dominated jet \citep{Zhang:2011ApJ}; (iii) dissipation occurring within a fuzzy photosphere; the photosphere
is specially 
invoked  to explain the quasi-thermal shape of the spectrum. The process manifests itself in such a way that it can produce a non-thermal shape of the spectrum as well \citep{Beloborodov2017}, (iii) gradual magnetic dissipation that occurs within the 
photosphere of a Poynting flux dominated jet \citep{Beniamini:2017} and (iv) Comptonization of soft corona photons off 
the electrons present in the outgoing relativistic ejecta \citep{Titarchuk:2012, Frontera:2013}.
These models are designed to explain the spectral properties of the prompt emission of GRBs and are successful to a certain extent.  Some prominent features like the evolution of the spectral parameters and the presence of correlations 
among GRB observables are not well understood and most of them remain unexplained within the framework of a single model. 

Another crucial information that can be added to the existing plethora of  observations of the  
prompt emission like the spectral and timing properties, presence of afterglows, presence of 
associated supernovae etc, is the polarization of the prompt emission.
Detection of polarization, therefore, provides an additional tool to test the theoretical models of the mechanism of GRB
prompt emission. 
This, however, has remained a scarcely  
explored avenue due to the  unavailability of dedicated polarimeters and reliable polarization measurements. 
The Cadmium Zinc Telluride Imager (CZTI) on board \asat, offers a new opportunity to reliably measure polarization of bright GRBs in the hard X-rays \citep{Chattopadhyay:2014, Vadawale:2015, Chattopadhyay:2017}.
The polarization expected from different models of prompt emission is not only different in magnitude but also in the expected pattern of temporal variability, depending on the emission mechanism, jet morphology and view geometry (see e.g. \citealt{Covino:2016}). 
A study of the time variability characteristics is very important because bright GRBs whose spectra  have been studied in great detail often found to have spectra  deviating significantly from the standard Band function 
conventionally used to model the 
GRB spectra \citep{Abdo:2009, Izzo:2012, Ackermann:2010, 
Vianello:2017, Wang:2017}. 

\thisgrb\ is a bright GRB and it presents an opportunity for a multi-pronged
approach to understand the GRB prompt emission. It has been observed by both \fermi\ and \asat-CZTI. Afterglows in
gamma-rays (Fermi/LAT), X-rays (\swift-XRT) and optical have been detected, and an associated 
supernova SN 2017htp has been found on the tenth day of the prompt emission. A redshift $z=0.33$ has been measured spectroscopically by the {\it extended Public ESO Spectroscopic Survey for Transient Objects} (ePESSTO) optical observations 
\citep{Kankare:2017}. We present here a comprehensive analysis of this
GRB using the \fermi\ observation for
spectral properties and attempt to relate the prompt spectral properties 
to the  detection of variable high polarization using \asat-CZTI. 
We present a summary of the observations in Section \ref{sec:observations}. The \fermi\ light curves and spectra are 
constructed in various energy bands and time bins respectively and they are presented in 
Sections \ref{sec:observations}, \ref{sec:lightcurves} and \ref{sec:spectral_analysis}. The polarization measurements in different time intervals 
and energies are presented in Section \ref{sec:polarization}. We discuss our results and derive conclusions 
in Section \ref{sec:conclusion}.
The cosmological parameters chosen were $\Omega_\lambda = 0.73$, $\Omega_m = 0.27$
and $H_0 ~= ~70$ $km~Mpc^{-1}~sec^{-1}$ \citep{Komatsu:2009}.

\section{\thisgrb}\label{sec:observations}
\thisgrb\ triggered the \fermi-LAT and \fermi-GBM at 19:00:50.58 UT $(T_0)$ on 2017    
October 10 \citep{Omodei:2017, Poolakkil:2017}. 
The observed high peak flux generated an autonomous
re-point request $(ARR)$ in the GBM flight software and the \fermi\ telescope slewed to the GBM in-flight location.
A target of opportunity observation was carried out by 
the {\em Niel Gehrels Swift Observatory} \citep{Evans:2017} and the \swift-XRT localized the burst to $RA(J2000):~ 66^\circ.58092$, and $Dec:~ -10^\circ.46325$ \citep{DAi:2017}. \swift-XRT followed the 
burst for $\sim 2\times10^6$  s\footnote{\url{http://www.swift.ac.uk/xrt_curves/00020778/}}. The prompt emission was also observed 
by Konus-Wind \citep{Frederiks:2017}. The fluence observed in the \fermi-GBM 10 - 1000~keV band 
from $T_0+5.12$ sec to $T_0+151.55 $ s is $( 6.42 \pm 0.05 ) \times 10^{-4}$ $erg~cm^{-2}$ \citep{Poolakkil:2017}.  
Here $T_0$ is the trigger time in the \fermi-GBM.
The first photon in 
LAT $(>100~MeV)$ with a probability 0.9 of its association with the source is received at $\sim T_0+374$ s and has an energy $\sim 194 ~MeV$ and a photon with energy close to 1 GeV (930 MeV) is detected at $\sim 404$ s.
The highest energy photon ($\sim 19~GeV$) in the \fermi-LAT is detected at $\sim 2890~s$. The rest frame energy of this photon is $\sim25 ~GeV$.

\section{Light curves}\label{sec:lightcurves}
 We   examined the light curves of the
GRB as obtained from all 12 NaI detectors and found that 
detectors n7, n8 and nb  (using the usual naming conventions) 
have significant detections (source angles are $n7:69^\circ$, $n8:31^\circ$ and $nb:40^\circ$). We have used data from all three detectors to examine the broad emission features of the GRB. For wide band temporal and spectral analysis, however, we chose  
n8 and nb, which have higher count rates than the other detectors  and 
the angles made with the source direction for these two detectors are less than 50 degrees.  The time 
scale used throughout this paper is relative to the \fermi-GBM trigger time $i.e.$ $t=T-T_0$.
Of the two BGO scintillation detectors, the detector BGO 1 (b1), which is positioned on the same side on the satellite as the selected NaI detectors, is chosen for further analysis.  

\begin{figure}
\centering
\includegraphics[scale=0.35]{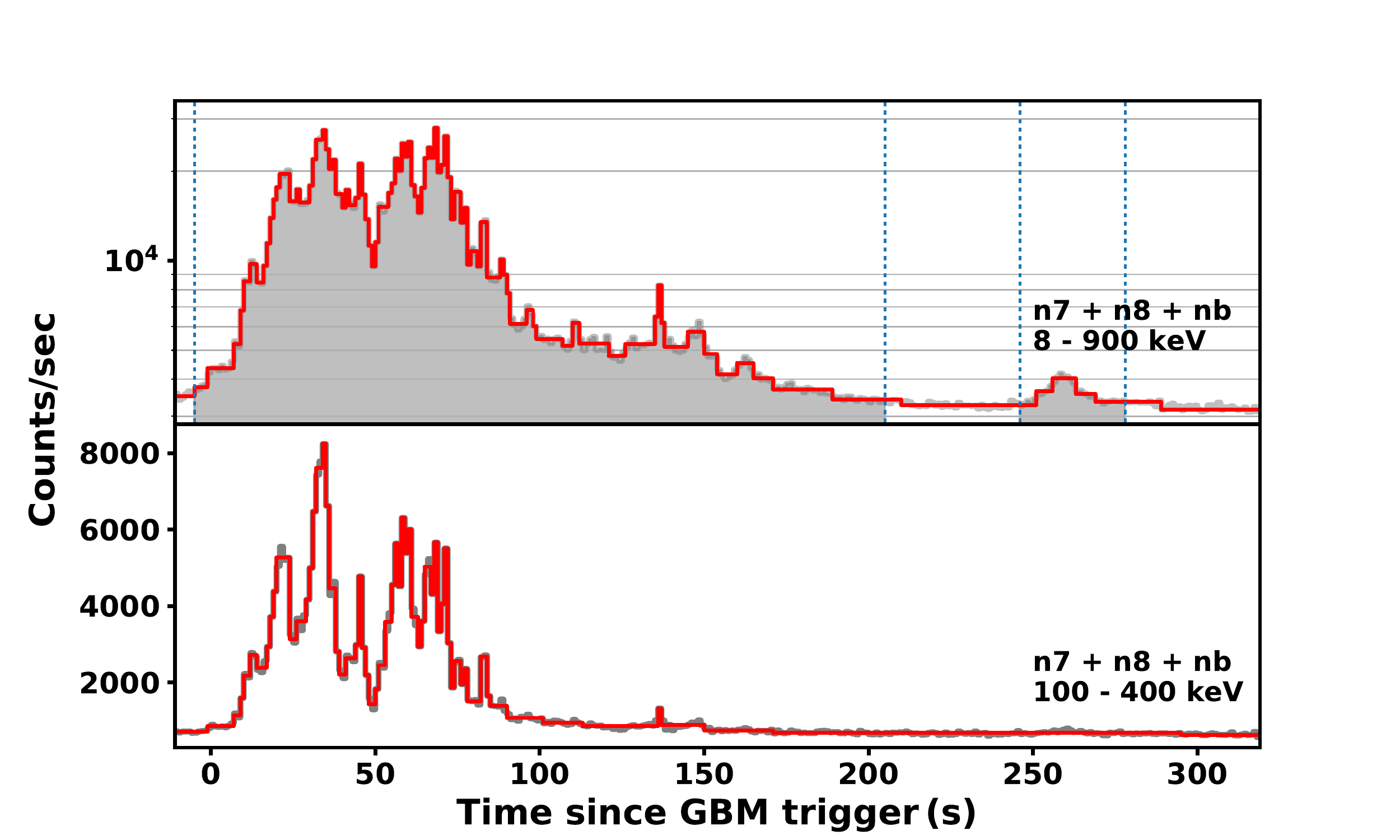}
\caption{\fermi-GBM light curve of \thisgrb\ obtained from three NaI detectors n7, n8 and nb. \\
\textit{Upper Panel}: Log-scale light curve in the full energy range (8 -- 900~keV), binned in 1~s intervals. The two emission episodes used for time-integrated spectral analysis ($-5$~s to 205~s, 246~s to 278~s) are shaded and demarcated with vertical dotted lines. The red envelope shows the Bayesian blocks representation of the light curve. \\
\textit{Lower panel}: Linear-scale light curve and Bayesian blocks representation for the 100 -- 400~keV energy range used in time-resolved spectral analysis.}
\label{fig:GBM_ti_tr}
\end{figure}

The light curve is shown in Figure \ref{fig:GBM_ti_tr} and it shows two episodes of emission (a) beginning at $\sim -5$s of the main burst and lasts up to $\sim 205$ s followed by a quiescent 
state when the emission level meets the background level and (b) the emission becomes active  again after a short period at $\sim 246$ s and radiation from 
the burst is detected till $\sim 278$ s. We name these emission phases 
as Episode 1 $\&$ 2. In Figure \ref{fig:GBM_ti_tr} we also show the 
light curve in the 100 - 400~keV region (the energy range of \asat-CZTI observations) and we use this light curve to divide the data for time resolved spectral analysis. The blocks of constant rate (Bayesian blocks) are constructed from the light curves and over-plotted on the count rate light curves to show statistically significant changes in them \citep{Scargle:2013}. 

\begin{figure*}
\centering
\includegraphics[scale=0.5]{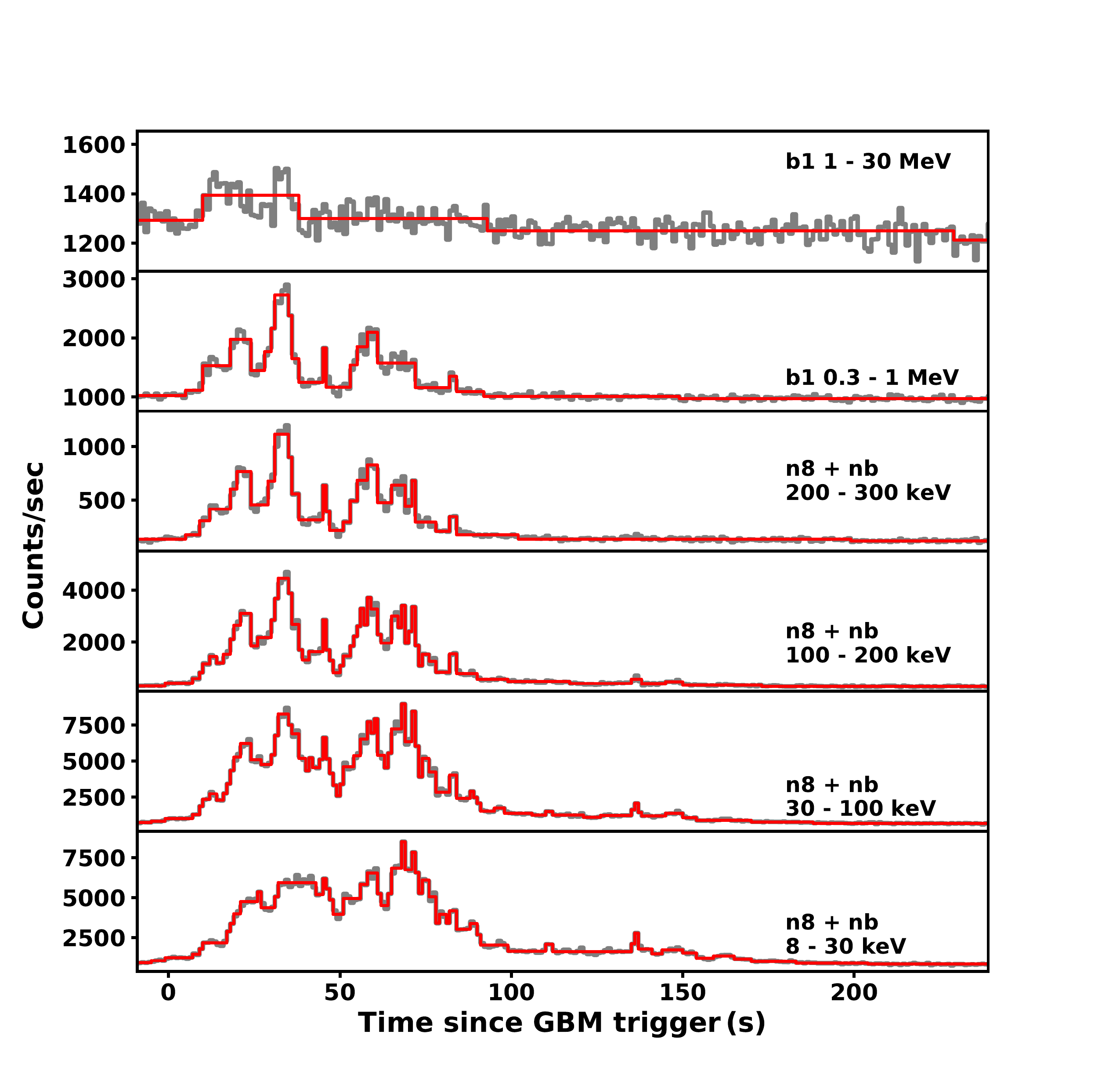}
\caption{\fermi-GBM light curve of \thisgrb\ in six different energy ranges, with Bayesian blocks  over-plotted in red. At the highest energies (1 -- 30~MeV, top panel), the emission is low after $\sim 50$~s.}
\label{fig:GBM_LC}
\end{figure*}

The light curves summed over the chosen GBM detectors (n8 and nb) in six different energy ranges 
are shown in Figure \ref{fig:GBM_LC}. The GRB has complex structure with multiple peaks and the
high energy emission (above 1 MeV) is predominant till about 50 s as seen in the light curve for $>1 ~MeV$.
Bayesian blocks are constructed from the light curves and over-plotted on the count rate.

\section{Spectral analysis} \label{sec:spectral_analysis}
The burst is very bright and even with a signal to noise ratio of 50, we are able to construct more than 150 time bins. We created time bins, therefore, from the Bayesian blocks constructed from the light curve in the energy range 100~keV to 400~keV where CZTI is sensitive for polarization measurements. These bins (large number of Bayesian blocks) track significantly varying features in the lightcurve and will also be sufficient to give a substantial idea about the energetics of the burst. The spectra are reduced 
using \fermi\ Science tools software {\it rmfit}\footnote{\fermi\ Science Tools, \sw{rmfit}: \url{https://fermi.gsfc.nasa.gov/ssc/data/analysis/rmfit}}
by standard methodology described in the tutorial\footnote{\fermi\ tutorial: \url{https://fermi.gsfc.nasa.gov/ssc/data/analysis/scitools/rmfit_tutorial.html}}.

GRB prompt emission spectra generally show a typical shape described by the Band function \citep{Band:1993}, however, for a number of bursts additional components are observed in the spectra such as quasi-thermal components modeled by blackbody function \citep{Ryde:2005, Page:2011, Guiriec:2011, BR:2015}
or they show additional non-thermal components modeled 
by a power-law or a power-law with an exponential cut-off \citep{Abdo:2009, Ackermann:2013}. Additionally, for a few cases, spectra with a top-hat shape are also 
observed and modeled by a Band-function with an exponential roll-off at higher energies \citep{Wang:2017, Vianello:2017}. 

The above models are driven by data and phenomenology. The physical models devised to explain the GRB prompt emission radiation have the synchrotron emission or the Compton scattering as the core processes. The synchrotron spectrum for a powerlaw distribution of electrons consists of power-laws joined at energies which depend upon cooling frequency ($\nu_c$), minimum-injection frequency ($\nu_m$) and absorption frequency ($\nu_a$).
The Comptonization models include inverse Compton scattering as a primary mechanism \citep{Lazzati:2000, Titarchuk:2012}.
Comptonization signatures ($XSPEC$ \sw{grbcomp} model) have been detected for a set of GRBs observed by Burst and Transient Source Experiment 
($BATSE$) on board $Compton ~Gamma ~Ray ~Observatory$ \citep{Frontera:2013}. The \sw{grbcomp} model differs from the other Comptonization model: the Compton drag model. In the Compton drag model the single 
inverse Compton scatterings of thermal photons shape the spectrum, while in the \sw{grbcomp} multiple Compton scatterings are assumed. Another major difference is that the 
scattering is off the outflow which is sub-relativistic in case of \sw{grbcomp} model, while relativistic in the Compton drag model.

We apply this prior knowledge of spectral shapes from the previous observations of GRBs to study \thisgrb. 
The spectra are modeled by various spectroscopic models available in $XSPEC$ \citep{Arnaud:1996}. The statistics \sw{pgstat} is used for optimization 
and testing the various 
models\footnote{\url{https://heasarc.gsfc.nasa.gov/xanadu/xspec/manual/node293.html}}.
We start with the Band function, and use the other models based upon the residuals of the data and model fit. 
The functional form of the Band model used to fit the photon spectrum is given in Equation \ref{eq:Band_function}
\citep{Band:1993},

\begin{widetext}
\begin{equation}\label{eq:Band_function}
  \mathop N\nolimits_{{\rm{Band}}}(E)  = \left\{ {\begin{array}{*{20}{c}}
  {K_{B}{{\left( {E/100} \right)}^\alpha }\exp \left( { - E\left( {2 + \alpha } \right)/{E_{{\rm{p}}}}} \right)}&{{\rm{,\ }}E < {E_{\rm{b}}}}\\
  {K_{B}{{\left\{ {\left( {\alpha  - \beta } \right){E_{{\rm{p}}}}/\left[ {100\left( {2 + \alpha } \right)} \right]} \right\}}^{\left( {\alpha  - \beta } \right)}}\exp \left( {\beta  - \alpha } \right){{\left( {E/100} \right)}^\beta }}&{{\rm{,\ }}E \ge {E_{\rm{b}}}}
\end{array}} \right.\ ,
\end{equation}
\end{widetext}

Other models include 
\sw{blackbody}\footnote{\sw{blackbody} model: \url{https://heasarc.gsfc.nasa.gov/xanadu/xspec/manual/node136.html}} (BB) 
and a power law with two breaks (\sw{bkn2pow})\footnote{\sw{bkn2pow} model: \url {https://heasarc.gsfc.nasa.gov/xanadu/xspec/manual/node141.html}} to model the broad band spectrum.

\begin{widetext}
\begin{equation}\label{eq:bkn2pow_function}
  \mathop N\nolimits_{{\rm{bkn2pow}}}(E)  = \left\{\begin{array}{*{20}{c}}
  {K {E}^{-\alpha_{\rm{1}}}} & {{\rm{,\ }}E \le {E_{\rm{1}}}}\\
  {K {E_{\rm{1}}}^{\alpha_{\rm{2}} - \alpha_{\rm{1}}} {E}^{-\alpha_{\rm{2}}}} &{{\rm{,\ }}  {E_{\rm{1}} \le E \le {E_{\rm{2}}}}}\\
{K {E_{\rm{1}}}^{\alpha_{\rm{2}} - \alpha_{\rm{1}}}{E_{\rm{2}}}^{\alpha_{\rm{3}} - \alpha_{\rm{2}}} {E}^{-\alpha_{\rm{3}}}} & {{\rm{,\ }E \ge {E_{\rm{3}}}}}\\
\end{array} \right.\ ,
\end{equation}
\end{widetext}

For the Comptonization model proposed by \citealt{Titarchuk:2012}, \sw{XSPEC} local model \sw{grbcomp} is fit to the photon spectrum\footnote{\sw{grbcomp} model: \url{https://heasarc.nasa.gov/docs/xanadu/xspec/models//grbcomp.html}}. The 
pivotal parameters of this model are temperature of the seed blackbody spectrum ($kT_s$), the 
bulk outflow velocity of the thermal electrons ($\beta$), the electron temperature of the sub-relativistic outflow 
($kT_e$), and the  energy index of the Green's function with which the formerly Comptonized spectrum is convolved ($\alpha_b$).
Using the normalization of the \sw{grbcomp} model, the apparent blackbody radius can be obtained. To avoid degeneracy in the parameters 
or the case when parameters are difficult to constrain, we froze some of the parameters.

To fit the spectra with synchroton radiation model we implemented a table model for XSPEC. We 
assumed a population of electrons with a power-law energy distribution (as a result of acceleration) 
$dN/d\gamma \propto \gamma^{-p}$ for $\gamma>\gamma_{m}$, where $\gamma_{m}$ is the minimum Lorentz factor of electrons. The cooling of electrons by synchrotron radiation is considered in slow and fast cooling regimes depending on ratio between $\gamma_{m}$ and the cooling Lorentz factor $\gamma_{c}$ \citep{Sari1998}. We computed the resulting photon spectrum of population of electrons assuming that the average electron distribution at a given time is $dN/d\gamma \propto \gamma^{-2}$ for $\gamma_{\rm c}<\gamma<\gamma_{\rm m}$ and $dN/d\gamma \propto \gamma^{-p-1}$ for $\gamma>\gamma_{\rm m}$ in fast cooling regime while $dN/d\gamma \propto \gamma^{-p}$ for $\gamma_{m}<\gamma<\gamma_{c}$  and $dN/d\gamma \propto \gamma^{-p-1}$ for $\gamma>\gamma_{\rm c}$ in slow cooling regime. The synchrotron model is made of four free parameters: the ratio between characteristic Lorentz factors $\gamma_{\rm m}/\gamma_{\rm c}$, the peak energy of $F_{\nu}$ spectrum $E_{\rm c}$ which is 
simply the energy which corresponds to the cooling frequency, the power-law index of electrons distribution p and the normalization. We built the table model for the range of $0.1 \le \gamma_{\rm m}/\gamma_{\rm c} \le 100$ and $2 \le p \le 5$.

\begin{figure*}
\centering
\includegraphics[scale=0.325,angle=270]{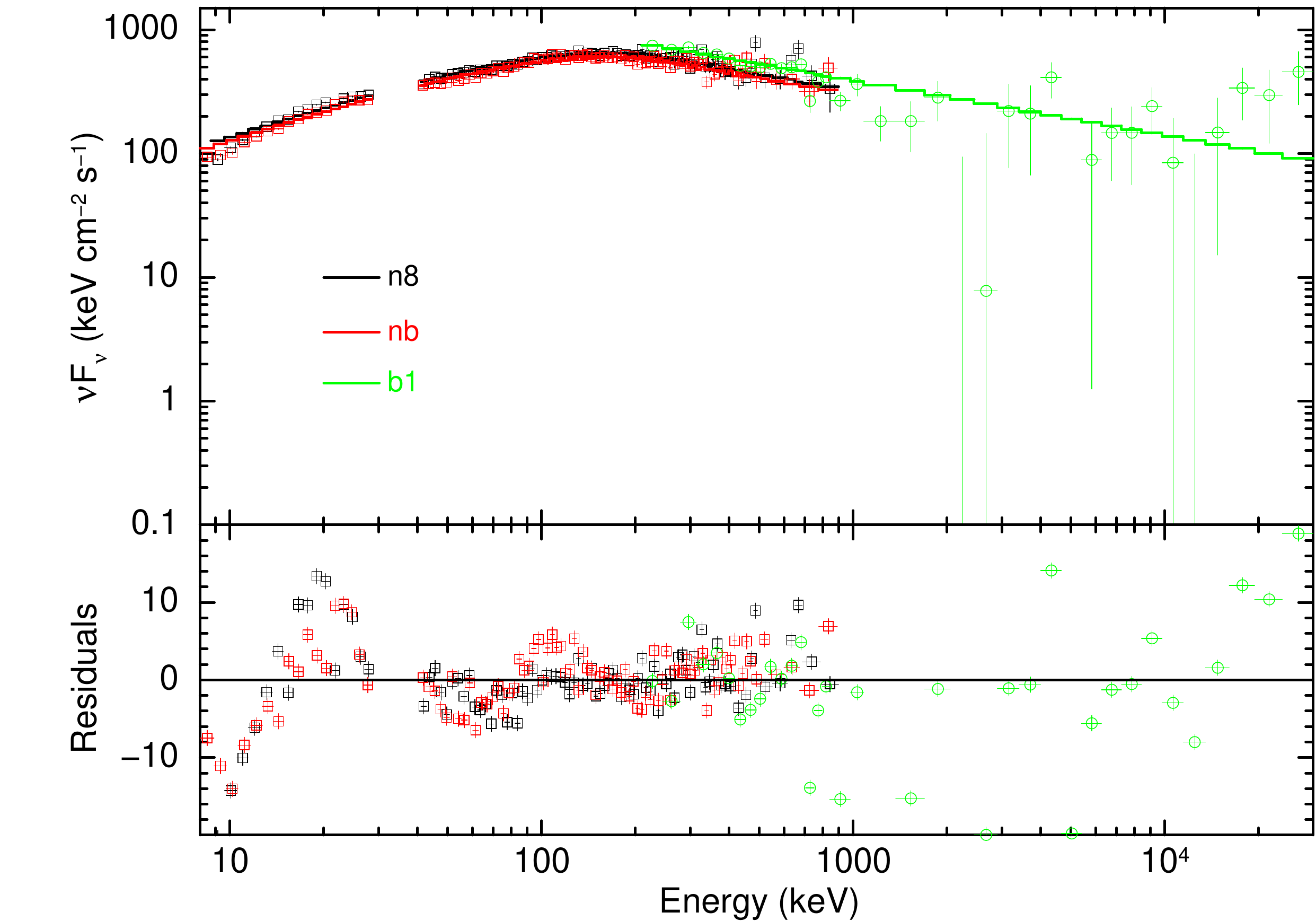}
\includegraphics[scale=0.325,angle=270]{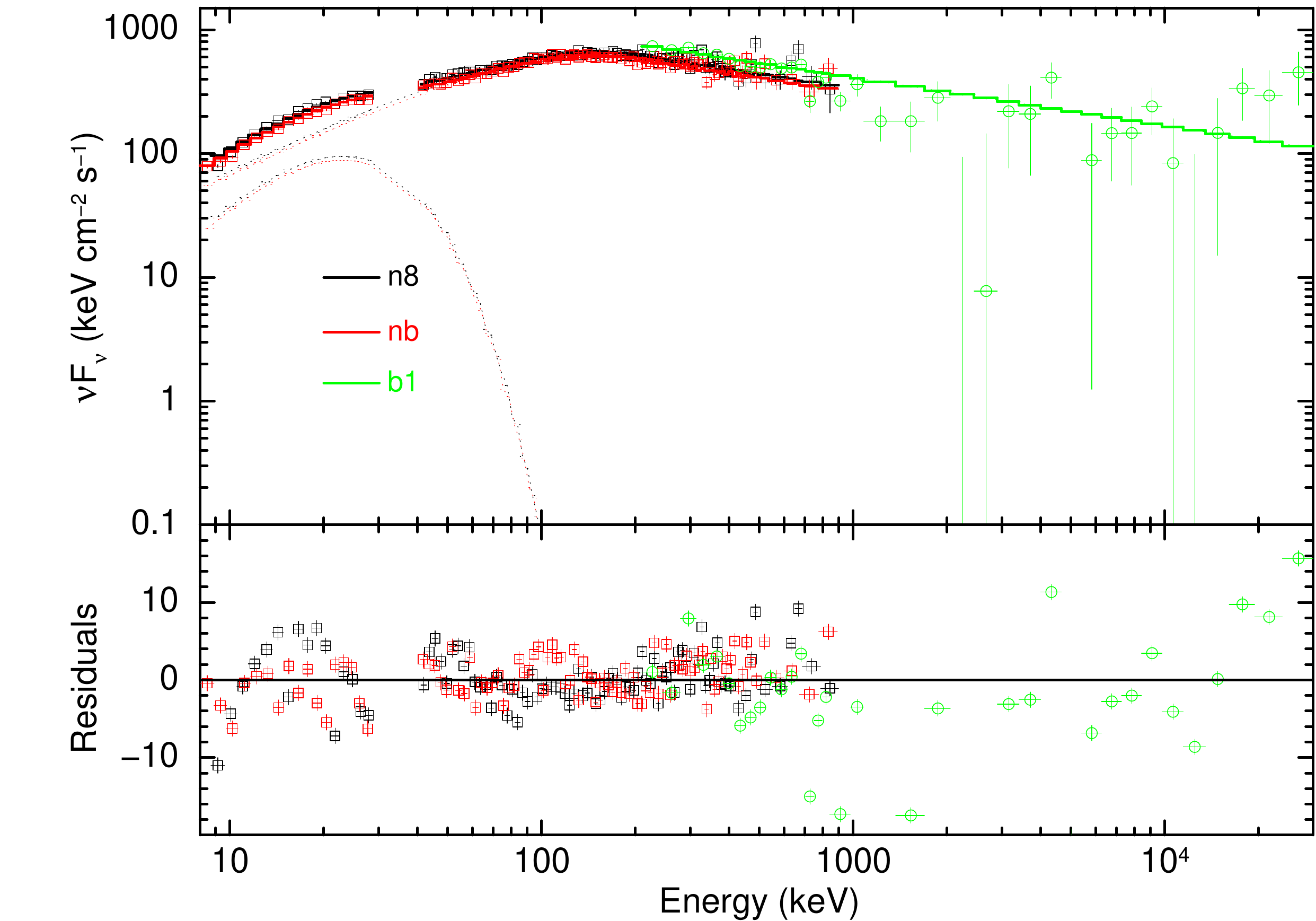}

\includegraphics[scale=0.325,angle=270]{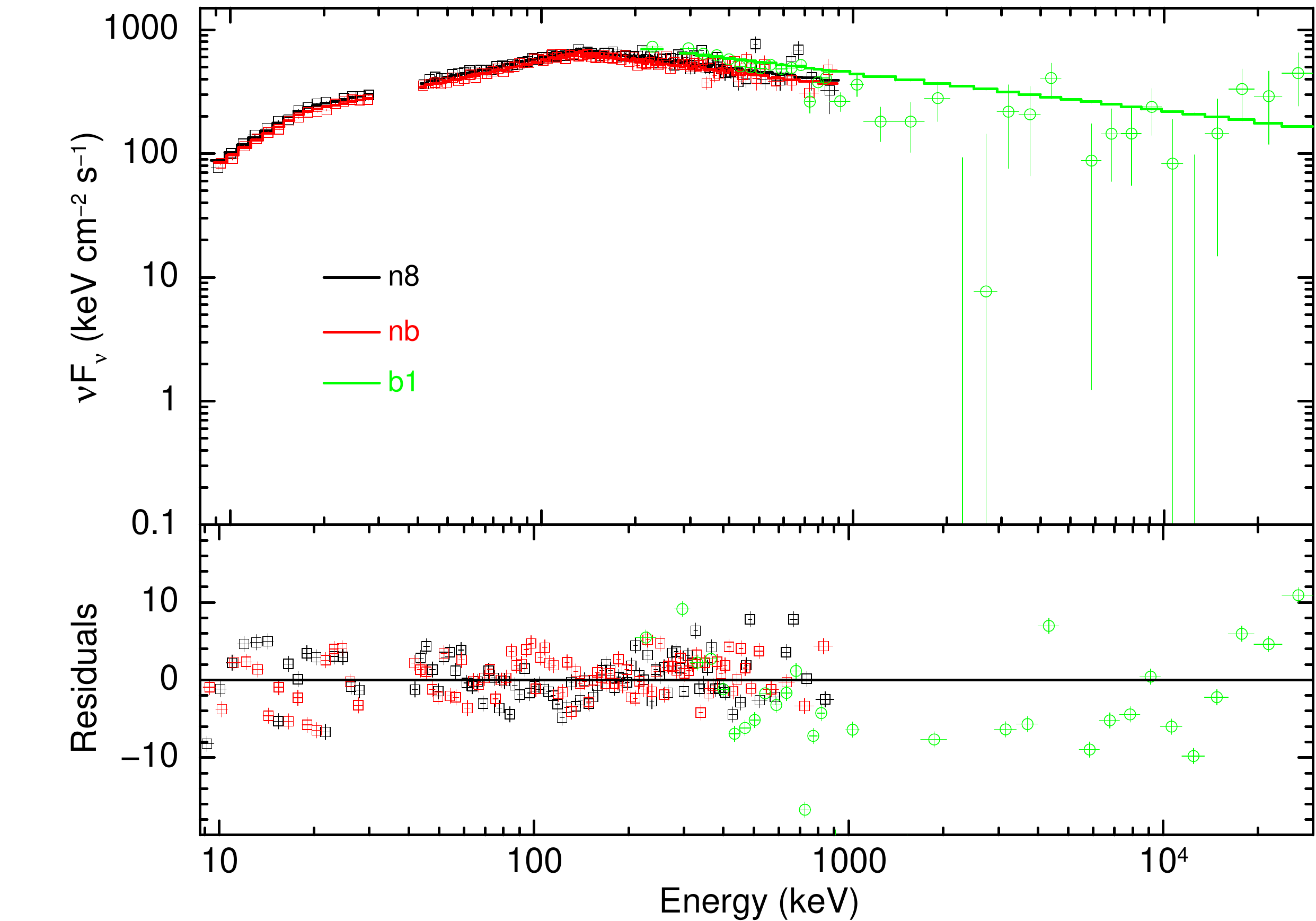}
\includegraphics[scale=0.325,angle=270]{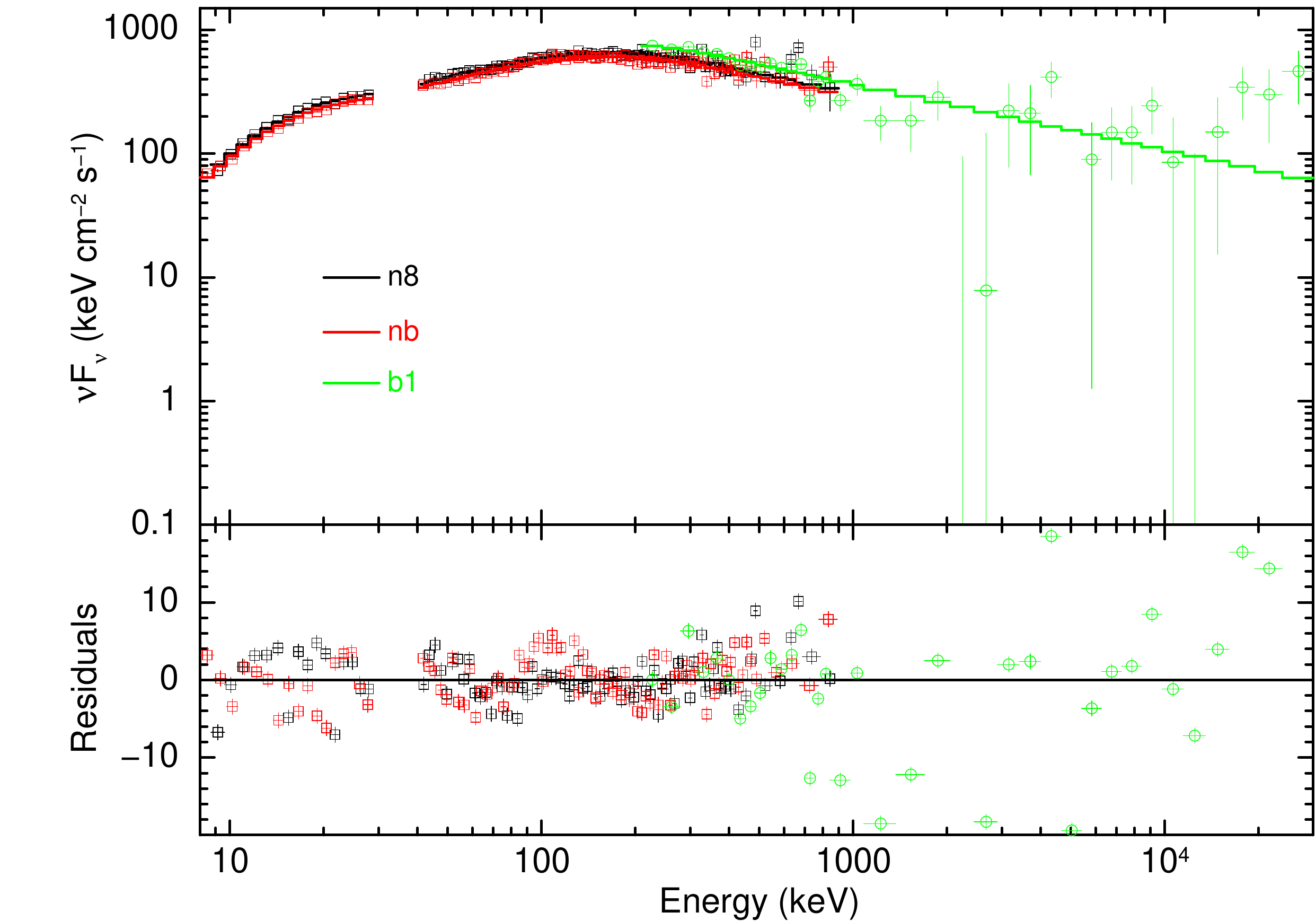}
\includegraphics[scale=0.325,angle=270]{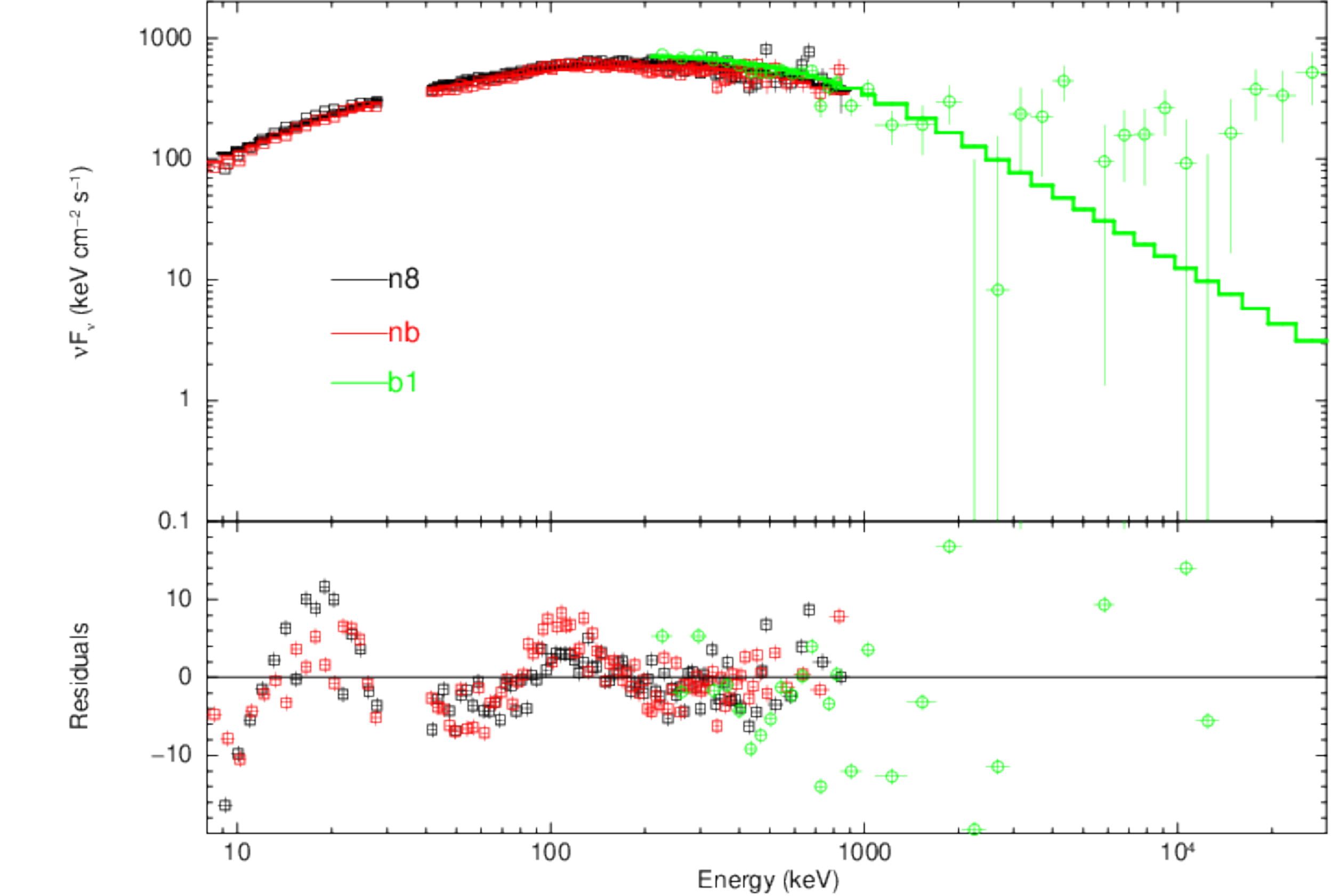}
\caption{
The time integrated $\nu F_\nu$ plot of Episode 1 of \thisgrb\ (-5 s to 246 s) along with the best fit models. Residuals to  the fit are shown in the bottom panels. The  Band model (top left) shows a low energy feature and  can be
either modeled by a blackbody (top right) or by adding another powerlaw at the low energy (middle left).
The best fit model is the Comptonization model \sw{grbcomp} (middle right). The synchrotron model is shown in bottom panel.}
\label{fig:Episode_1_spectra}
\end{figure*}

To fit the time integrated spectrum of  Episode 1, we use the four models described above: (i) Band, (ii) Band + blackbody (BB), (iii) broken power law (\sw{bkn2pow}), and (iv) the Comptonization model (\sw{grbcomp}).
The best fit parameters of the tested models are reported in Table \ref{tab:episode1_specfitting}. 
The $\nu F_\nu$ plots, along with the residuals, are shown in Figure \ref{fig:Episode_1_spectra} for the four models used.  The Band fit shows  deviations at lower energies signifying deviations  from the  power law that 
is used to model the spectra from the low energy threshold of  8~keV to the peak energy of the Band function ($\sim$150~keV). We have examined these residuals in detail by
progressively  raising the low energy threshold to 40~keV (greater than the energy in which the deviations are seen) and extrapolating  the 
model to the low energies. The low energy features could still be seen. 
The presence of such an anomaly is found in previous studies \citep{Tierney:2013} and we conclude that 
a separate low energy feature in the spectra can exist. We included a blackbody (BB) along with the Band model for higher energies, but the residuals
still show a systematic hump.  We also tried another model for this feature of the spectrum: a powerlaw with two breaks (\sw{bkn2pow} model). 
The 
\sw{bkn2pow} model is preferable as the \sw{pgstat} is much less for the same number of parameters ($\Delta pgstat = 41$).
The presence of a narrow residual hump also necessitate  the need of a sharper break than a smooth blackbody curvature. This also hints that
the break does not evolve much in time and thus, is not smeared out. The Comptonization model (\sw{grbcomp}), however,  gives the  best fit for the time integrated spectrum of Episode 1.
All the parameters of this model other than $\beta$, \sw{fbflag} and $log(A)$ were left free. The parameter $\beta$ was frozen to the value $0.2$ to 
ignore terms $O(\sim\beta^2)$ and 
\sw{fbflag} was set to zero to include only the first order bulk Comptonization term. A value of 3.9 for  $\gamma$ shows 
deviation for the seed photons from the seed blackbody spectrum. However, during the time resolved spectral analysis as reported in the next subsection we kept 
it fixed to 3 to consider a blackbody spectrum for the seed photons.

The isotropic energy ($E_{\gamma, iso}$) is calculated in the cosmological frame of the GRB by
integrating the observed energy spectrum over $1~\mathrm{keV}/(1+z)$ to $10 MeV/(1+z)$. The tested models differ significantly only at 
low energies  and yield similar $E_{\gamma, iso}$. We have $E_{\gamma, iso} = 2.2\times10^{53}$~erg for the
best fit model \sw{grbcomp}. The $\Gamma_0~-~E_{\gamma,~iso}$ correlation between the initial Lorentz factor and the isotropic energy can be used to estimate  $\Gamma_0$ of the
fireball ejecta \citep{Liang:2010}. The estimated $\Gamma_0 $ is $ 392^{+38}_{-34}$.
The errors are propagated from the normalization and slope of the
correlation.
The episode 2 could be spectrally well described by a simple power-law and the best fit power-law index is $1.90_{-0.06}^{+0.07}$ for \sw{pgstat} 274 for 229 degrees of freedom.

\begin{table*}
\caption{The best fit parameters for the  time integrated spectrum of \thisgrb, Episode 1}
\centering
\label{tab:episode1_specfitting}
\begin{tabular}{c|ccccccccc}
\hline
&Model  &    &    &  & &  \\ \hline
&Band &$ \alpha $&  $\beta$  & $E_p$ (keV) & \sw{pgstat}/dof \\ \hline
&     &$-1.15_{-0.01}^{+0.01}$&$-2.40_{-0.03}^{+0.03}$ &$163_{-3}^{+3}$&  $2786 / 335$& &\\ \hline
& BB+Band & $\alpha$ &$\beta$ & $E_p$ (keV) &  $kT_{BB}$ (keV) &  \sw{pgstat}/dof\\ \hline
&         &  $-0.75_{-0.03}^{+0.04}$ & $-2.40_{-0.03}^{+0.04}$ &$150.0_{-2.6}^{+2.5}$ & $5.8_{-0.1}^{+0.1}$ & 1465/333         \\\hline
&\sw{bkn2pow}& $\alpha_1$ &$\alpha_2$ & $\alpha_3$ & $E_1$ (keV) &  $E_2$ (keV)  & \sw{pgstat}/dof \\ \hline
&         &   $0.48_{-0.07}^{+0.06}$    & $1.464_{-0.005}^{+0.006}$  &  $2.33_{-0.02}^{+0.02}$ &  $17.2_{-0.4}^{+ 0.4}$ & $132.7_{-2.3}^{+ 2.3}$ &1421/333\\ \hline
&\sw{grbcomp} & $kT_{s}$ (keV)      &   $kT_{e}$ (keV)      &   $\tau$  &  $\alpha_b$  & $R_{ph}$ ($10^{10}$ cm) & \sw{pgstat}/dof\\\hline
& & $4.8_{-0.4}^{+0.5}$  & $55_{-3}^{+3}$ & $4.15_{-0.15}^{+0.18}$ &   $1.52_{-0.04}^{+0.04}$ & $8.0_{-2.0}^{+2.2}$   &1281/332 \\\hline
&\sw{Synchrotron model}& $\gamma_{m}/\gamma_{c}$  & $E_{c}$ (keV) & p & \sw{pgstat}/dof &    &  \\ \hline
&  & $4.06_{-0.08}^{+0.03}$ & $21.6_{-0.4}^{+0.2}$ & $>4.82$ & $2520/335$ \\\hline

\end{tabular}
\end{table*}

\subsection{Parameters evolution}
\begin{figure*}
\centering
\includegraphics[scale=0.35]{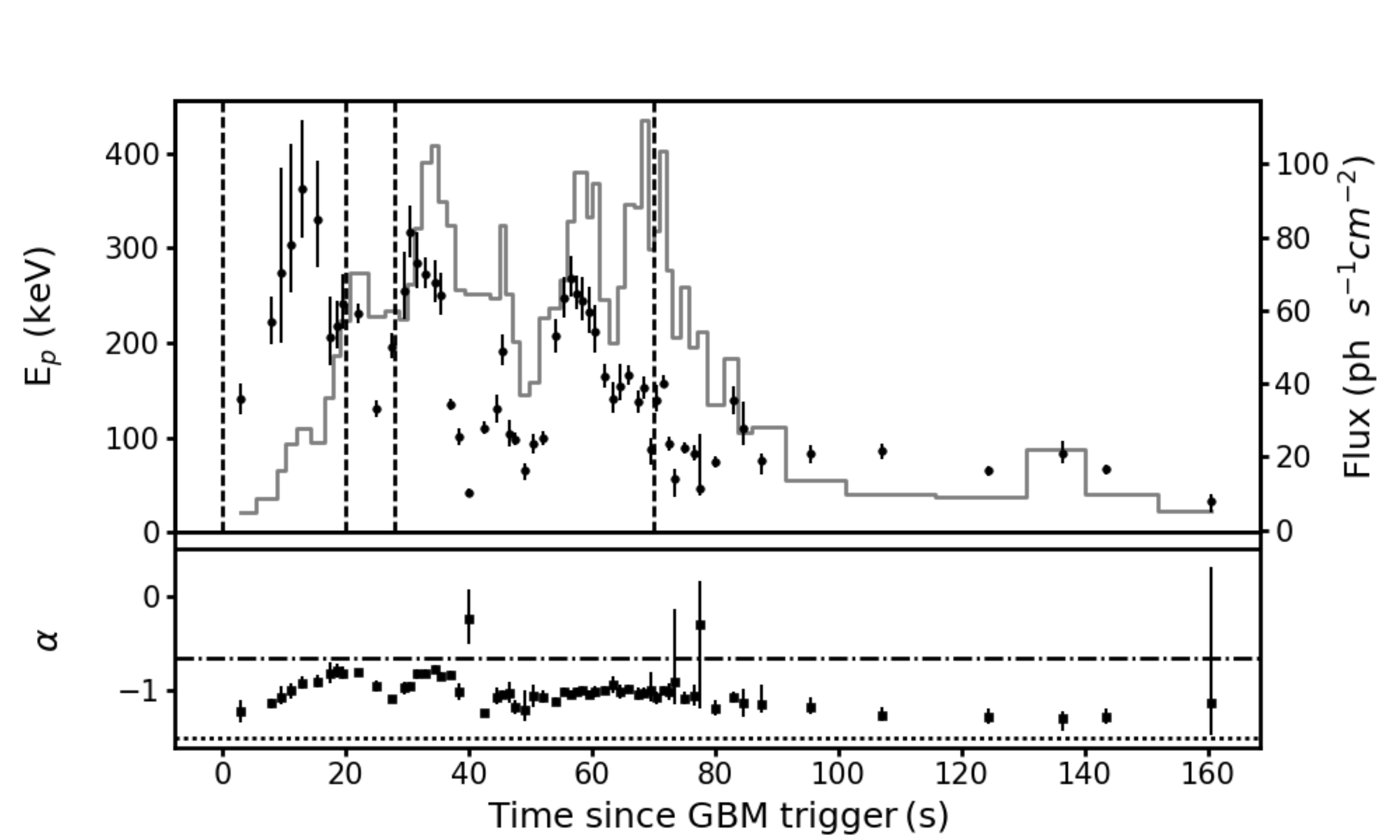}
\includegraphics[scale=0.35]{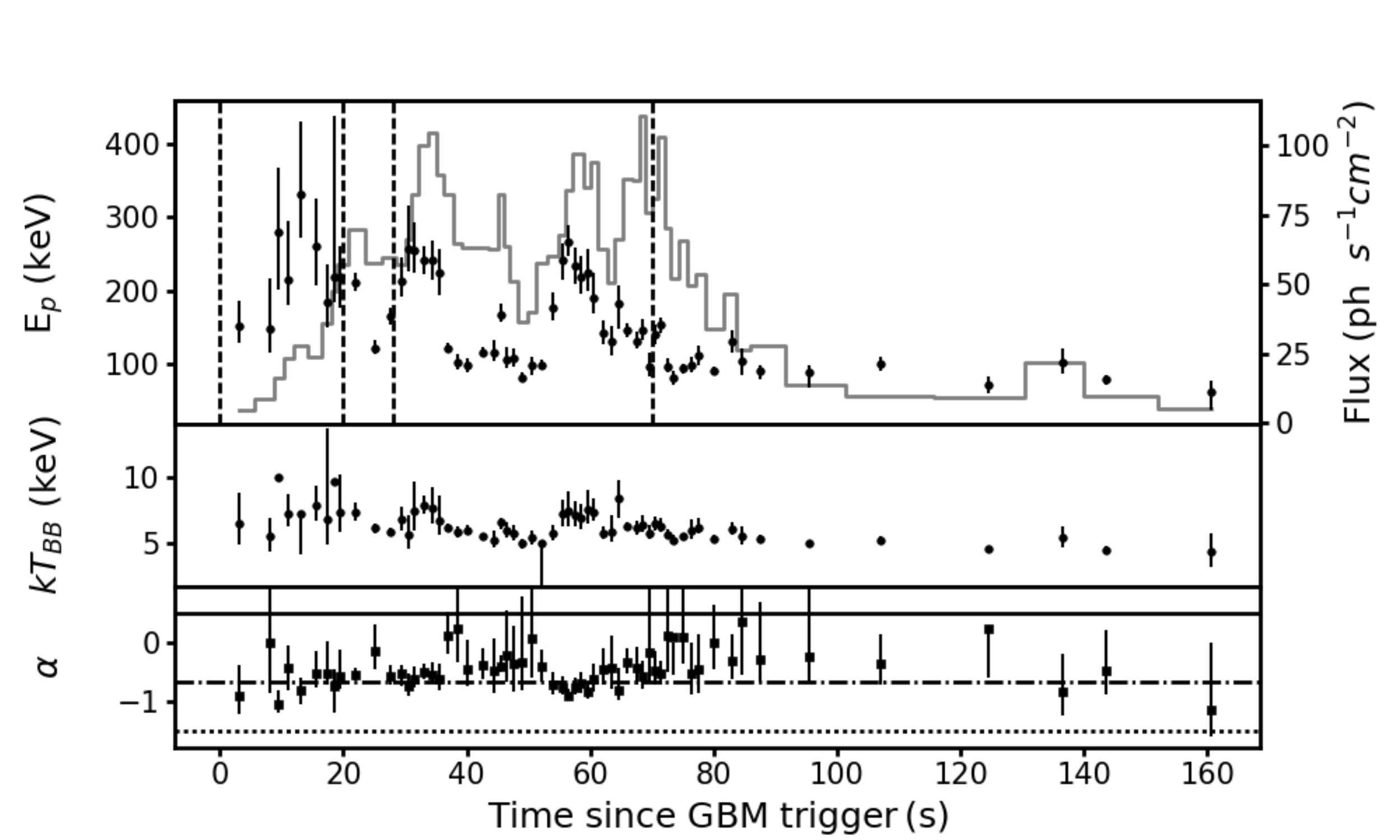}
\includegraphics[scale=0.35]{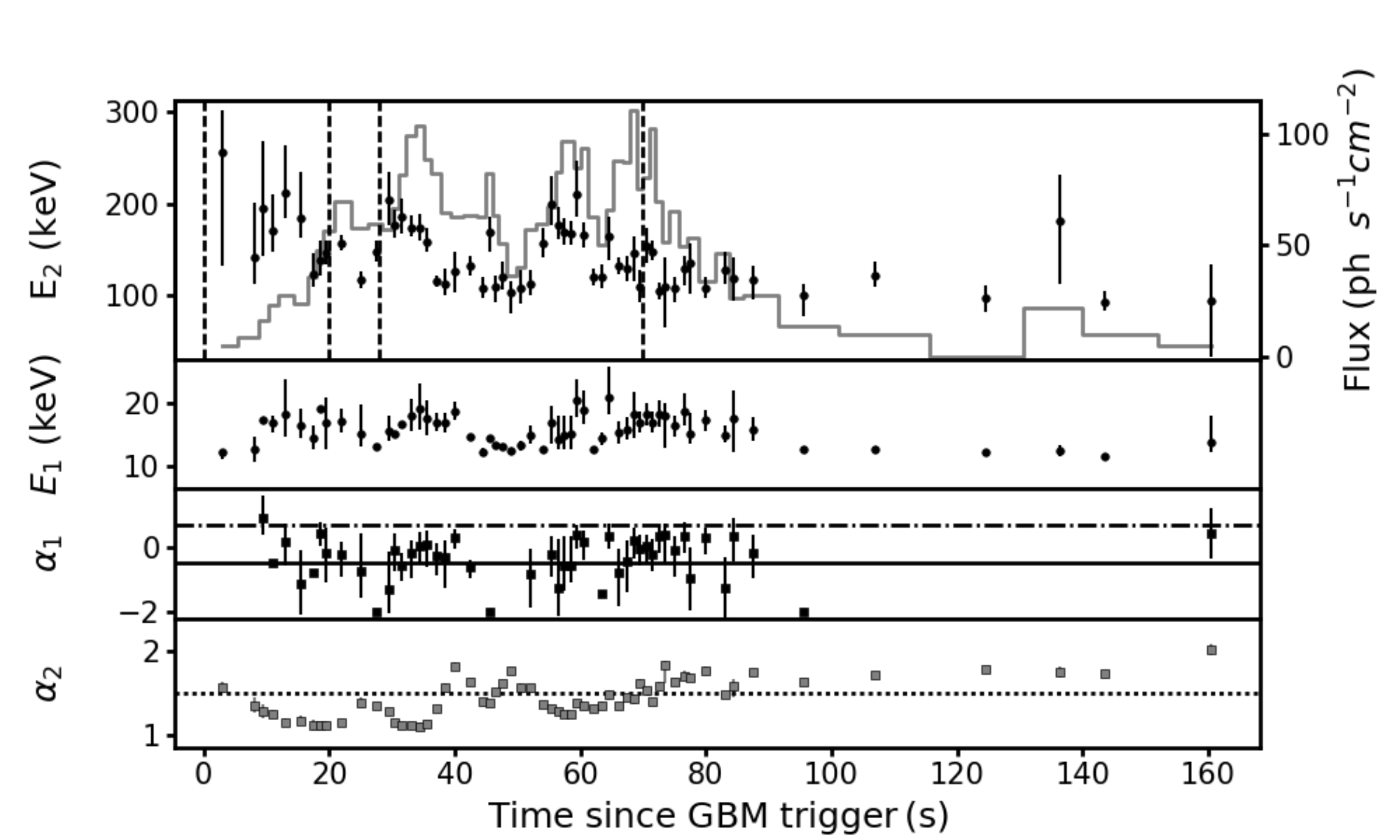}
\includegraphics[scale=0.35]{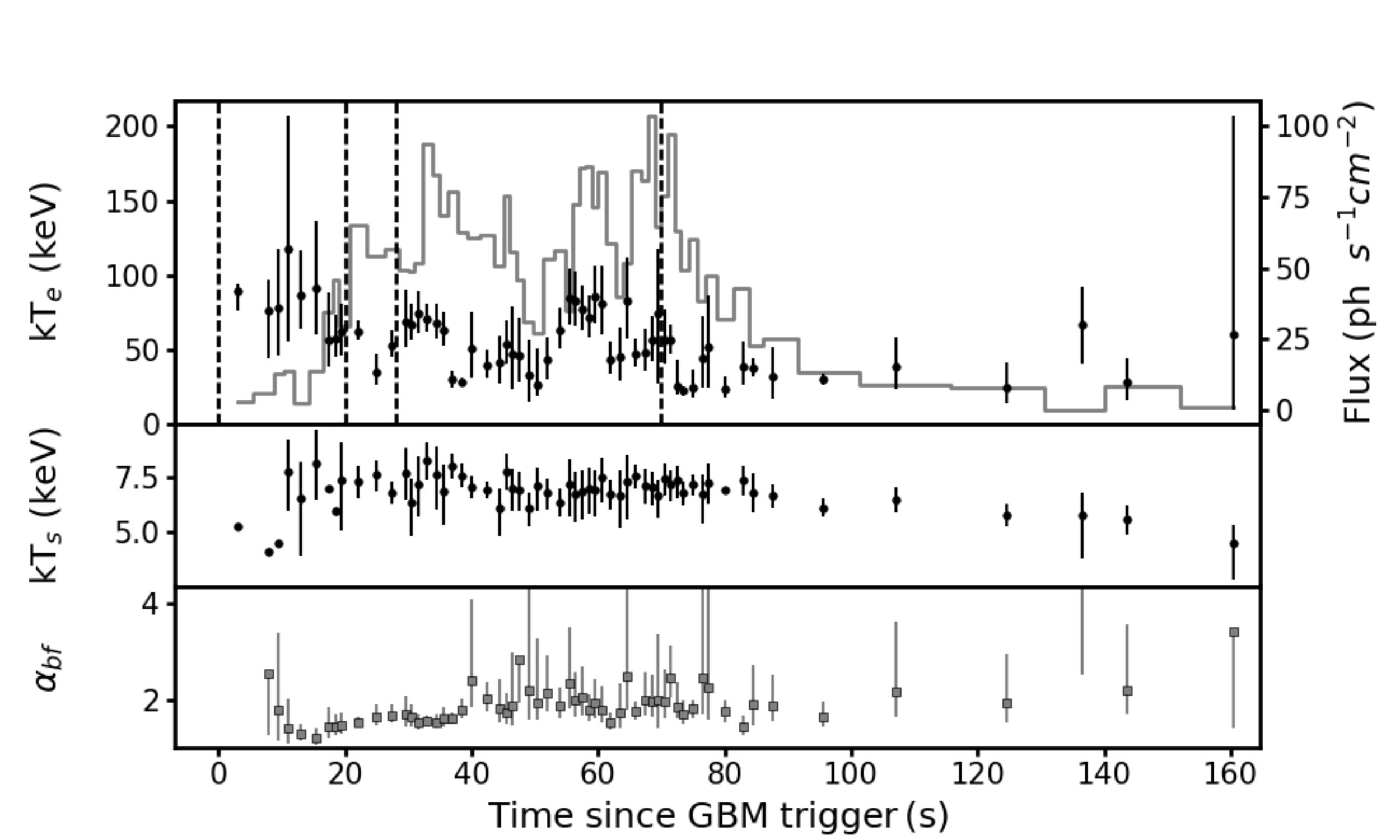}
\caption{$Upper ~ Left:$   The evolution of peak energy  and the low energy spectral index of the Band model fit to time resolved spectra. The low energy spectral indices lie
between fast   and slow cooling   limits of synchrotron radiation
(indicated by dotted line and dot-dashed line, respectively). {\bf Intervals used for polarization measurement are segregated by vertical dotted lines, to guide the eye.} The photon flux is shown as a grey histogram (right hand side scale). $Upper ~ Right$:  The evolution of peak energy,  blackbody temperature and low energy index 
when the low energy anomaly seen in the  Band function fit is modeled by an additional blackbody. 
Low energy indices are now  above the slow cooling limit, but below the limit of 
jitter radiation (black solid line).  $Lower ~ Left$:  The Evolution of two break energies and indices for a broken power law fit.    The index $-\alpha_2$ in this case shifts closer to the fast cooling line or even softer than this limit. The energy index $\alpha_1$
below the low energy break is harder with a median value 1/3. The break energy is also low and falls in the range between 15 - 25~keV. $Bottom ~Right:$  Evolution of  electron temperature, seed photon temperature and
high energy index $\alpha_{bf}$   of the Comptonization model \sw{grbcomp}.  }
\label{fig:params_evolution}
\end{figure*}

We performed time resolved analysis to capture
the variations in the  spectral properties by dividing  the data into
time segments based on the Bayesian blocks made from the 100 - 400~keV light curve. 
We test the models that were used for the  time-integrated analysis. The deviation from Band spectrum at low energies observed in the time-integrated spectrum is also present in the  time-resolved 
bins, thus justifying the need to use more complex models than the simple Band function.  We, however, find that the three additional models (Band+BB, bknpwl, and grbcomp) represent the time resolved spectral data equally well.  The evolution of the model parameters is shown in Figure \ref{fig:params_evolution}.
In \fermi-GBM, energy range 
8-20~keV is divided into 12 energy channels and thus it provided sufficient data points to constrain the 
power-law or BB temperature whenever BB peak or low energy break falls well within these energy channels.
In case of \sw{bkn2pow}, the low energy index is very steep ($\sim-3$) when the break energy $E_1$ found to be 
close to the lower edge of the detection band. For such cases, we froze the index to $-3$ to obtain constraints on the other parameters of the 
model. When we fit with the Band function,  the peak energy $E_p$ shows three pulse like structures in their temporal evolution  and these can be identified as showing a hard to soft (HTS) evolution for peaks in the photon flux. The first structure in E$_p$ (at $\sim$10 s) can be associated with the enhancement seen in the $>$1 MeV flux (Figure \ref{fig:GBM_LC}, top panel). 
When the low energy
feature is modeled using \sw{bkn2pow}, almost all the peak energies fall below  200~keV. 
The variation in the low energy break ($E_1$) remains concentrated in a very narrow  band ($10 ~-~ 20$~keV). The best fit parameters
for all the models are presented in Table \ref{tab:specfitting}. 

\begin{figure}
\includegraphics[scale=0.5]{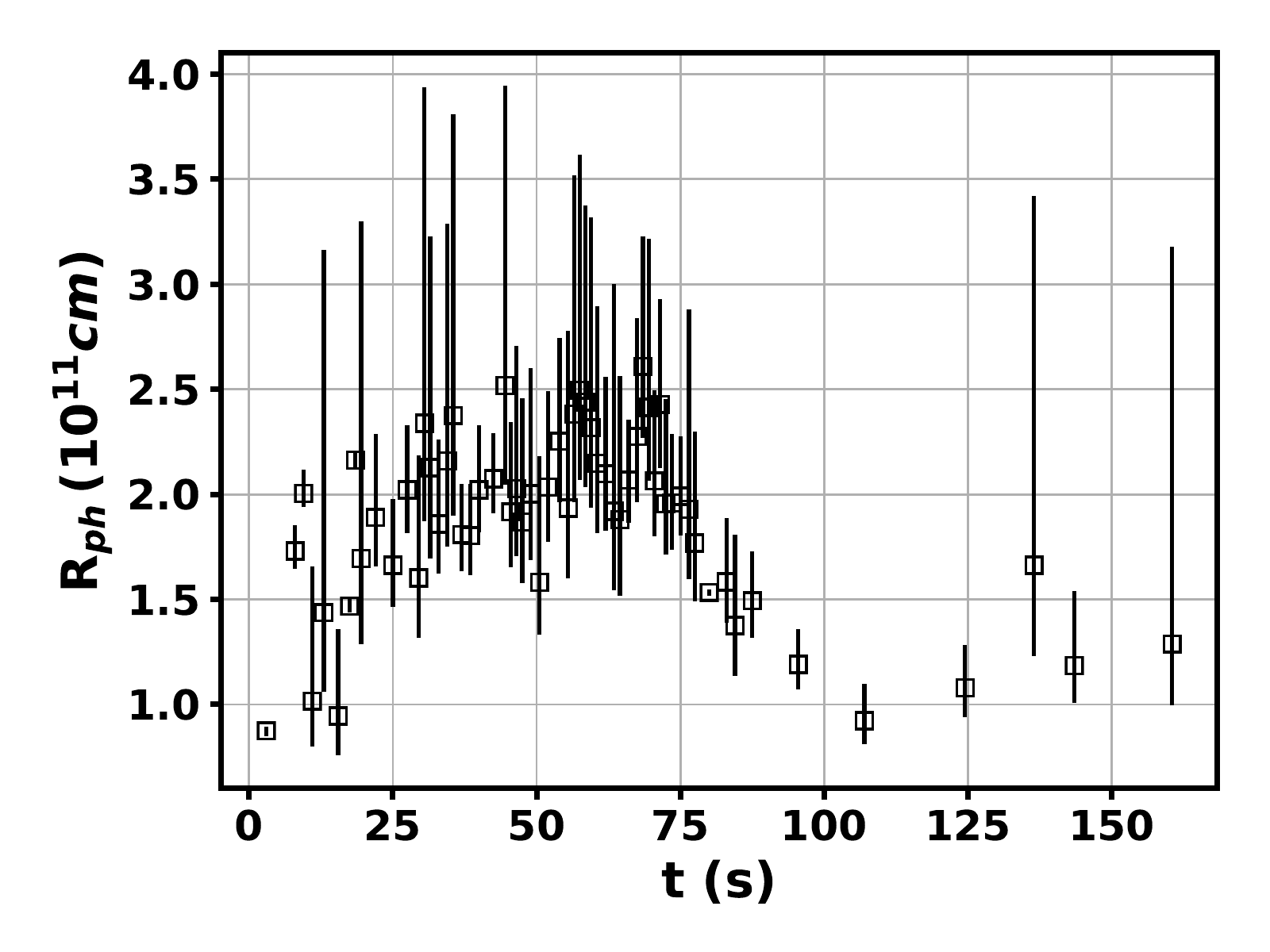}
\includegraphics[scale=0.5]{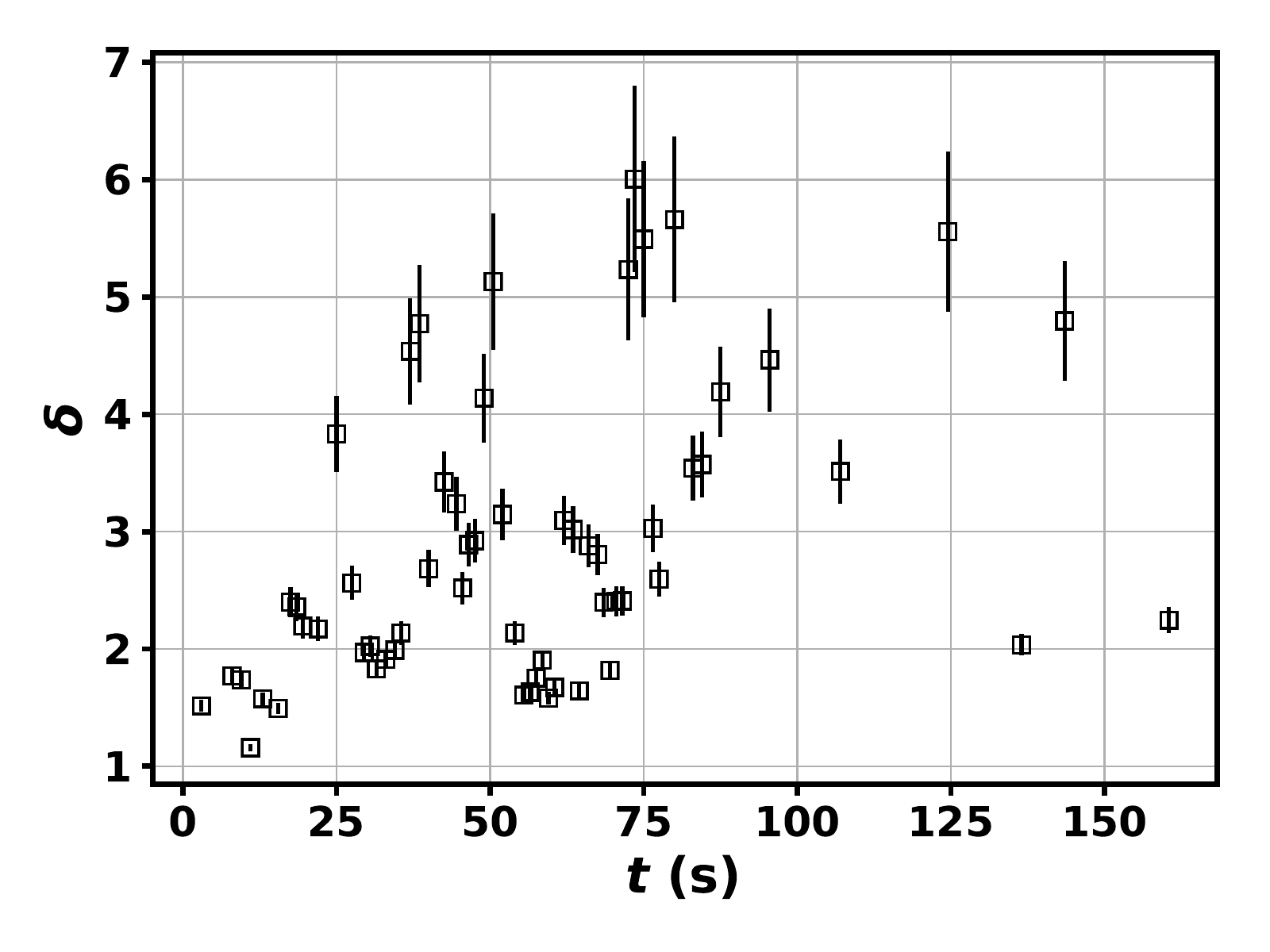}
\caption{ The evolution of the photospheric radius (upper panel) and the bulk parameter (lower panel) obtained from the \sw{grbcomp} model.   }
\label{fig:grbcomp_Rph_delta}
\end{figure}

The evolution of the derived parameters 
of \sw{grbcomp} model such as photospheric radius ($R_{ph}$) and bulk parameter ($\delta$) is shown in Figure \ref{fig:grbcomp_Rph_delta}. 

\begin{figure*}
\includegraphics[scale=0.50]{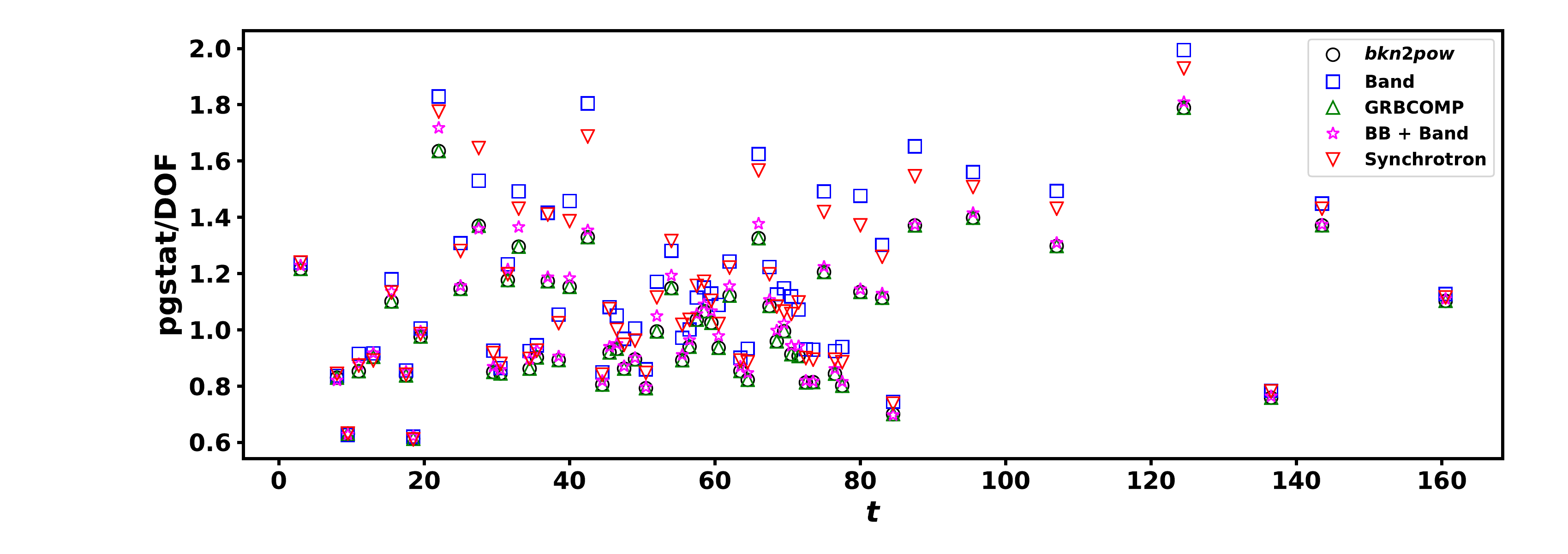}
\caption{The variation of the statistic \sw{pgstat} with time. A low energy break is preferable in almost all  the bins. All three models: \sw{BB+Band}, \sw{bkn2pow}, synchrotron and \sw{grbcomp} have comparable \sw{pgstat} values and provide acceptable fits to the data in the time-resolved analysis. However, in time-averaged analysis, the \sw{grbcomp} model provides the best fit.}
\label{fig:pgstat_dof}
\end{figure*}

\subsubsection{Distribution of parameters}
\begin{figure*}
\centering
\includegraphics[scale=0.8,angle=0]{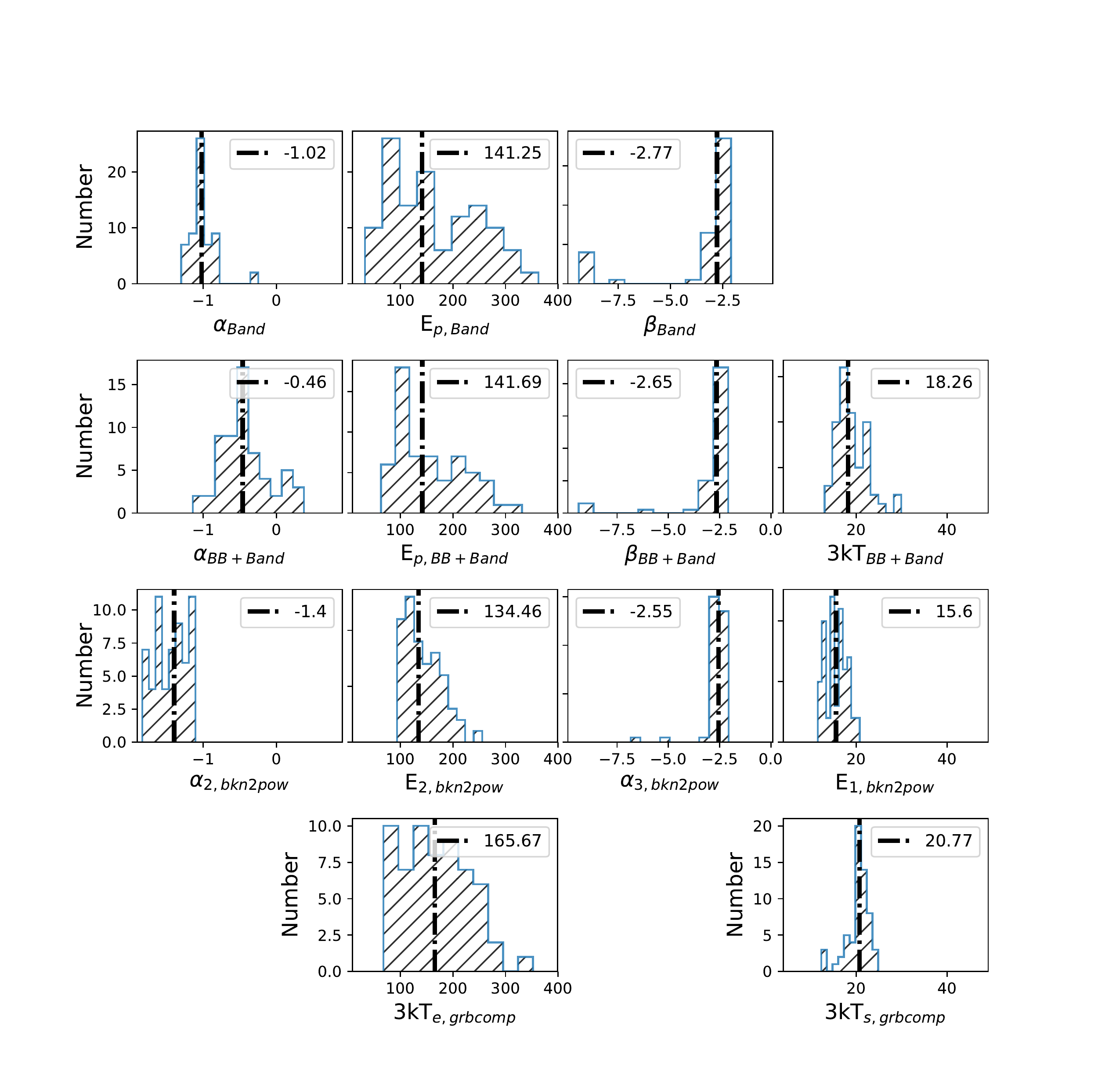}
  \caption{The distributions of best-fit parameters for \sw{Band}, \sw{bkn2pow} and \sw{grbcomp}  models. Each histogram consists of ten bins between minimum and maximum values. Thick vertical dot-dashed lines represent the median of the observed distribution of the parameters.}
\label{fig:Episode_1_parameter_distribution}
\end{figure*}

Despite the fact that the best fit to the time integrated spectrum is the Comptonization model (\sw{grbcomp}),
we have tested the models (i) Band, (ii) \sw{bkn2pow}, (iii) BB+Band and (iv) \sw{grbcomp} for the time resolved spectra. The distribution of the parameters is presented  
in Figure \ref{fig:Episode_1_parameter_distribution} and are summarized here. 

For the Band function, the  low energy index $\alpha$ is distributed with
a mean value $-1 ~(\sigma=0.13)$. The maximum value observed is $-0.25$ for a few time bins and 
in the other cases $\alpha$ remains mostly below $-0.8$. The values of low energy index 
agree with what is typically observed for long GRBs ($<\alpha> = -1$) which is, however, known to be
harder than the synchrotron radiation in the fast cooling regime. The high energy index follows a bimodal
distribution with a mean value of $-2.7 ~(\sigma =0.3)$ for the major chunk concentrated around $-3<\beta<-2$ 
and can be interpreted as values normally observed for GRBs. The peak energy $E_p$ is between 32~keV and 365~keV with a median 
$\sim 140~(\sigma = 83$)~keV. Its distribution also shows a bimodal nature which reflects the evolution of $E_p$.

In the case of \sw{bkn2pow}, we observed that when the lower break energy $E_1$ is near the lower edge of the detection band, the power-law index 
below $E_1$ is very steep. This is expected because there are only a few channels here and the fit probably gives an  unphysical result. We thus fixed the values to $-3$ as they are difficult to constrain by the fits and also
hamper the overall fitting process. Note that \sw{bkn2pow} model is defined with an \textit{a-priori} negative sign with it and we have to
be cautious while comparing it with the Band function where the indices are defined without a negative sign. For the
purpose of comparison we will explicitly reverse the signs of the \sw{bkn2pow} indices. The mean value of 
 $-\alpha_1$
is $\sim 0.3 (\sigma~= ~0.65$). Here, we have ignored values $>$ 1.5 as they form another part of the bimodal distribution
with low $E_1$, during the time when $E_1$ is near the lower edge of the GBM energy band. The mean value of $-\alpha_2$
is $\sim -1.4 ~ (\sigma = 0.22)$. This forms the second power law from $E_1$ to $E_2$. The third 
segment of the emission has power law index
$-\alpha_3$ with a mean of $\sim -2.6 ~(\sigma = 0.23)$.
The low energy break $E_1$ has a mean   $\sim 16 ~(\sigma = 2.3)$~keV and it falls in a narrow range of 11 to 20~keV. 
The mean $E_2$ 
is $\sim 140  ~(\sigma = 32)$~keV, comparable to the mean of Band function peak energy with a uni-modal distribution. The $E_2$ values are concentrated
in the range 93 to 350~keV.

When a blackbody is added, the fits have an improved statistics compared to  the Band function throughout the burst (see Figure \ref{fig:pgstat_dof}).
The blackbody temperature vary between $4~-~10$~keV with a median of $\sim 6.3 (\sigma = 1.12)$~keV. The $\alpha$ distribution 
has harder values, while $E_p$ and $\beta$ are similar to the  distributions of their counterparts in the Band function. The presence of
blackbody does not seem to move the peak energies significantly, albeit an upward shift is noticeable. 
In the \sw{grbcomp} model, the average temperature of the seed photons was 6.8 $(\sigma = ~0.8)$~keV. The electron temperature $kT_e$ was found to be 55 $(\sigma = 21)$~keV. We derived the 
photospheric radius from the normalization of the model. We found the photospheric radius $\sim 10 ^{11}$~cm. The \sw{grbcomp} model parameters are reasonable and a photospheric radius of $10^{11}$~cm is consistent with its predictions \citep{Frontera:2013}.

\subsubsection{Correlations of parameters}
\begin{figure*}
\centering
\includegraphics[scale=0.335,angle=0]{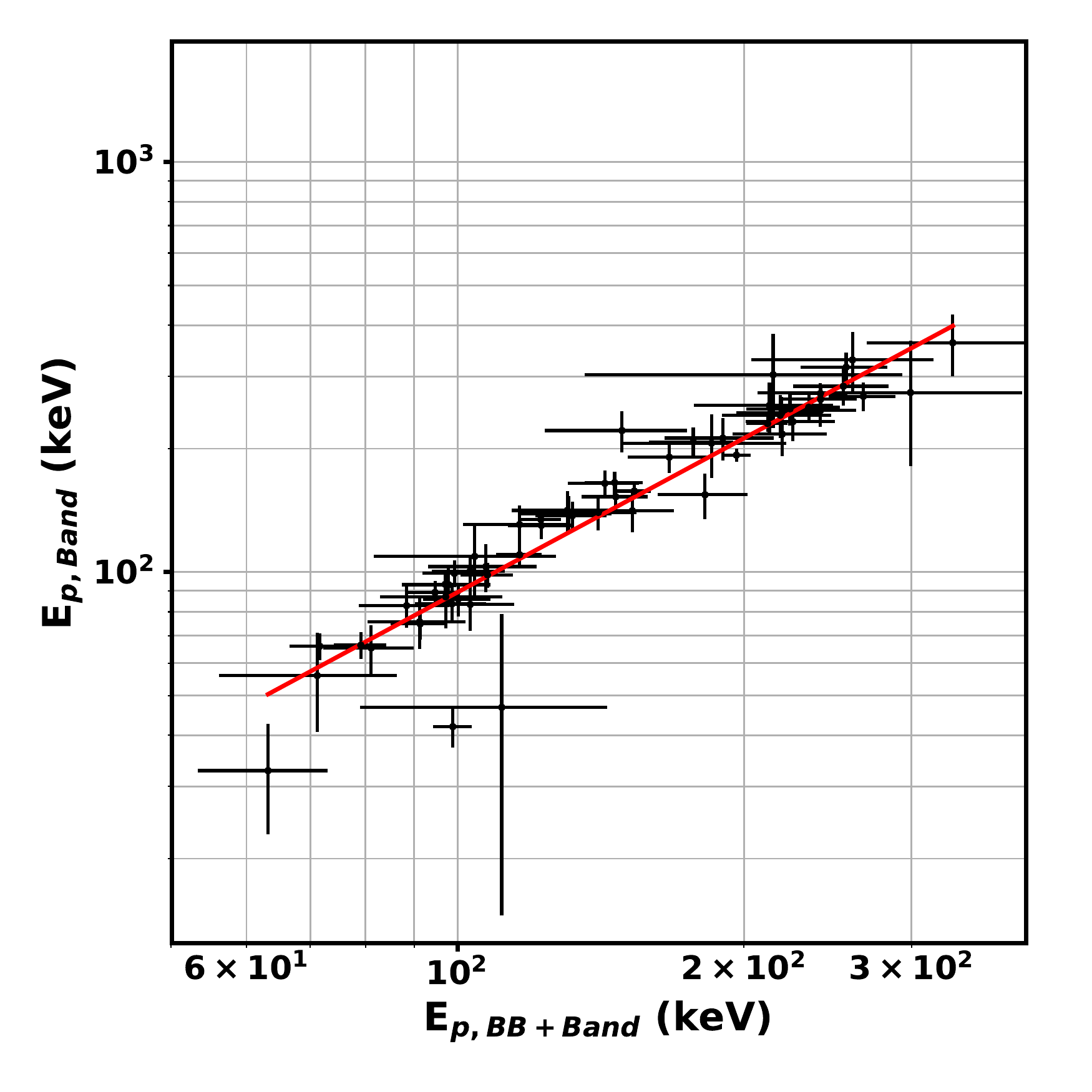}
\includegraphics[scale=0.335,angle=0]{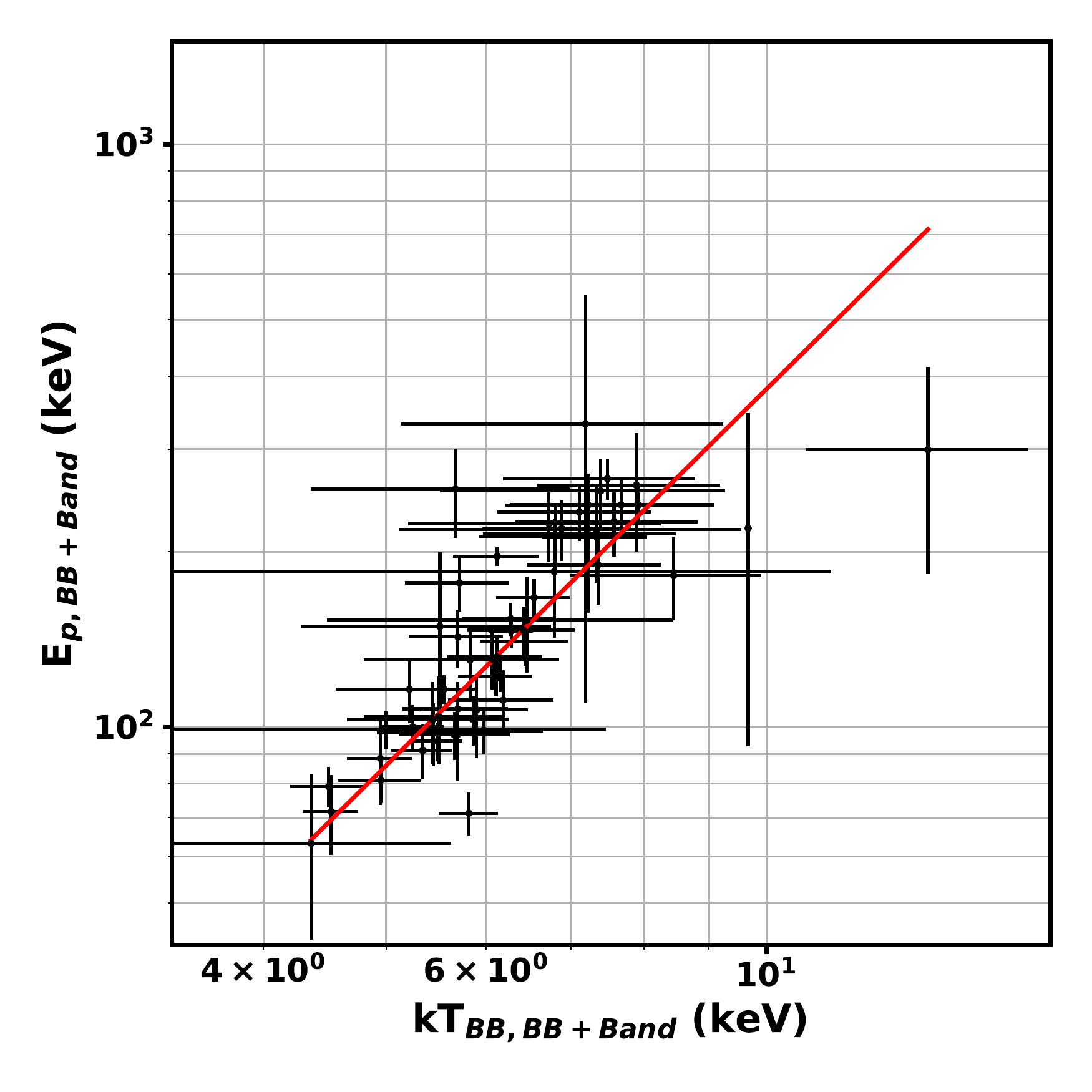}
\includegraphics[scale=0.335,angle=0]{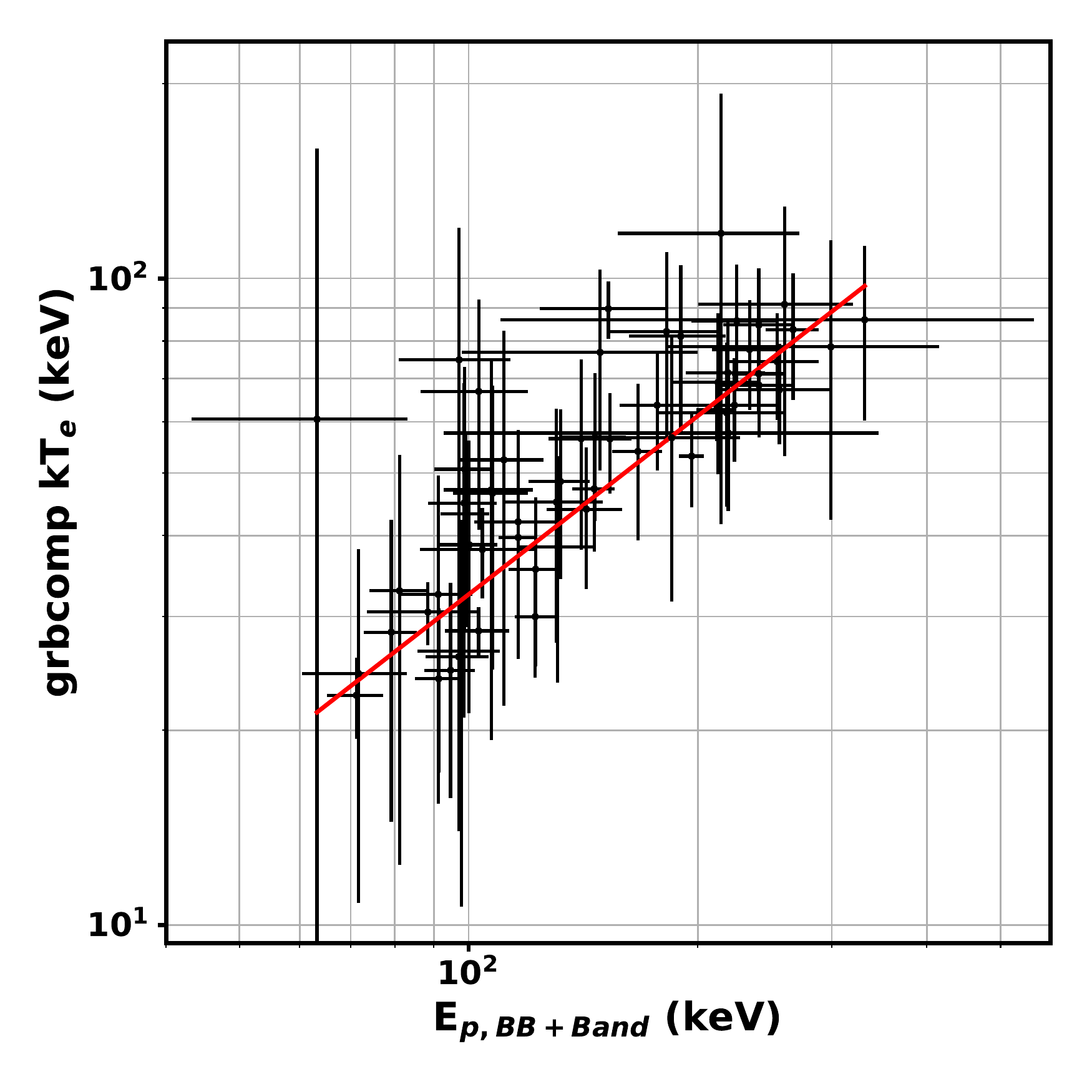}
\includegraphics[scale=0.335,angle=0]{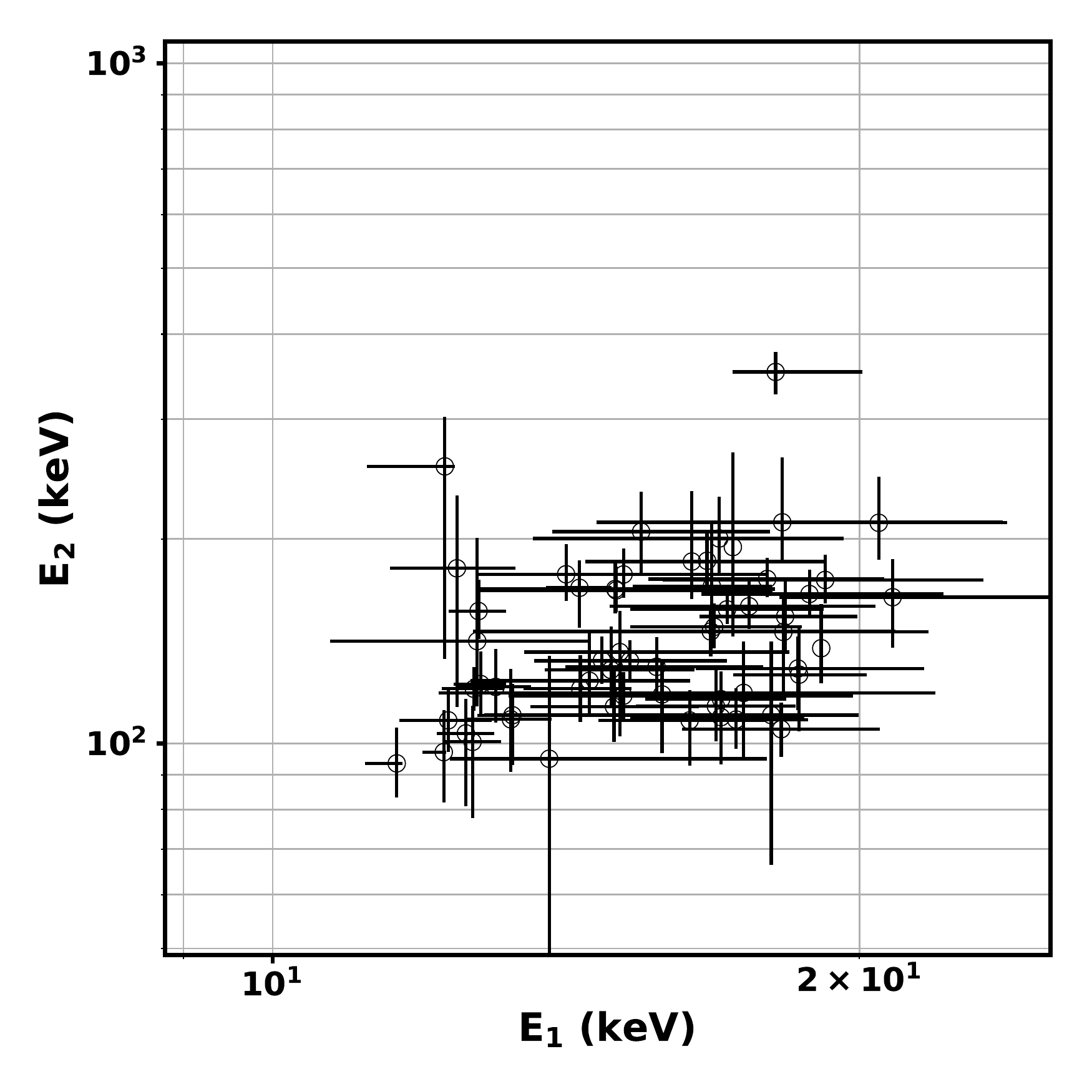}
\includegraphics[scale=0.34,angle=0]{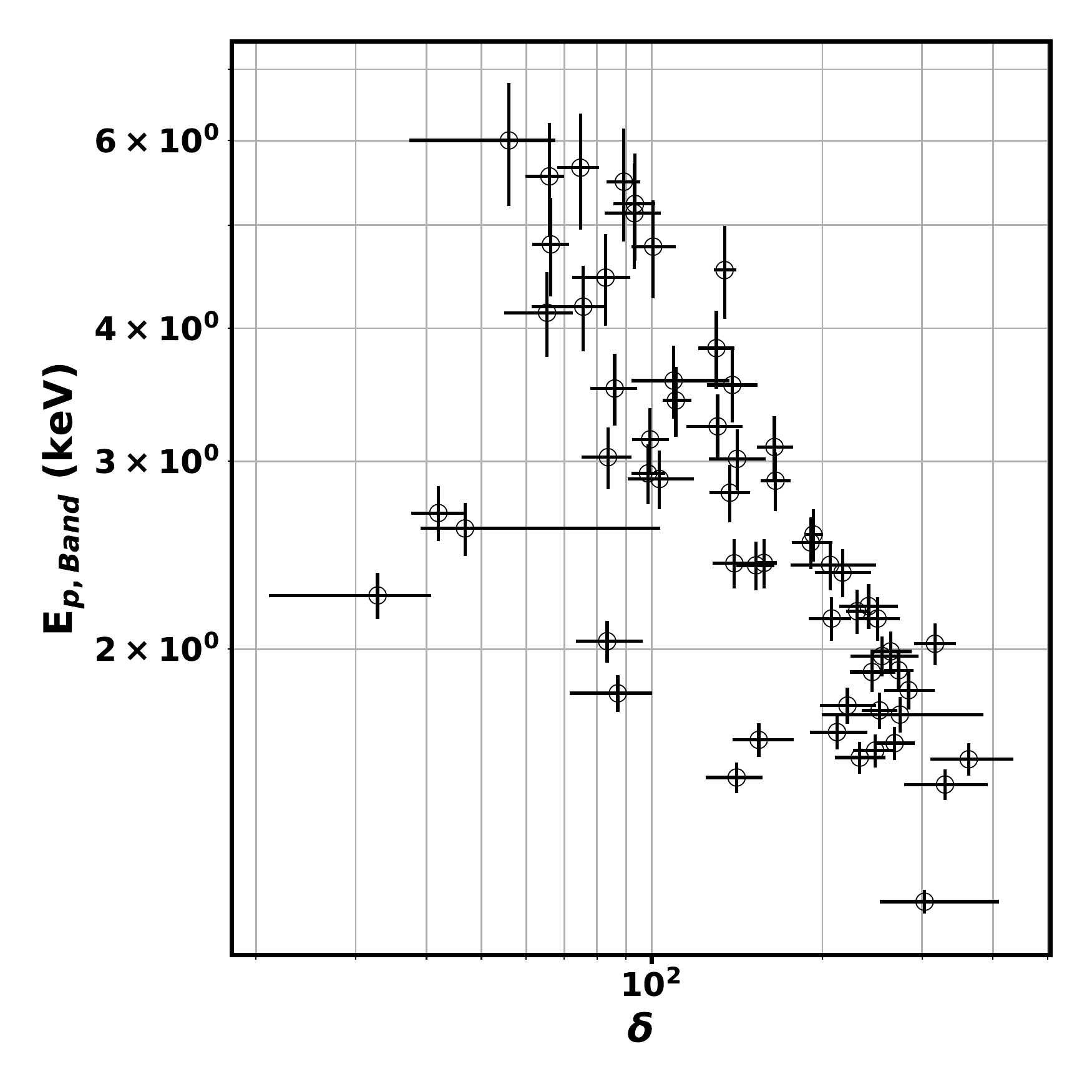}
\includegraphics[scale=0.335,angle=0]{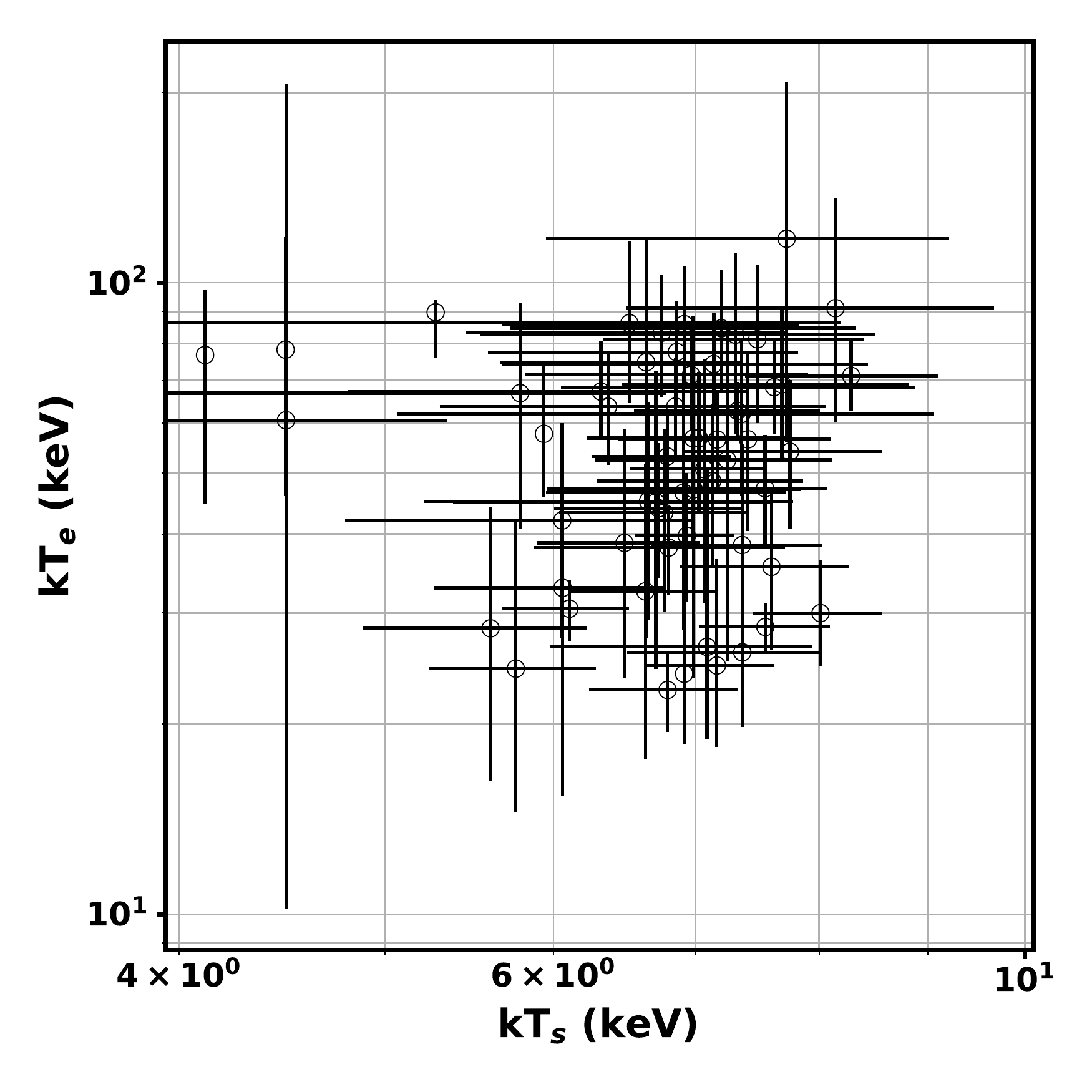}
\includegraphics[scale=0.335,angle=0]{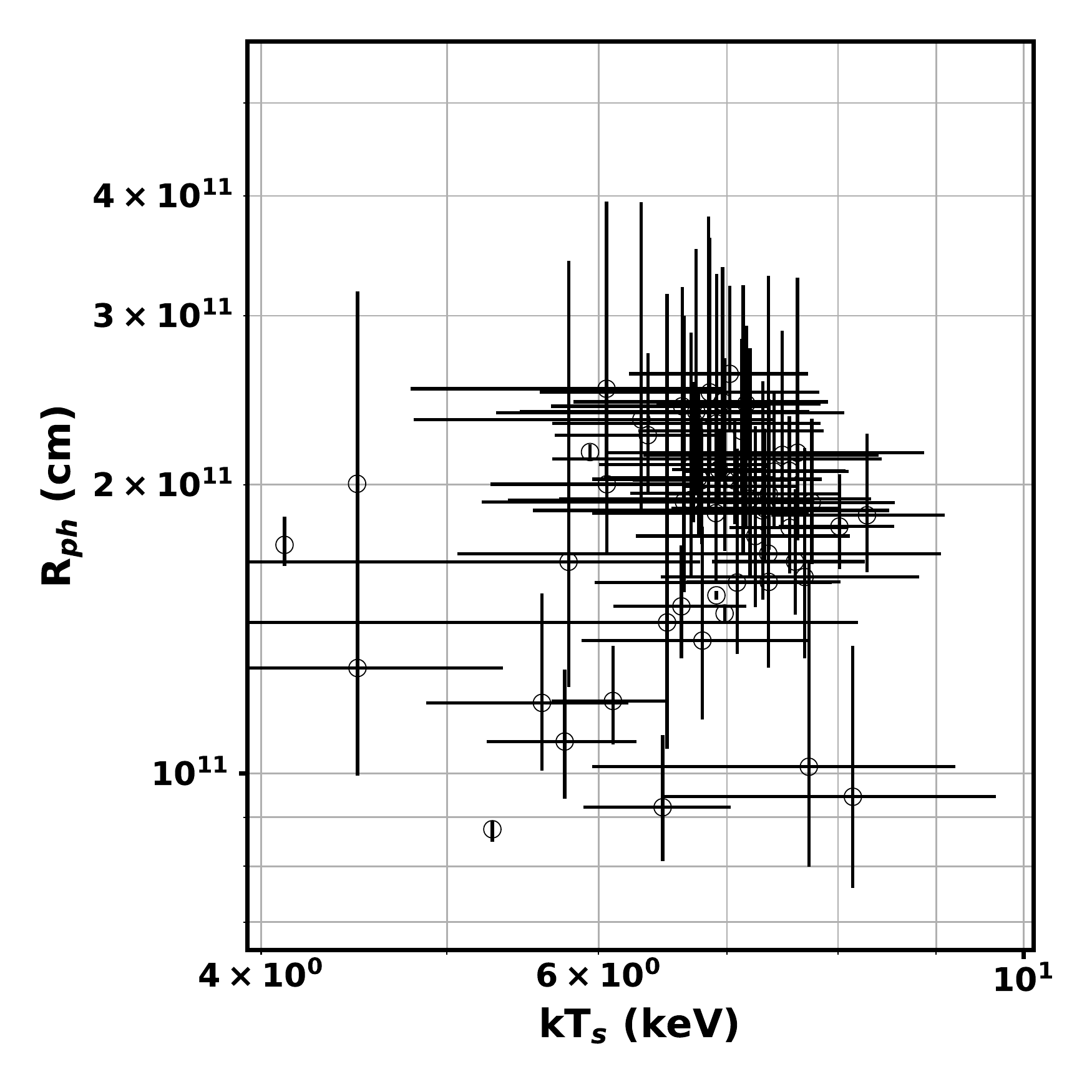}
\caption{Correlations obtained for various model parameters and derived parameters.} 
\label{fig:Episode_1_correlations}
\end{figure*}

We explore two parameter correlations and present the graphs in Figure \ref{fig:Episode_1_correlations}.

(i) {\it $kT_{BB}$ vs $E_{p}$:} Thermal components observed in a set of bright \fermi\ single pulse GRBs are correlated to
the peak of non-thermal emission \citep{Burgess:2014}. The exponent of the power-law relation ($E_p ~\propto ~T^{q}$) indicates the position
of the photosphere if it is in the coasting phase or
acceleration phase. The jet is dominated by magnetic fields
when   $q$ is 2 and baryonic if q is nearly 1. 
However, these results were found on GRBs with single pulses and for GRBs made up of overlapping pulses such criteria may not be valid.
For \thisgrb\ a positive correlation ($E_p\propto T^{2.2}$) between $E_p$ and $kT_{BB}$ is found with a log linear Pearson correlation
coefficient of 0.81, p-value $\sim10^{-15}$.
The index points out, therefore, a magnetically dominated jet with photosphere below the saturation radius. We will discuss this result coupled with the polarization results in Section \ref{sec:conclusion}.
The peak energy from Band function and BB + Band function are also found to be correlated, $E_{p, ~Band} \propto E^{1.2}_{p, ~BB+Band}$.

(ii) {\it Correlations among \sw{grbcomp} parameters:} Correlations between \sw{grbcomp} model parameters and
other model parameters are reported in \citet{Frontera:2013}. We found a strong correlation between 
peak energy of the Band function and the bulk parameter ($\delta$)
with log linear Pearson coefficient and p-value r(p) of -0.68 ($\sim10^{-9}$).
The seed photon temperature 
$kT_{BB}$ and photospheric radius $R_{ph}$ are uncorrelated contrary to a strong anti-correlation
reported in \citealt{Frontera:2013}. The seed blackbody temperature 
and the electron temperature are not correlated, consistent with the predictions of \sw{grbcomp} model. The parameter
$kT_e$ and peak energy of BB + Band model are correlated, with $kT_e\propto E^1_{p, ~BB+Band}$.

\section{Afterglows}
We have analyzed the afterglow data of \thisgrb\ in gamma-rays (\fermi-LAT) and X-rays (\swift-XRT) and the results are
shown in Figure \ref{fig:LAT_SED} and \ref{fig:XRT_SED}. 

\subsection{$\gamma$-ray afterglows}  \label{subsec:LAT_emission}
A $12^\circ$ region of interest (ROI) was selected around the refined \swift-XRT coordinates
and an angle of $100^\circ$ between the GRB direction and \fermi\ zenith was selected based on the navigation plots. 
The zenith cut is applied to reduce the contamination from $\gamma$-rays of Earth albedo. Transient event class and its
instrument response function $P8R2\_TRANSIENT020E\_V6$ is used as it is appropriate for GRB durations.
Details about LAT analysis methods and procedures can be found in the \fermi-LAT GRB catalog \citep{Ackermann:2013}.

A simple power-law
temporal decay is observed for the LAT light curve with a hint of momentary increase or steady emission in both energy and photon fluxes
during the first and second time bins:  $346 - 514$ s \& $514-635$ s. The LAT photon flux varies with time as a power law with an index 
$-1.56 \pm 0.40$ and the energy flux varies as a power law with an index $-1.37\pm0.45$. The photon index of LAT-HE is
$\Gamma_L$ = $-2.0\pm 0.1$  from a  spectral fit obtained by fitting the first $10^{5}$ sec data. 
This gives spectral index $\beta_L = \Gamma_L + 1$ to be $-1.0\pm0.1$. The time resolved spectra do not show variation in the photon index in the first four bins. 

\begin{figure*}
\centering
\includegraphics[scale=0.5]{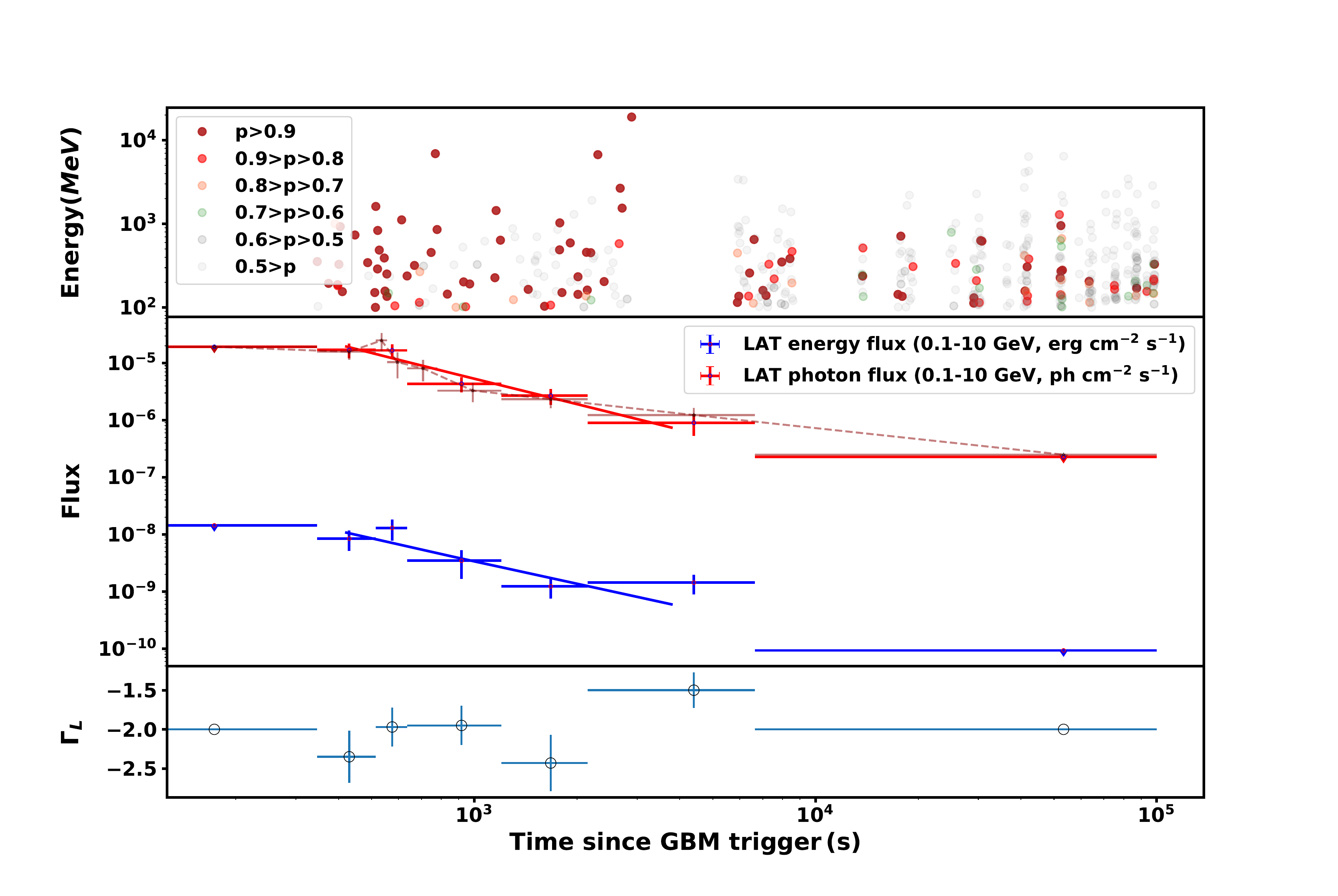}
\caption{\textit{Top panel}: Energies of individual \fermi-LAT photons ($> 100$~MeV), detected more than 345~s after the trigger. The color and transparency of circles depend on the probability (p) of their association with source. 
\textit{Middle panel}: \fermi-LAT photon and energy fluxes in 0.1 --- 10~GeV range. The photon index in the first time bin was fixed to $-2$ to get an upper limit on the fluxes. The dashed curve in the background shows the evolution of the LAT photon flux assuming a constant photon index, $-2$.
The photon flux rises fast initially, peaks at $\sim530$~s, and then declines. This flux peak was used to obtain a lower limit on the Lorentz factor of the ejecta (\S\ref{sec:spectral_analysis}).
\textit{Bottom panel}: Photon indices for \fermi/LAT in the 0.1 --- 10~GeV range.}
\label{fig:LAT_SED}
\end{figure*}

In the external shock model, for $\nu~>~max \{ \nu_m,~\nu_c \}$ which is generally true for reasonable shock parameters we can derive
the power law index of the shocked electrons by $f_\nu \propto \nu^{-p/2}$. We 
have synchrotron energy flux $f_L \propto \nu^{-\beta_L} t^{-\alpha_L} $ (see the LAT lightcurve in Figure \ref{fig:LAT_SED}). We found $\alpha_L = 1.37\pm0.45$ and $\beta_L = 1.0\pm0.1$. The value of  $\beta_L$ gives us 
$p=2.0\pm0.2$. Thus, the power-law index for the energy flux decay can be predicted by using $f_L\propto t^{(2-3p)/4}$. 
The calculated value of 
$\alpha_L$ is $-1.0\pm 0.2$, which agrees well with the observed value of  $-1.37\pm0.45$.  Hence, 
we can conclude that for \thisgrb, the LAT high energy afterglows are formed 
in an external forward shock.

For the thin shell case in a homogeneous medium and assuming that the peak of the LAT afterglow ($t_{p}$) occurs either before or in the second time bin gives  $t_{p} < 530$ s and  we can constrain the 
initial Lorentz factor ($\Gamma_0$)  of the GRB jet (see $e.g.$ \citealt{Sari:1999}, \citealt{Molinari:2007}) using 
\begin{equation}
 \Gamma_0 > 193(n\eta)^{-1/8}\times\left(\frac{E_{\gamma,52}}{t^3_{p, z,2}}\right)^{1/8}
\end{equation}
where $m_p$ is the mass of proton, $\eta$ is the radiation efficiency, $t_p$ is the time when afterglow peaks, $E_{\gamma,52}$ is the
k corrected rest frame energy of the GRB in the $1 ~-~ 10,000$~keV band. For a typical density $n~=~0.1~cm^{-3}$ of homogeneous
ambient medium
and $\eta=0.5$, we can constrain the initial Lorentz factor $\Gamma_0 > 330$. This limit is consistent with $\Gamma_0$ found  in Section \ref{sec:spectral_analysis} from $\Gamma_0~-~E_{\gamma,~iso}$ correlation.

The photon index hardens in the 
$2150~-~6650$ s time bin and the flux also seems to deviate from the power law fit. This hints at a contribution
from an inverse Compton component. The highest energy photon with a rest frame energy of $\sim 25 ~GeV$ is also observed during this interval.

\subsection{X-ray afterglows} \label{subsec:XRT_emission}
\begin{figure}
\centering
\includegraphics[scale=0.45]{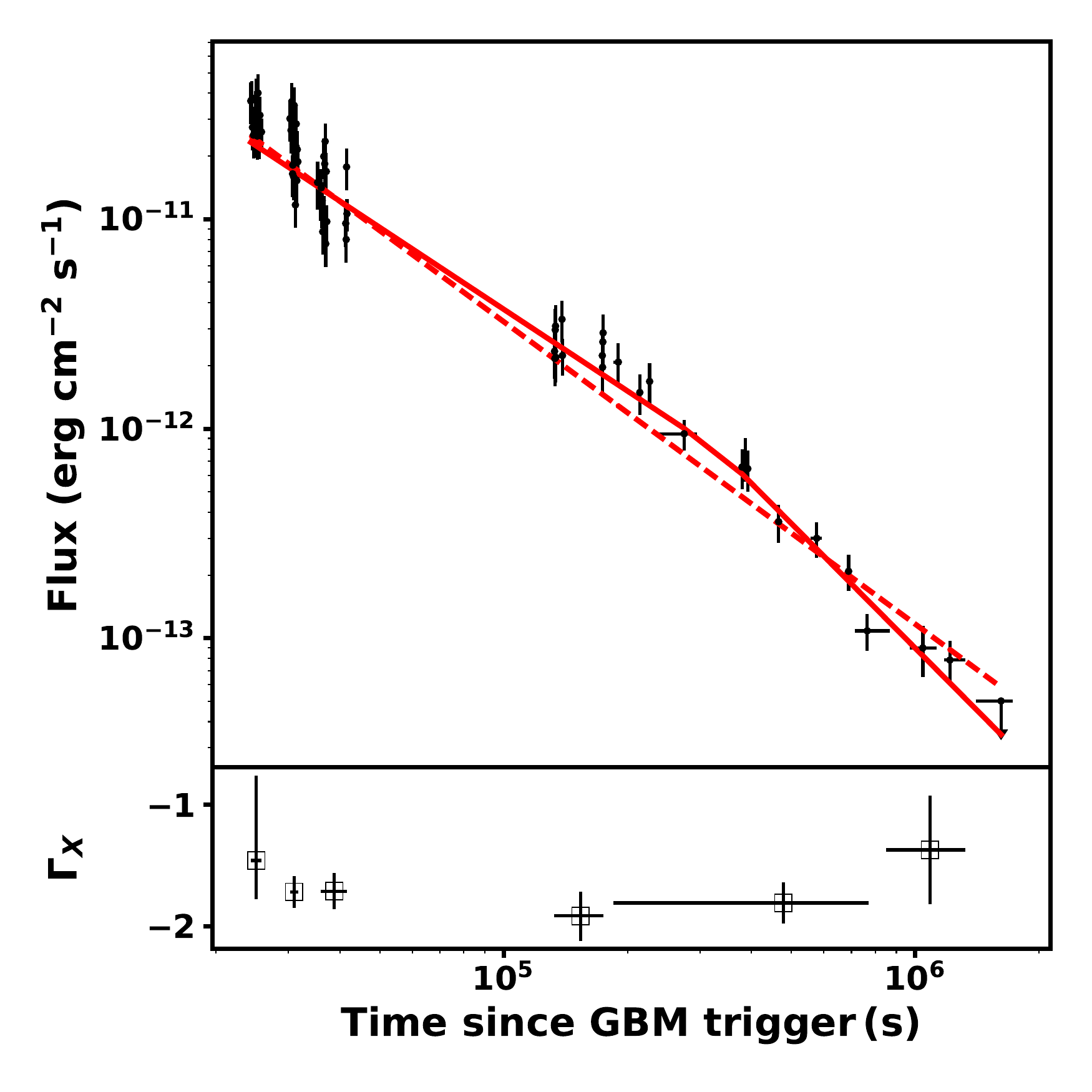}
\caption{\swift-XRT energy flux in the range 0.3 - 10~keV (upper panel). A single power law fit (red dotted line) to the flux gives $f_X\propto t^{-1.44 \pm 0.03}$.
When a break is introduced at $t_b$ (red solid line), the powerlaw indices are $1.29 \pm0.05 $ and $1.97\pm0.24$ before and after the break, respectively. The break observed is at $(3.37\pm1.03)\times 10^5$ s.
The photon indices for the \swift-XRT spectra are shown in the bottom panel.}
\label{fig:XRT_SED}
\end{figure}

The primary goal of the {\it Swift} satellite is to detect GRBs and observe the afterglows in X-ray and optical wavelengths.  If a GRB is not detected by {\it Swift} but is detected by some other mission (like {\it Fermi} etc.),  {\it target of opportunity} (ToO) mode observations can be initiated in {\it Swift} for very bright bursts. For such bursts the prompt phase is missed by {\it Swift} and afterglow observations are also inevitably delayed. However, the detection of X-ray afterglows with {\it Swift}-XRT not only allows us to study the delayed afterglow emissions but also to precisely localize the GRB for further ground and space based observations at longer wavelengths. The X-ray afterglow of \thisgrb\ was observed by {\it Swift} $\sim24300$ s ($\sim0.3$ days) after the burst. We have used the XRT products and lightcurves from the XRT online repository\footnote{\url{http://www.swift.ac.uk/xrt_products/index.php}} to study the light curve and spectral characteristics. The statistically preferred fit to the count rate light curve in the $0.3 ~- ~10$~keV band has three power-law segments with two breaks.  The temporal power-law indices and the break times in the light curve, as given in the GRB online repository,  are listed in Table \ref{tab:XRT_temporal}.  The data are also consistent with 3 breaks (Table \ref{tab:XRT_temporal}).  The light curve with three breaks resembles the canonical GRB light curves observed in XRT \citep{Zhang:2006ApJ, Nousek:2006ApJ}. We have analysed the XRT spectral data and generated the energy flux light curve shown in Figure \ref{fig:XRT_SED}. The flux light curve in the 0.3 - 10~keV energy band is fit with a single powerlaw and a broken powerlaw with a single break at time $t_b$. The best fit single powerlaw shows a decay index of $f_X\propto t^{-1.44\pm0.03}$, while in the broken powerlaw the pre- and post-break decay indices are $1.29\pm0.05$ and $1.97\pm0.24$ respectively with the break at $t_b=(3.37\pm1.03)\times10^5$ s. These values are consistent with the fits for the count rate light curves given in  Table \ref{tab:XRT_temporal}. 

We name the two segments of the light curve with three breaks as phase 1 (see Table \ref{tab:XRT_temporal}): before $t_{b, 2}$ and phase 2: after $t_{b, 2}$. The phase 1 is subjectively divided into 3 time bins to track the evolution of spectral parameters during this phase. 
The phase 2 is also divided at the $t_{b, 3}$ for inspecting the spectral change with the apparent rise in the light curve after this point.  

We froze the equivalent hydrogen column density ($n_H$) to its Galactic value \citep{Willingale:2013}. Another absorption model $tbabs$ in $XSPEC$ is used to model intrinsic absorption and $n_H$ 
corresponding to it is frozen to the value obtained from the time integrated fit.
A simple power law fits the spectrum for both the phases.

The photon index in the  XRT band is found to be $\Gamma_X = -1.9\pm0.1$ for phase 1 and $\Gamma_X = -1.7_{-0.6}^{+0.3}$ for phase 2. The evolution of
energy flux in $0.3-10~keV$ and the photon index are shown in Figure \ref{fig:XRT_SED}. Similar to LAT, it also predicts $p \sim 2$ when 
both $\nu_m$ and $\nu_c$ evolve to energies
that lie below the  XRT band. If we consider $p\sim2.2$ (See e.g. \citealt{Zhang:2006ApJ}, \citealt{Nousek:2006ApJ}) which 
will be nearly consistent with the $p$ inferred from LAT afterglow and also near to XRT-afterglow value.
Thus for the late time decay we have $\alpha_X \sim -1.15$. When the light curve is modeled by a broken power-law, we get $f_{X} \propto t^{-1.29\pm 0.05}$ before the break
and $f_{X} \propto t^{-1.97\pm0.24}$, thereafter.  Therefore, predicted $\alpha_X \sim -1.15$ is nearly 
consistent with $\alpha_X \sim 1.29 \pm 0.05$ (three times the error bar). The $\alpha_X$ before and after the break are consistent with the late time decay in the external shock model. And, the break observed 
can be identified as a jet break ($t_{break} = 3.4\times10^{5}$ s) and used for finding the opening angle ($\theta_j$) of the jet as given by Eq. \ref{eq:theta_j} \citep{Frail:2001}.
However, an achormatic break at all wavelengths in the lightcurves is required to claim it as a jet break. The absence of a break till the last data point observed in $XRT$ can also be utilized to put a lower limit on the jet break (see for example \citealt{Wang:2018ApJ}). So any break before corresponding to last data point in XRT will also respect that limit because for a given initial Lorentz factor ($\Gamma_0$) the jet break will occur later for a wider jet.

\begin{widetext}
\begin{equation}\label{eq:theta_j}
\theta_j \approx 0.057\left(\frac{t_j}{86,400 ~s}\right)^{3/8}\left(\frac{1+z}{2}\right)^{-3/8}\left(\frac{E_{iso}}{10^{53} ~erg}\right)^{-1/8}\left(\frac{\eta}{0.2}\right)^{1/8}\left(\frac{n}{0.1 ~cm^{-3}}\right)^{1/8}
\end{equation}
\end{widetext}
We get $\theta_j = 0.11$ $rad$  ($= ~6.3^\circ$) by using eq. \ref{eq:theta_j}. The beaming angle can be estimated by using Lorentz factor $\Gamma_{0}$ using $\theta_{beam} = 1/\Gamma_0$.
For $\Gamma_0=330$, we have $\theta_{beam} = 0.003$ $rad$ ($= ~0.17^\circ$). Thus, we have a wide bright jet with a narrow beaming angle. The Lorentz factor estimated above 
is for the final merged shells propagating to the circum-burst medium (CBM) and forming an external shock.  The Lorentz factor for individual shells can even be higher than our estimate.
The jet energy corrected for collimation is $1.37\times10^{51} ~erg$.

\section{Polarization measurements}\label{sec:polarization}
\thisgrb\ is one of the brightest GRBs detected in CZTI 
with an observed fluence $>$ 10$^{-4}$ erg/cm$^2$. This makes the GRB 
one of potential candidates 
for polarization measurement. CZTI works as a Compton polarimeter where 
polarization is estimated from the azimuthal angle distribution of
the Compton scattering events between the CZTI pixels at energies 
beyond 100~keV.
Polarization measurement capability of CZTI has been demonstrated 
experimentally during its ground calibration 
\citep{Chattopadhyay:2014,Vadawale:2015}. First on board verification of
its X-ray polarimetry measurement capability came with the detection of 
high polarization in Crab in 100 --- 380~keV \citep{Vadawale:2017}. 
Crab was observed for $\sim$ 800 ks in two years after its launch and this
was statistically the most significant 
polarization measurement till date in hard X-rays. 
Polarimetric sensitivity of CZTI for off-axis
GRBs is expected to be less than that for ON-axis sources 
(e.g. Crab), but
the high signal to background ratio for GRBs and availability of pre-GRB and 
post-GRB background makes CZTI equally sensitive to polarization measurements
of GRBs. Recently, we reported systematic polarization 
measurement for 11 GRBs, with $\sim$3$\sigma$ detection for 5 GRBs and 
$\sim$2$\sigma$ detection for 1 GRB, and upper limit estimation for the 
remaining 5 GRBs. \thisgrb\ is a bright GRB registering around 
$\sim$ 2000 Compton events in CZTI. It is the second brightest GRB in 
terms of number of Compton events after GRB 160821A, and therefore is a 
potential candidate for polarization analysis.

The details of the method of polarization measurement of GRBs 
with CZTI can be found in \citet{Chattopadhyay:2017}. Here we briefly
describe the different steps involved in the analysis procedure.
\begin{itemize} 
\item The first step is to identify and select the valid Compton events. 
We select the double pixel events during the prompt emission by 
identifying events happening within a 40 $\mu$s time window. 
The double pixel events are then filtered against various Compton 
kinematics criteria to finally obtain the Compton scattered events. 
\item The selection of Compton events is confined within the 
3$\times$3 pixel block of CZTI modules, which results in an 8 bin 
azimuthal scattering 
angle distribution. We compute the azimuthal scattering angles for events 
during both the GRB prompt emission and the background before and after the prompt emission. The 
combined pre and post-GRB azimuthal distribution is used for background 
subtraction to finally obtain the azimuthal distribution for the GRB.
\item The background corrected prompt emission azimuthal distribution 
shows some modulation due to (a) asymmetry in the
solid angles subtended by the edge and the corner pixels to the
central scattering pixel, and (b) the off-axis viewing angle of the burst. 
These two factors are corrected by normalizing the azimuthal angle distribution
with a simulated distribution for the same GRB spectra at the same off-axis
angle assuming the GRB is completely unpolarized. The simulation is performed 
in Geant4 with a full mass model of CZTI and \asat. 
\item We fit background and geometry corrected azimuthal angle distribution
using MCMC simulation to estimate the modulation factor and polarization 
angle (in CZTI plane). This 
is followed up by detailed statistical tests to determine whether 
the GRB is truly polarized or not. This is an
important step, particularly because, there can be systematic effects
which can produce significant modulation in the 
azimuthal angle distribution even for completely unpolarized photons. 
These effects are even more prominent in cases where the GRB is not very 
bright.
\item If the statistical tests suggest that the GRB is truly polarized, 
we estimate the polarization fraction by normalizing
the fitted modulation factor with $\mu_{100}$, which is the 
modulation factor for 100$\%$ polarized photons, obtained by simulating
the GRB simulated in Geant4 with the \asat\ mass model. 
If the GRB is found to be unpolarized, we estimate the upper limit 
of polarization (see \citet{Chattopadhyay:2017}).
\end{itemize}

\begin{figure}
\centering
\includegraphics[scale=0.38,angle=0]{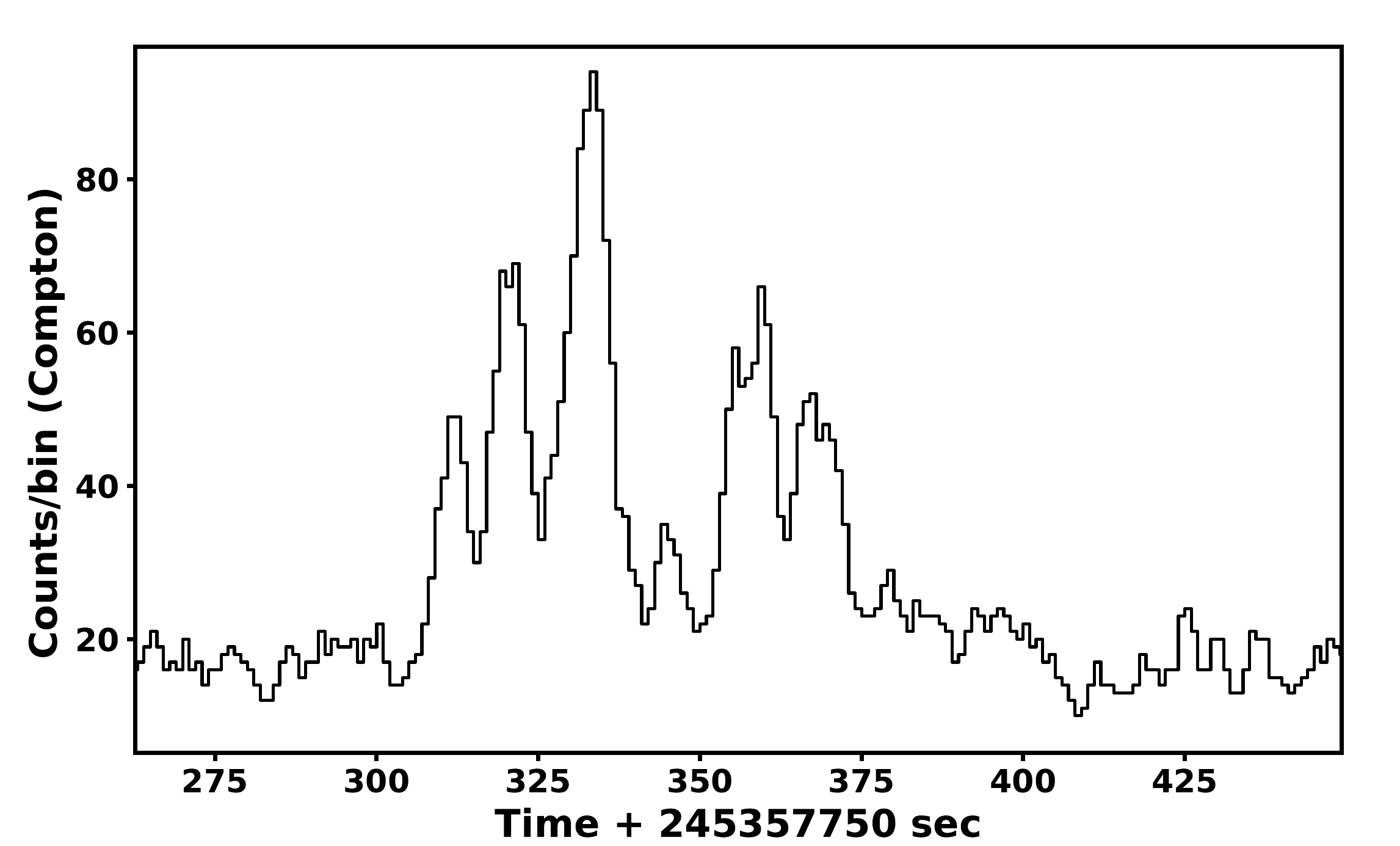}
\caption{Observed Compton event light curve for \thisgrb\ in 
100 --- 300~keV band with 1 second binning. 
The events in the time interval between 300---370 seconds
in the light curve are used for polarization analysis.}
\label{fig:light_curve_Compton}
\end{figure}

\begin{figure}
\centering
\includegraphics[scale=0.5,angle=0]{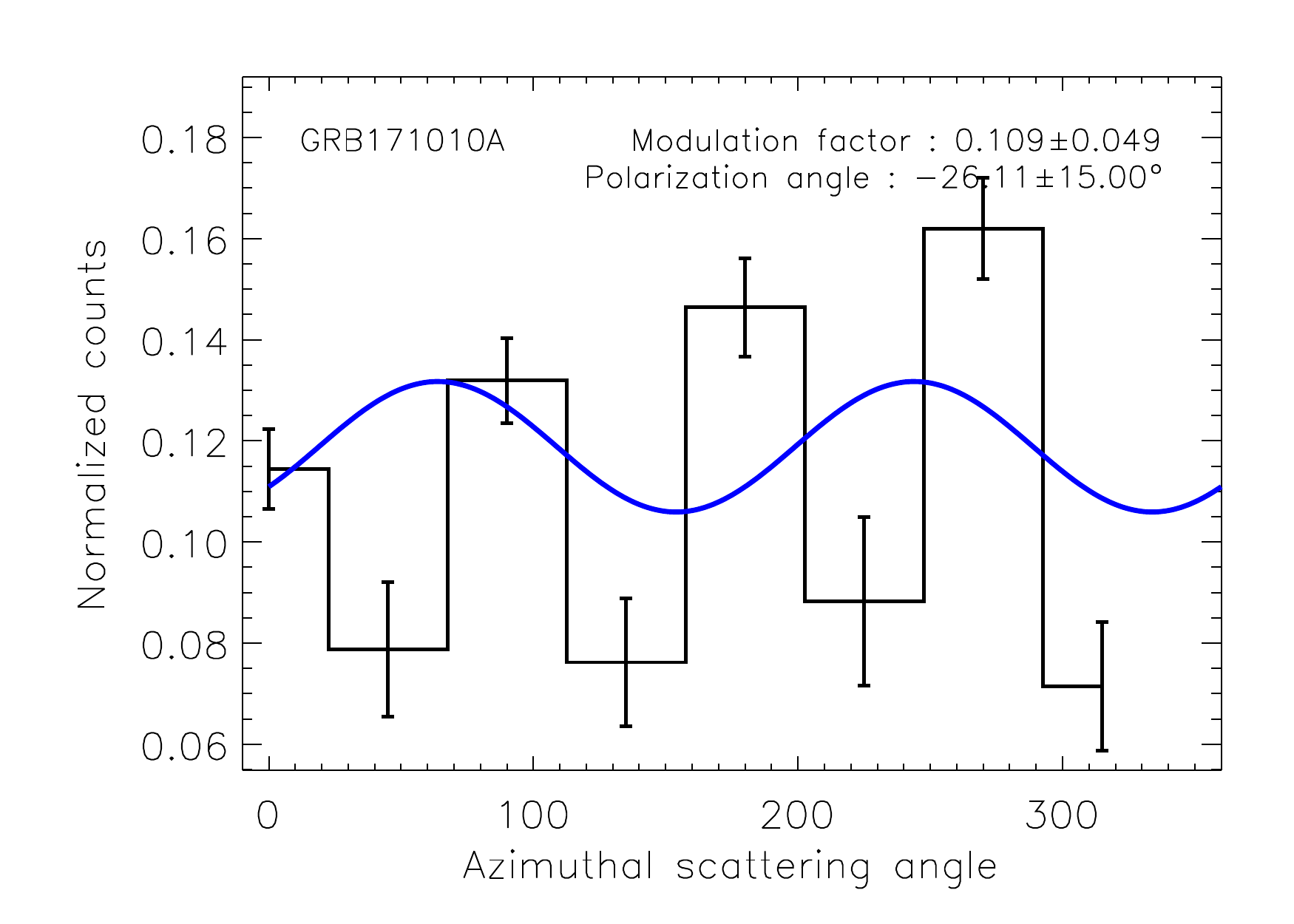}
\caption{The background subtracted and geometry corrected modulation 
curve for \thisgrb\ in 100 --- 300~keV. The blue solid line is the 
sinusoidal fit to
the modulation curve. We find no clear sinusoidal variation in the 
azimuthal angle distribution.}
\label{fig:time_integrated_polarization_100_300}
\end{figure}

\begin{figure}
\centering
\includegraphics[scale=0.37,angle=0]{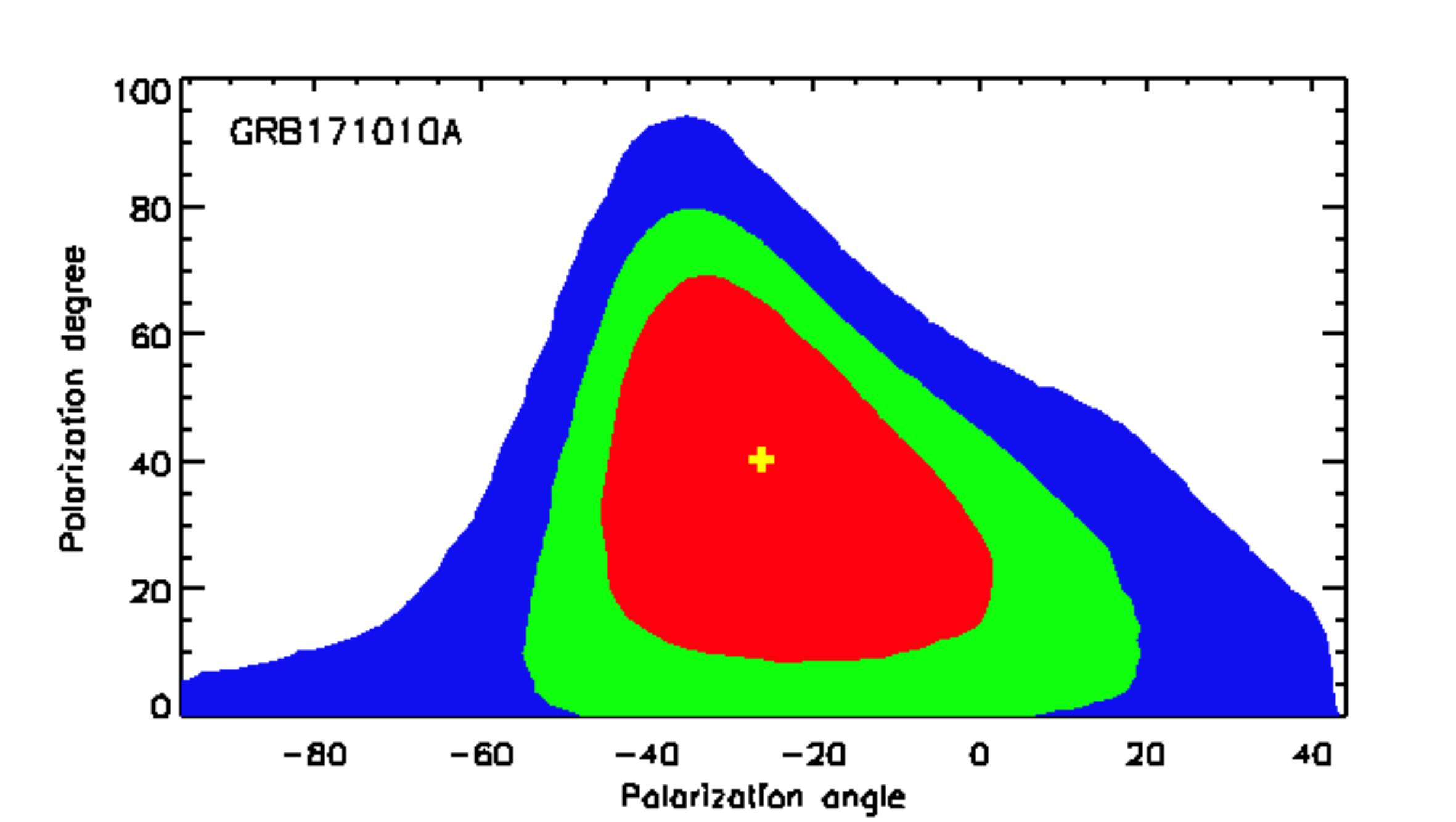}
\caption{Contour plot of polarization angle and fraction for \thisgrb\ 
in 100 --- 300~keV as obtained from the MCMC method. The red,
green and blue represent the 68 \%, 90 \% and 99 \% confidence levels 
respectively.}
\label{contour_100_300}
\end{figure}

Figure \ref{fig:light_curve_Compton} shows the light curve of \thisgrb\ in
Compton events in 100 --- 300~keV. The burst is clearly seen in Compton
events.   
We used a total of 747 seconds of background before and after the burst for
the final background estimation. After background subtraction, 
the number of Compton events
during the prompt emission is found to be $\sim$ 2000. 
Figure \ref{fig:time_integrated_polarization_100_300} shows the 
modulation curve in the 100 --- 300~keV band
following background subtraction and geometry correction as discussed in the
last section. We do not see any clear sinusoidal modulation in the 
azimuthal angle distribution. 
Figure \ref{contour_100_300} shows the corresponding contour plot for 
\thisgrb. 
Both polarization fraction and angle
are poorly constrained signifying that the burst is either unpolarized 
or polarization fraction is below the sensitivity level of CZTI.

In order to verify this we measure Bayes factor for sinusoidal (for polarized
photons) and constant model (unpolarized photons) by utilizing 
`Thermodynamic Integration' method on the MCMC parameter space
(for details refer to \citet{Chattopadhyay:2017}). This yields a value
less than 2 which is assumed to be the threshold value of 
Bayes factor for claiming detection of polarization. We therefore 
estimate the upper limit of polarization for \thisgrb\ which is done in
two steps. First step involves estimation of polarization detection 
threshold (P${_{thr}}$) by limiting the probability of false detection 
(to 0.05 for $\sim$2$\sigma$ or 0.01 for $\sim$3$\sigma$). The false 
polarization detection
probability is estimated by simulating \thisgrb\ for the observed number of
Compton and background events with 100 \% unpolarized photons. 
The second step involves measurement of the probability of detection of 
polarization such that probability of detection of a certain level of
polarization (P${_{upper}}$) being greater than the polarization 
detection threshold (P${_{thr}}$) is $\ge$ 0.5 (see 
\citet{Chattopadhyay:2017} for more details). 
The 2$\sigma$ upper limit (5 \% of false detection 
probability) for \thisgrb\ is found to be $\sim$42 $\%$. It is to be noted 
that in the sample of bursts used for polarization analysis in 
\citet{Chattopadhyay:2017}, GRB 160821A was found to possess maximum number of
Compton events ($\sim$ 2500). The next brightest burst was GRB 160623A with
$\sim$ 1400 Compton events. We estimated $\sim$ 50$\%$ polarization at 
$>$ 3$\sigma$ detection significance for GRB 160821A. In comparison, 
\thisgrb\ is found to have $\sim$ 2000 Compton events. 
The GRB is detected at an off-axis angle of 55$^\circ$. We expect 
CZTI to have significant polarmetric sensitivity at such off-axis angles. 
Therefore, polarization for this GRB should be detected at 
a significant detection level provided the GRB is at least $\sim$50 \% 
polarized. This is consistent to our estimation of 2$\sigma$ polarization 
upper limit of $\sim$42 \%. 

\begin{figure*}
\centering
\includegraphics[scale=0.5,angle=0]{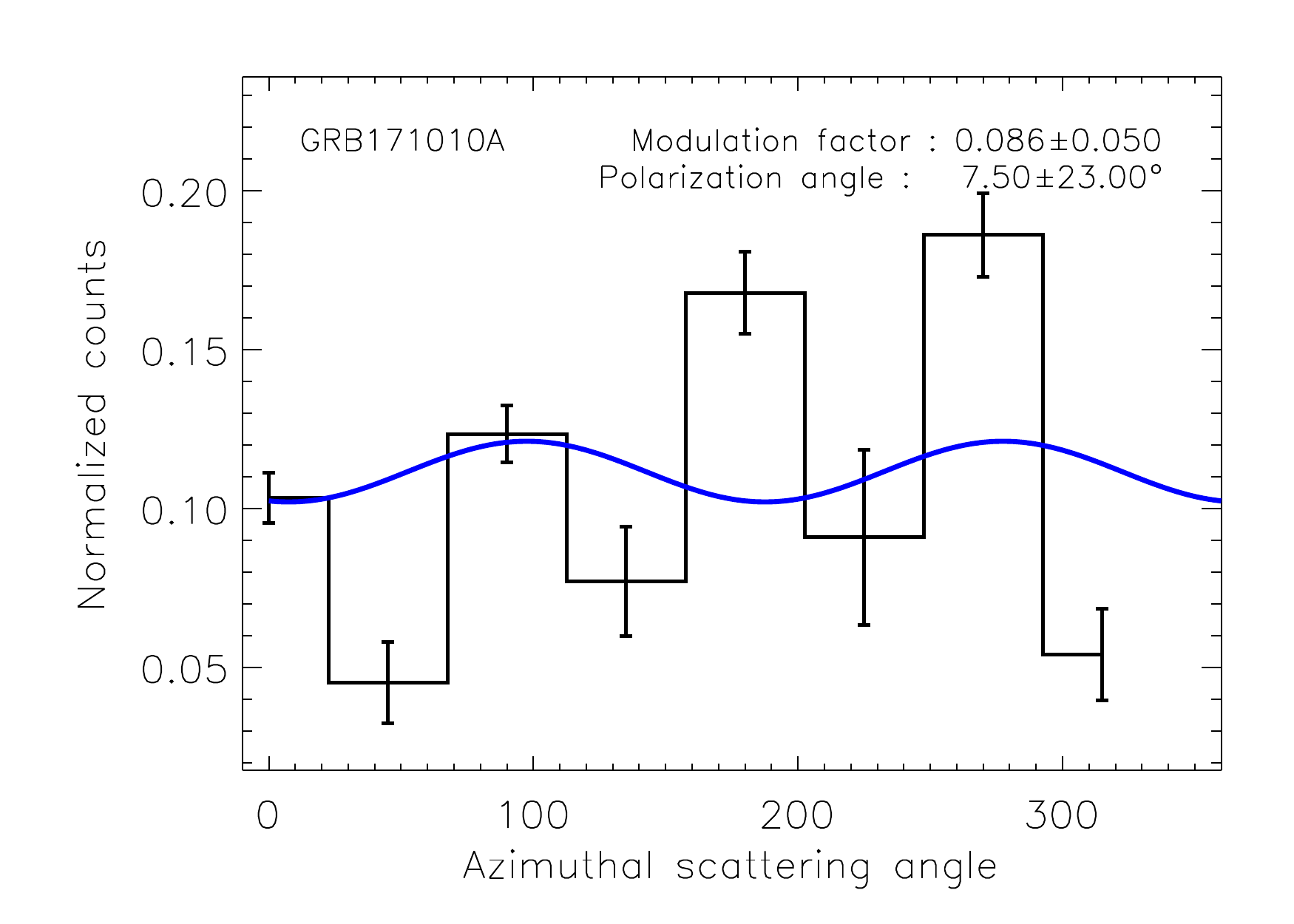}
\includegraphics[scale=0.5,angle=0]{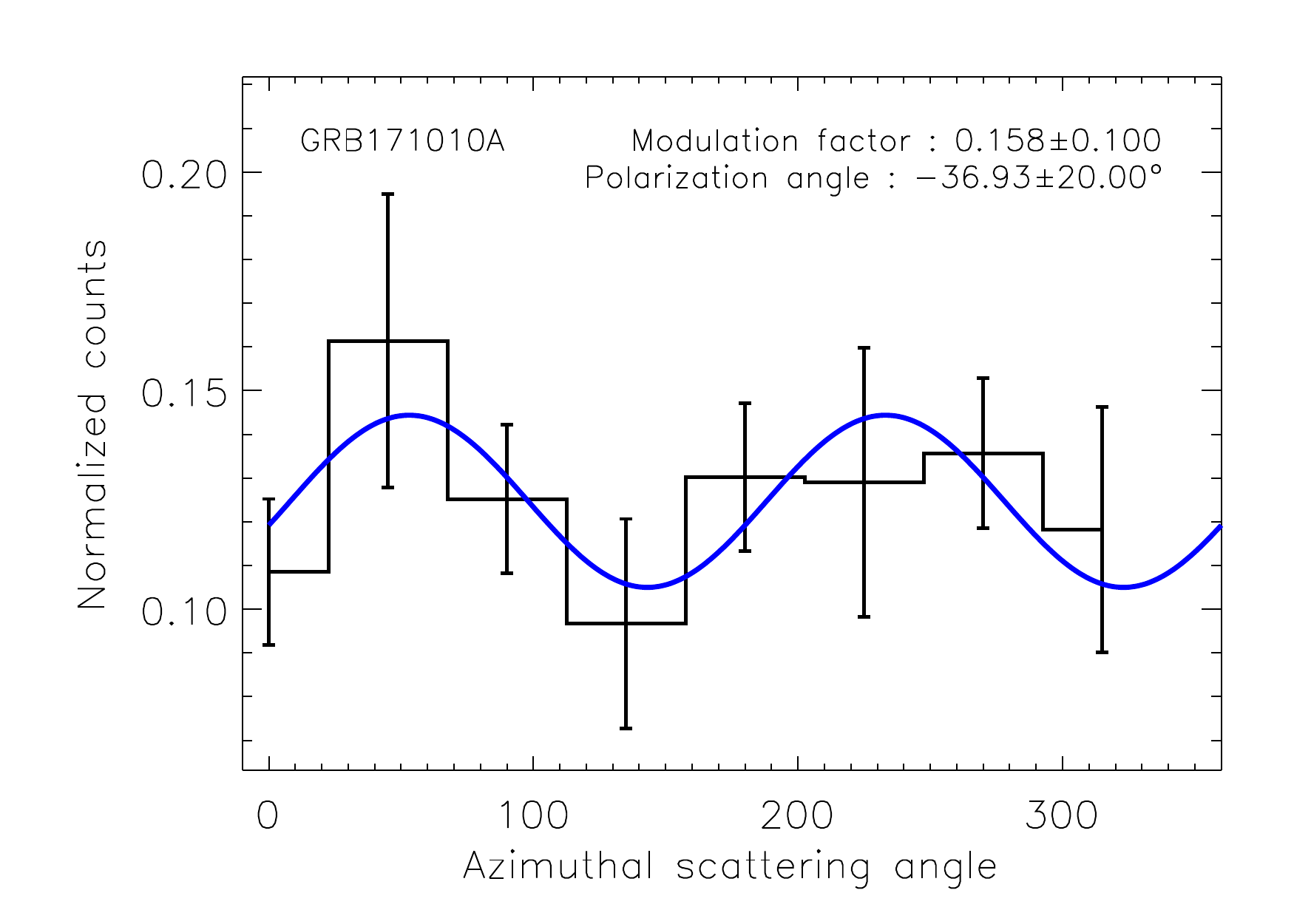}
\caption{Same as Figure \ref{fig:time_integrated_polarization_100_300} but
for different energy ranges --- 100 --- 200~keV (left) and 200 --- 300~keV
(right). We see a sinusoidal modulation in the 200 --- 300~keV modulation 
curve.}
\label{fig:Episode_1_polarization_variation}
\end{figure*}

This is an interesting result considering the fact that the spectral 
analysis suggests the time integrated peak energy for all the models
is less than 200~keV which falls within the energy range of polarization 
analysis. Therefore, in order to see the variation of polarization 
below and above the peak energy, we estimate polarization in two different
energy ranges --- 100 --- 200~keV and 200 --- 300~keV (see Figure 
\ref{fig:Episode_1_polarization_variation}). 
There is no clear modulation at the lower energies, whereas we see a 
sinusoidal variation in the modulation curve in 200 --- 300~keV, which
is beyond the peak energy of the GRB. However, it is to be noted that
the Bayes factors for both the energy ranges are less than 2 signifying
that there is no firm detection of polarization. The 
modulation at higher energies, therefore, is just a hint of polarization,
which is still an interesting result. 

\begin{figure*}
\centering
\includegraphics[scale=0.5,angle=0]{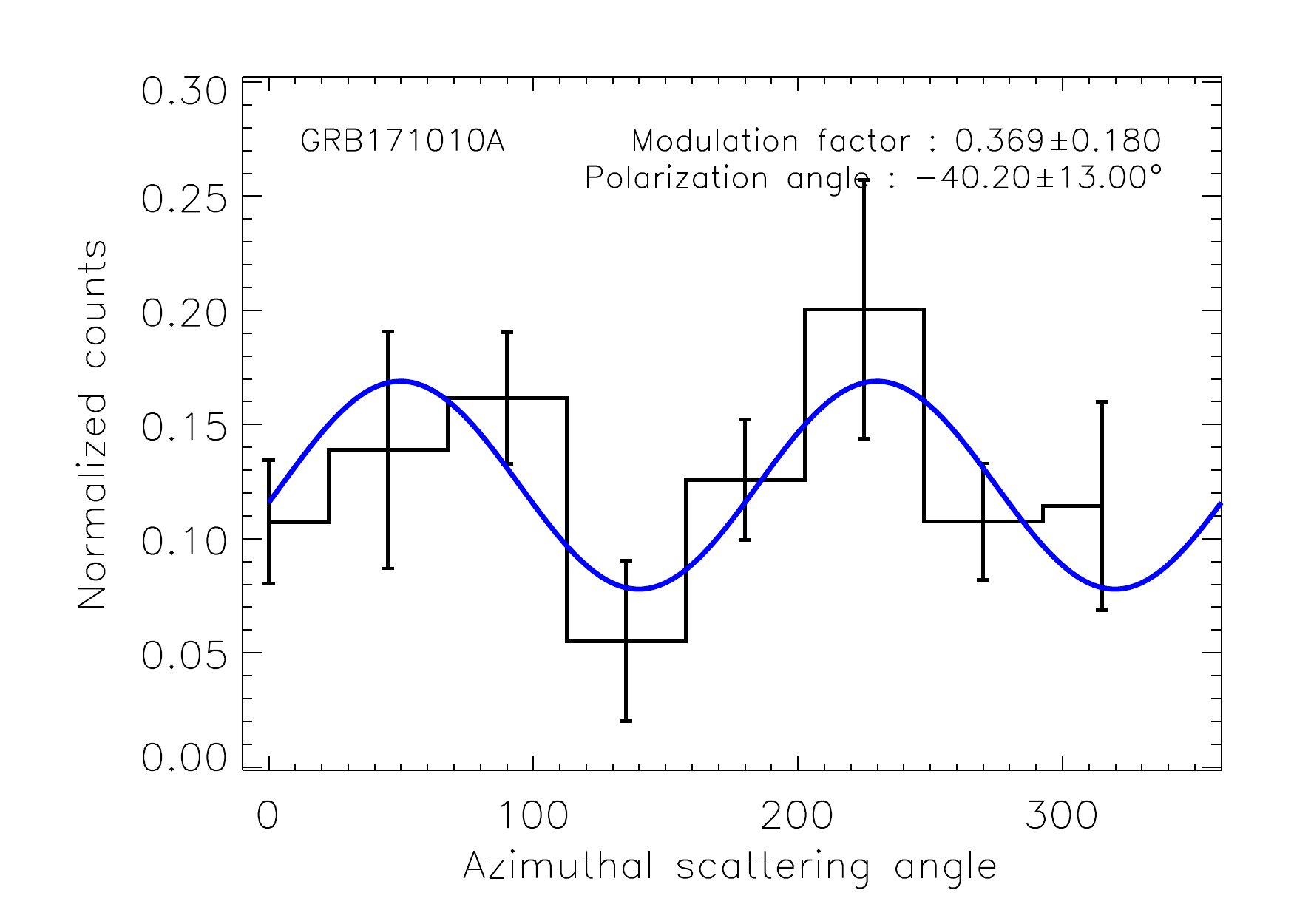}
\caption{Azimuthal angle distribution for the third time interval (28 --- 70
seconds at 200 --- 300~keV). We see a clear and high modulation 
signature in the azimuthal distribution. The Bayes factor is found to be 
around 2 with a false polarization detection probability by chance less than 1 \%.}

\label{modcurve_3rd_interval_200_300}
\end{figure*}

One major difference between GRB 160821A (or GRB 160802A, GRB 160910A 
\citep{Chattopadhyay:2017}) and
\thisgrb, is that the later lasts longer and has multiple pulses. 
We also see significant variation of peak energy with time (see Figure \ref{fig:params_evolution}).
These pulses might exhibit different polarization signatures resulting
in a net zero or low polarization when integrated in time. We therefore 
divided the whole burst in 3 different time intervals --- 0 --- 20 seconds,
20 --- 28 seconds and 28 --- 70 seconds, where `0' is the onset of the 
burst. Since we have already seen a hint of polarization
signature in 200 --- 300~keV, we further divided the signals 
in 100 --- 200~keV and 200 --- 300~keV.  
\begin{figure*}
\centering
\includegraphics[scale=0.6,angle=0]{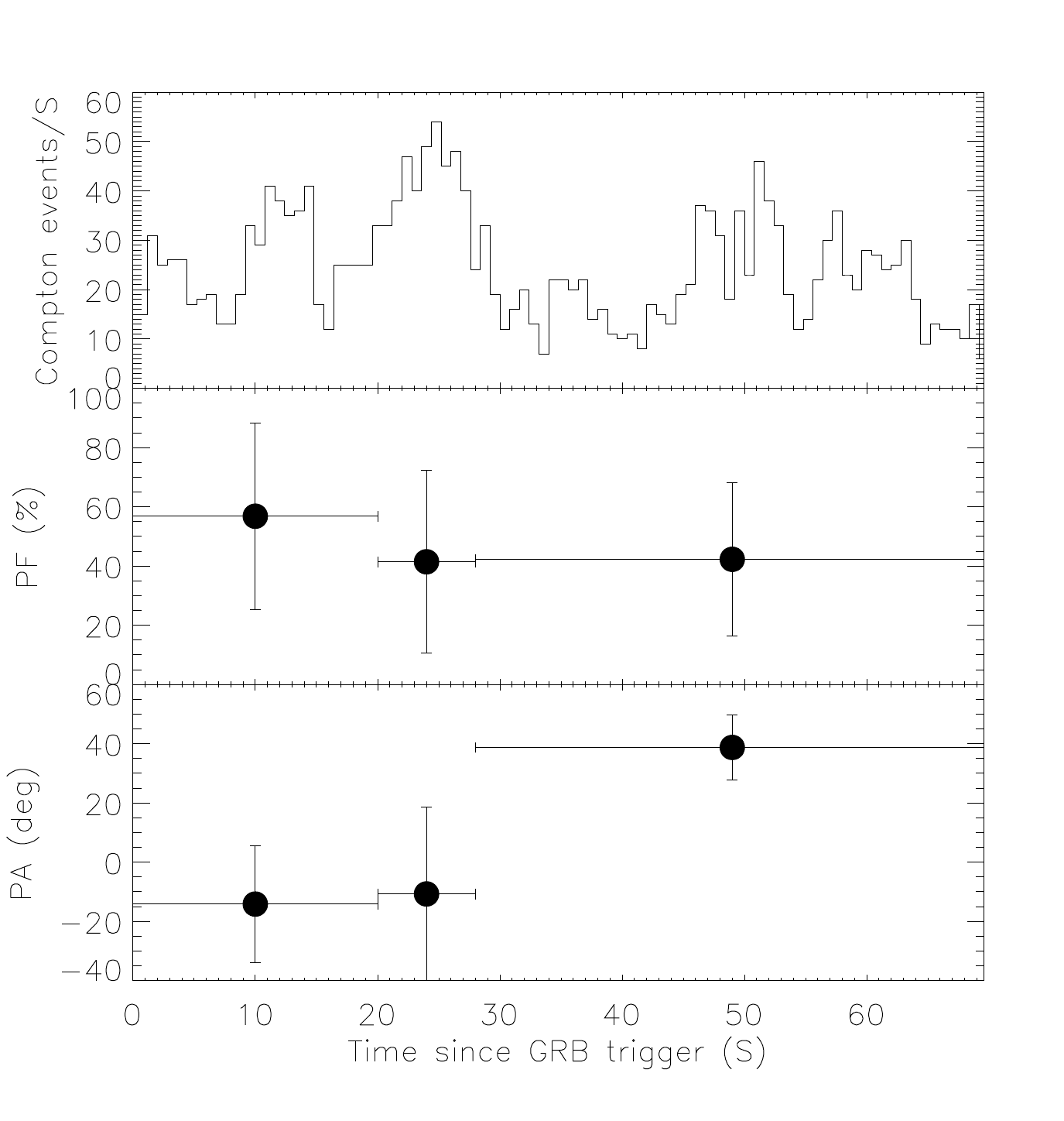}
\includegraphics[scale=0.6,angle=0]{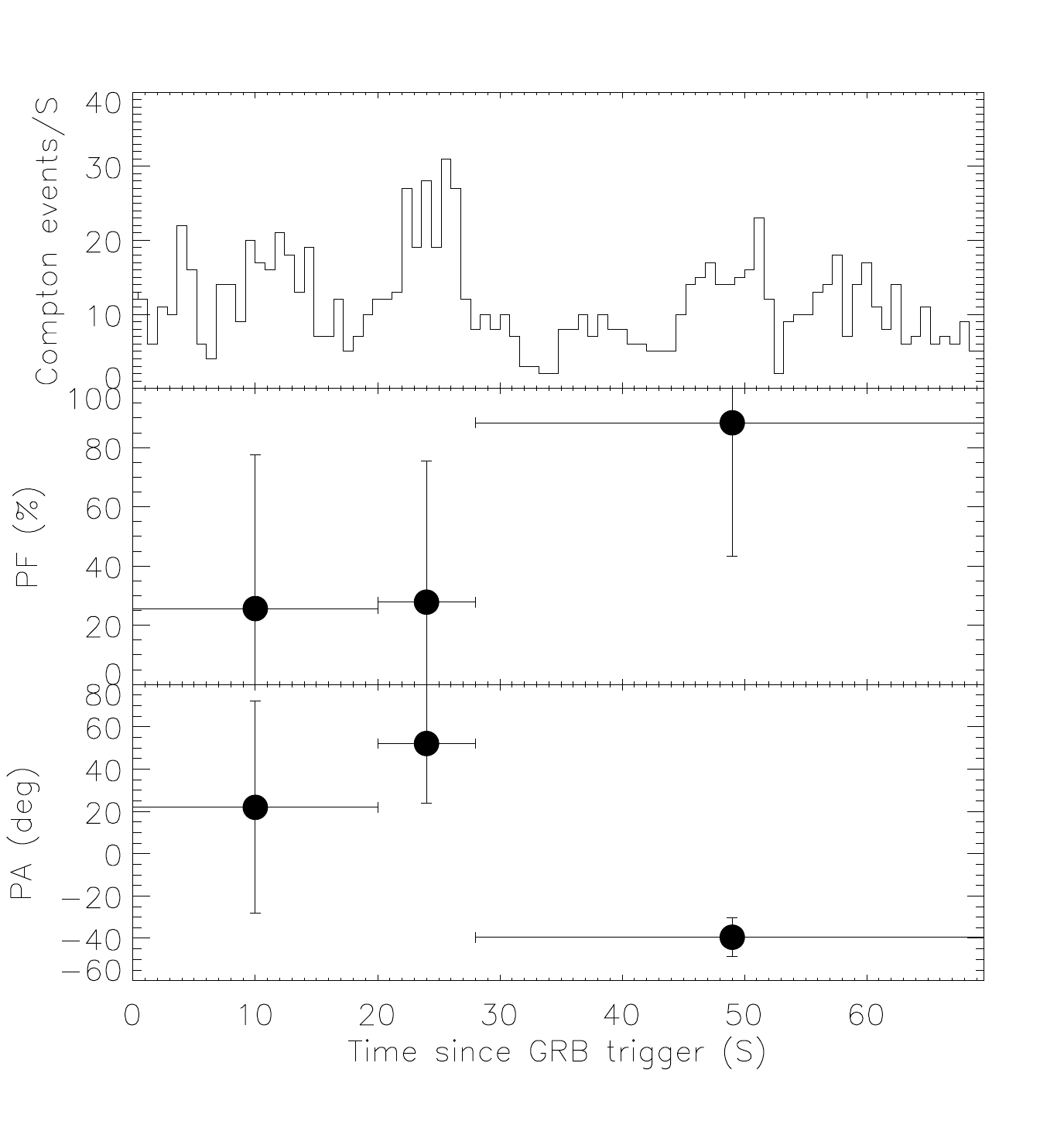}
\caption{Compton light curve (top), polarization fraction (middle) and angle
(bottom) for three different time intervals during the prompt emission
of \thisgrb\ in 100 --- 200~keV (left) and 200 --- 300~keV (right).}
\label{fig:Episode_1_polarization_variation_time}
\end{figure*}
Figure \ref{fig:Episode_1_polarization_variation_time} shows the variation
of polarization fraction (middle panel), and polarization angle (bottom panel)
in three time intervals for 100 --- 200~keV (left) and 200 --- 300~keV 
(right). The errors in polarization angle in the first two intervals 
in both the energy ranges are quite
large with no significant modulation in azimuthal angle distribution
consistent to being unpolarized. This is independently verified with the
estimation of low values of Bayes factor. The third interval, on the 
other hand shows high polarization fraction with a very clear sinusoidal
modulation in the azimuthal angle distribution 
in 200 --- 300~keV (see Figure \ref{modcurve_3rd_interval_200_300}). 
The polarization angle is also constrained within 13$^\circ$ uncertainty. 
The Bayes factor for this interval in 200 --- 300~keV is found to be
$\sim$2 with a false polarization detection probability by chance $<$ 1 \% 
clearly signifying that the GRB is polarized at the later part of
the emission at higher energies.  

\section{Discussions and conclusions}\label{sec:conclusion}
We presented the spectral, timing and polarization analysis of 
 \thisgrb\ which has an observed fluence $>10^{-4}$ $ergs~cm^{-2}$. We found that the spectrum integrated over the 
duration of the burst is peculiar as it shows a low energy break and can be modeled by either a BB or another power law. Some GRBs have 
shown the presence of such a component which was modeled by a BB with  peaks ranging up to 40~keV \citep{Guiriec:2011}. {\bf In 
a comprehensive joint analysis of X-rays and higher energies in the prompt emission, a break was found in the XRT-energy window \citep{Zheng2012,Oganesyan:2017a, Oganesyan:2017} and also in the GBM energy range \citep{Ravasio2018}.} To study the detailed evolution of the spectral parameters we sliced the spectrum into multiple time  bins and found that Band $E_p$ shows a bimodal 
distribution. Inclusion of a low energy component (modeled as a BB or a separate power-law) shows that the distribution of the peak energy remains $>~ 100$~keV and also falls in a concentrated region of between $100~ -~ 200$~keV. The mean value of the break energy in the   power-law with two breaks model ($E_2$) is $\sim 140$~keV. The power law index above the break is softer than the 
index  $\alpha$ obtained from the Band function. The mean value of $-\alpha_2$ is -1.45 and we are tempted to identify it with the synchrotron fast cooling process. 
$E_2$ can therefore be identified with the minimum energy injected to the electrons and the corresponding frequency is $\nu_m$. On the 
other hand, $-\alpha_1$ is 0.3, harder than the value of spectral index $-2/3$ allowed
below the cooling frequency $\nu_c$  of synchrotron radiation. However, absorption frequency which can be near the GBM lower energy 
window has
expected index to be +1, thus it is possible  that the cooling frequency falls very near to the  absorption frequency. In such a
case $E_1$ can be either identified as $\nu_c$ or $\nu_a$ as the ratio $\nu_c/\nu_a$ or $\nu_a/\nu_c$ will be closer to 1. Now, relating 
to the fast cooling scenario $-\alpha_3$ can be set equal to $-p/2 - 1$ to get the
index of power-law distribution of electron energies.  We get $p\simeq 3.2$ and for the distribution shown by $-\alpha_3$ allowed values of $p$ fall in the range 2 to 4.
The component at the low energy end can also be interpreted as being of photospheric origin, generated in a region 
that is optically thick to the radiation, and radiation escaping at a radius where it becomes transparent.
{\bf We have tested also the synchrotron model since 
\sw{bkn2pow} gives a possibility that synchrotron model 
in marginally fast cooling regime could in principle 
work. The direct fitting of the prompt emission spectra by the synchrotron 
model have been performed in number of studies \citep{Tavani1996,Lloyd00,Burgess2014,Zhang2016,Zhang2018,Burgess2018} that 
give the preference to the slow or moderataly fast cooling regimes of radiation.  
In our analysis, the synchrotron model returns small ratio between $\gamma_m$ and $\gamma_c$, 
and the range of \sw{pgstat}/dof 
that is similar to other models and 
it has same number of parameters as Band function.}
Our results show that the spectral data can also be well described by a Comptonization (\sw{grbcomp}) model. 

From the afterglows observed in LAT ($>$100 MeV) and XRT (0.3 -10~keV), we concluded that the afterglows are produced in an external shock propagating into the ambient medium.
By assuming the ambient medium to be homogeneous, we estimated the initial Lorentz factor of the ejecta. However, the Lorentz factor thus obtained is for the merged ejecta.  
The sub-MeV emission possibly arises from multiple internal shocks and a stratification in the Lorentz factor can not be ruled out. From the observed break 
in the XRT light curve, we obtained the jet opening angle. \thisgrb\ consists of a jet with initial beaming angle much narrower than the jet opening angle. 

The burst shows high polarization, however significant variation in the polarization fraction (PF) and angle (PA) can be seen with energy and time.
Such variations in PA can arise due to ordered magnetic field or random magnetic fields produced in shocks and when within the beaming angle, the net magnetic field 
can be oriented in any direction independent of the other shells. For a high polarization, the coherence length of the magnetic field ($\theta_B$) 
should be larger than or comparable to the beaming angle $1/\Gamma$ $i.e.$ $\theta_B \gtrsim 1/\Gamma$. {\bf In case of a Poynting flux dominated jet a pulse is produced in each ICMART event. 
The peak energy and polarisation decreases in each ICMART event. The polarisation degree and angle can vary \citep{Zhang:2011ApJ, Deng:2016ApJ}. 
The spectrum can be a hybrid of blackbody and Band function. We have observed a low energy feature and
the spectrum deviating from Band function which was also modelled by a blackbody. This can be contribution from a photosphere \citep{Gao:2015ApJ}. The lower polarisation at energies below 200 keV can be due to multiple events superimposing and decreasing the net degree of polarization. The change is polarisation angle with energy also supports this argument. Therefore, the spectrum and polarization measurents are consistent with ICMART model.}

In the Comptonization model multiple Compton scatterings are
assumed, producing higher energy photons in the jet.
The photons below the peak energy in the \sw{grbcomp} model are produced by the Comptonization of the seed blackbody
photons during the sub-relativistic phase. This component is, therefore, expected to be unpolarized. The high energy 
photons above the peak energy are mostly generated by further inverse Compton scattering off the non-thermal relativistic electrons of the
relativistic outflow. This can give a polarization up to $100 \%$ when the emission is observed at an angle $1/\Gamma$ with the beaming axis and unpolarized when viewed on-axis \citep{Rybicki:1979}. 
As for polarization, the highest value is expected for emission in the jet because in this case a photon, originating in the Compton cloud, illuminates the jet (or outflow) and undergoes a single   up-scattering. Single scattering off jet electrons can result in a degree of polarization up to $100\%$. Change in the
polarization fraction and angle can be possible only if jet is fragmented and we are observing it at different viewing angles. The measured high polarization below the peak is in contradiction with the predictions of this model.
 
The photospheric component found in the spectra suggests that the overall spectral shape may also be explained by the
sub-photospheric dissipation model \citep{Vurm:2016}. The polarization expected in the sub-photospheric dissipation model is low and
tends to decrease at higher energies because the higher energy photons are produced deep within the photosphere and are reprocessed through multiple Compton scatterings before finally leaving the system at the photosphere \citep{Lundman:2016}. The observed increase in the polarization with energy is in contradiction with the predictions of this model.

Another interpretation independent of the synchrotron or inverse Compton origin is that the emission is from fragmented fireballs moving with different velocity vectors \citep{Lazzati:2009}. 
The fragments moving into the direction of the observer will have the least polarization. If the intrinsic brightness of all the fragments are the same then this would also be the brightest fragment.
The fragments making larger angle with the line of sight, on the other hand, will have higher polarization and lower intensity. The polarization angle can sweep randomly between different pulses
in this setting. Another prediction of this geometry is that the PF and PA should not change within a single pulse. 
In time integrated analysis, we also find a change in polarization with energy. One possible explanation for this is that given the 
variability seen in the low energy light curve (100 - 200~keV), the contribution in this energy range could be from
more fragments than that in the high energy light curve (200 - 300~keV). The averaging effect thus results in a relatively lower polarization fraction (maximum PF up to 100\% can be achieved for the fragments viewed at $1/\Gamma$, where $\Gamma$ is the Lorentz factor of a fragment) and a different position angle at lower energies in comparison to those in the high energy band. 
The change in polarization with time can be explained by temporally separated internal shocks produced in the different fragments. Therefore, the change in
polarization from high to low seen in the 100 - 200~keV energy can be explained by considering the first pulse produced off-axis and the other brighter pulses 
near the axis (line of sight). In the high energy part, an increase in polarization with time is observed as well as the polarization 
angle are found to anti-correlate with the low energy counterpart. This suggests that the high energy emission has increased contribution with time 
from energetic off-axis fragments.

In the fragmented fireball scenario the jet should be fragmented into small scale ($\theta_{fragment} < \theta_{beam}$). 
The Lorentz factor $\Gamma_0$ was calculated from the LAT high energy afterglows that were produced in the external shocks. 
If the internal shocks were produced before the saturation in Lorentz factor was achieved then we can have a lower value of Lorentz factor when 
the internal shocks occur. This will allow a little room to increase $\theta_{beam}$.
If the internal shocks are produced at small radii, then the velocity vector within the $\theta_{beam}$ for a radially expanding ejecta will also have more divergence in their direction.

We conclude that the polarization results when used with spectral and temporal information are highly constraining. Although
we cannot decisively select a single model, we find that models with a decrease in polarization with energy are either less probable or  contributions from multiple underlying shocks in different energy ranges are needed to explain the apparent modulation with energy. {Comptonization model has low polarization at low energies is contradiction with our observations. However independent of emission mechanisms, a geometric model consisting of multiple fragments can explain the data, but demands fragmentation at small angular scale, well within $\theta_{beam}$.

\begin{table*}
\begin{small}
\caption{LAT emission ($>100$ MeV) in different  time intervals for a  fit with  a power-law spectrum. }
\label{tab:lat_sed}
\begin{center}
\begin{tabular}{ccccccD}
\hline
Time   &Index           & Energy flux                        & Photon flux              & Test Statistic  \\
(s)    &                &( $10^{-9} ~ergs$  $cm^{-2}$ $s^{-1}$)   & ($\times 10^{-7} ~photons$   $cm^{-2}$ $s^{-1}$) &  (TS)\\ \hline
(0) 0 - 346   & $-2$ (fixed) & $ <14.4$& $ < ~193$& 0 \\
(1) 346 - 514 & $-2.35\pm0.33$ & $8.43\pm3.33$ &    $170\pm48$  &39\\ 
(2) 514 - 635 & $-1.97\pm 0.25$ &$13.00\pm5.25 $&$167 \pm46$ &85 \\
(3) 635 - 1200  &$-1.95 \pm0.25$& $3.46 \pm1.80$& $44\pm13$ &62\\
(4) 1200 - 2150 &$-2.43\pm 0.36$& $1.23 \pm0.48$& $27\pm  8$& 33 \\
(5) 2150 - 6650 &$-1.50 \pm0.23$&$1.43 \pm0.55$& $9.0\pm 1.4$& 58 \\
(6) 6650 - 100000 &$-2$ (fixed)& $ <0.093$& $ < ~2.2$& 9  \\\hline \hline
\end{tabular}
\end{center}
\end{small}
\end{table*}

\begin{table*}[!htbp]
\caption{Best fit parameters for power law fits with multiple breaks for the low energy
light curve from XRT.  Power law with two breaks 
is the best fit model.  }
\label{tab:XRT_temporal}
\begin{center}
\begin{tabular}{ccccccccD}
\hline\hline
breaks & $\alpha_1$ & $t_{b, 1}$ & $\alpha_2$ & $t_{b, 2}$ & $\alpha_3$ & $t_{b, 3}$ & $\alpha_4$\\
       &            &    ($10^4$ s)        &            &   ($10^5 $ s)      &            & ($10^5 $ s)    & \\
\hline

1 & $1.30^{+0.07}_{-0.08}$ & $34.0^{+0.0}_{-15.0}$ & $2.0^{+0.0}_{-0.0}$ \\
2 & $2.37^{+0.32}_{-0.17}$ & $8.6^{+1.6}_{-5.0}$ &$-0.95^{+2.13}_{-0.55}$ & $1.58^{+1.57}_{-0.14}$ & $1.85^{+0.14}_{-0.14}$ \\
3 & $2.37^{+0.39}_{-0.16}$ & $7.0^{+0.0}_{+3.0}$ & $0^{+1}_{-2}$ & $1.8^{+0.0}_{-0.4}$ &$2.0^{+6.0}_{-0.5}$ & $8.0^{+0.0}_{-6.0}$    &$ 1^{+7}_{-3}$ \\
\hline
\end{tabular}
\end{center}
\end{table*}

\begin{table*}[!htbp]
\caption{Spectral fit to the XRT spectra at different time intervals using the power law (PL) model}
\begin{center}
\label{tab:XRT_specfitting}
\begin{tiny}
\begin{tabular}{c|cccccc}
\hline
Time &Model & $n_{H, i}$ &$\alpha_{po}$&    CSTAT/dof\\ 
($10^4$ s)&PL &$10^{22}$ &             &    & \\ \hline
Phase 1 &&  & & &&\\
(2.43 - 17.5)& && $1.9_{-0.1}^{+0.1}$& & &\\ 
Phase 2 &&$0.27_{-0.05}^{+0.05}$  && 513/566 && \\
(17.5, 132.2)& &&$2.0_{-0.2}^{+0.2}$ & & &\\ \hline
\hline
Phase 1& PL& $\alpha_{po}$ &CSTAT/dof & &\\ 
(2.43, 2.58)&&$1.45_{-0.70}^{+0.32}$& $179  / 222$&&\\
(3.03, 3.17)&&$1.7_{-0.1}^{+0.1}$&  $211 /  246$ &&\\
(3.59, 4.16)&&$1.71_{-0.15}^{+0.15}$ & $204  / 205$&&\\
(13.3, 17.5)&&$1.9_{-0.2}^{+0.2}$& $109/ 135$  &\\
Phase 2 &&  & &&\\ 
(17.5, 77.2)&&$1.8_{-0.2}^{+0.2}$ & $151/188 $  &\\
(85, 132.2)&& $1.4_{-0.4}^{+0.5} $ & $45/41$   &\\\hline
\end{tabular}
\end{tiny}
\end{center}
\end{table*}

\begin{table*}
\caption{Time-resolved Spectral fitting for different models  }

\label{tab:specfitting}
\begin{tiny}
\begin{center}
\begin{tabular}{c|cccccccD}
\hline
Band&&&  &&\\
\hline
Sr. no. &(t$_1$,t$_2$) & $\alpha$ & $\beta$ & E$_p$ (keV) & $K_{B}$ &\sw{pgstat}/dof &\\
\hline
1& -1, 7&$-1.2_{-0.1}^{+0.1}$&$-9.3_{-\infty}^{+19.4}$&$141_{-19}^{+16}$& $0.016_{-0.003}^{+0.004}$&$420/339$\\
2& 7, 9&$-1.10_{-0.05}^{+0.05}$&$-8.8_{-\infty}^{+18.8}$&$222_{-23}^{+28}$&$0.029_{-0.005}^{+0.004}$ &$284/339$\\
3& 9, 10&$-1.08_{-0.07}^{+0.1}$&$-7.7_{-\infty}^{+17.7}$&$274_{-74}^{+111}$&$0.054_{-0.009}^{+0.011}$ &$212/339$\\
4& 10, 12&$-1.0_{-0.1}^{+0.1}$&$-2.2_{-0.5}^{+0.2}$&$303_{-51}^{+107}$&$0.081_{-0.011}^{+0.011}$ &$310/339$\\
5& 12, 14&$-0.90_{-0.06}^{+0.07}$&$-2.3_{-0.3}^{+0.2}$&$362_{-52}^{+72}$&$0.10_{-0.01}^{+0.01}$ &$311/339$\\
6& 14, 17&$-0.90_{-0.06}^{+0.07}$&$-2.15_{-0.17}^{+0.10}$&$329_{-50}^{+63}$&$0.09_{-0.01}^{+0.01}$ &$400/339$ \\
7& 17, 18&$-0.8_{-0.1}^{+0.1}$&$-2.4_{-0.4}^{+0.2}$&$206_{-31}^{+43}$&$0.18_{-0.03}^{+0.04}$ &$290/339$\\
8& 18, 19&$-0.8_{-0.1}^{+0.1}$&$-2.40_{-0.25}^{+0.17}$&$217_{-23}^{+27}$&$0.23_{-0.03}^{+0.03}$ &$210/339$\\
9& 19, 20&$-0.82_{-0.07}^{+0.07}$&$-2.5_{-0.3}^{+0.2}$&$241_{-27}^{+31}$&$0.26_{-0.03}^{+0.03}$ &$341/339$ \\
10& 20, 24&$-0.80_{-0.03}^{+0.03}$&$-2.50_{-0.01}^{+0.08}$&$230_{-10}^{+11}$&$0.33_{-0.01}^{+0.02}$ &$620/339$\\
11& 24, 26&$-1.00_{-0.06}^{+0.06}$&$-2.6_{-0.2}^{+0.2}$&$130_{-9}^{+10}$& $0.30_{-0.30}^{+0.04}$&$443/339$\\
12& 26, 29&$-1.10_{-0.03}^{+0.04}$&$-2.8_{-0.4}^{+0.2}$&$196_{-13}^{+14}$& $0.22_{-0.01}^{+0.01}$ &$519/339$\\
13& 29, 30&$-1.00_{-0.06}^{+0.07}$&$-2.8_{-1.6}^{+0.3}$&$255_{-30}^{+40}$  &$0.22_{-0.02}^{+0.03}$&$314/339$ \\
14& 30, 31&$-1.00_{-0.04}^{+0.04}$&$-9.4_{9.4}^{+19.4}$&$316_{-26}^{+28}$ &$0.24_{-0.01}^{+0.02}$&$293/339$\\
15& 31, 32&$-0.83_{-0.05}^{+0.05}$&$-2.5_{-0.3}^{+0.2}$&$284_{-26}^{+32}$  &$0.35_{-0.03}^{+0.03}$&$418/339$\\
16& 32, 34&$-0.82_{-0.03}^{+0.03}$&$-2.5_{-0.1}^{+0.1}$&$273_{-15}^{+17}$  &$0.43_{-0.02}^{+0.02}$&$506/339$\\
17& 34, 35&$-0.80_{-0.05}^{+0.05}$&$-2.5_{-0.2}^{+0.1}$&$264_{-20}^{+23}$& $0.47_{-0.03}^{+0.04}$ &$314/339$\\
18& 35, 36&$-0.90_{-0.05}^{+0.05}$&$-2.7_{-0.5}^{+0.2}$&$250_{-21}^{+24}$& $0.40_{-0.03}^{+0.03}$ &$320/339$\\
19& 36, 38&$-0.80_{-0.04}^{+0.04}$&$-2.6_{-0.1}^{+0.1}$&$134_{-6}^{+7}$& $0.49_{-0.04}^{+0.04}$ &$480/339$\\
20& 38, 39&$-1.0_{-0.1}^{+0.1}$&$-2.7_{-0.4}^{+0.2}$&$101_{-9}^{+10}$  &$0.35_{-0.05}^{+0.07}$&$357/339$\\
21& 39, 41&$-0.3_{-0.3}^{+0.3}$&$-2.80_{-0.06}^{+0.05}$&$42_{-4}^{+5}$&$2.59_{-1.21}^{+3.07}$  &$494/339$\\
22& 41, 44&$-1.25_{-0.04}^{+0.04}$&$-3_{-0.7}^{+0.3}$&$110_{-6}^{+7}$  &$0.23_{-0.02}^{+0.02}$&$612/339$\\
23& 44, 45&$-1.1_{-0.1}^{+0.1}$&$-3_{-\infty}^{+0.4}$&$131_{-16}^{+14}$ &$0.27_{-0.04}^{+0.05}$ &$288/339$\\
24& 45, 46&$-1.05_{-0.05}^{+0.05}$&$-2.8_{-1.1}^{+0.3}$&$191_{-14}^{+18}$ &$0.32_{-0.03}^{+0.03}$ &$366/339$\\
25& 46, 47&$-1.0_{-0.1}^{+0.1}$&$-2.5_{-0.5}^{+0.2}$&$103_{-12}^{+15}$  &$0.33_{-0.06}^{+0.09}$&$356/339$\\
26& 47, 48&$-1.20_{-0.07}^{+0.07}$&$-9.4_{-\infty}^{+19}$&$98_{-6}^{+7}$& $0.21_{-0.03}^{+0.03}$ &$328/339$\\
27& 48, 50&$-1.2_{-0.1}^{+0.2}$&$-2.8_{-1.2}^{+0.4}$&$65_{-10}^{+7}$ & $0.17_{-0.04}^{+0.10}$&$341/339$\\
28& 50, 51&$-1.1_{-0.1}^{+0.1}$&$-2.9_{-1.1}^{+0.4}$&$93_{-11}^{+11}$ & $0.21_{-0.04}^{+0.06}$&$291/339$\\
29& 51, 53&$-1.1_{-0.1}^{+0.1}$&$-2.8_{-0.3}^{+0.2}$&$99_{-7}^{+8}$&$0.28_{-0.03}^{+0.04}$ & $397/339$\\
30& 53, 55&$-1.10_{-0.04}^{+0.05}$&$-3.3_{-\infty}^{+0.6}$&$208_{-19}^{+17}$ &$0.21_{-0.01}^{+0.02}$ &$434/339$\\
31& 55, 56&$-1.00_{-0.04}^{+0.05}$&$-4.1_{-\infty}^{+1.2}$&$248_{-21}^{+22}$ &$0.25_{-0.02}^{+0.02}$ &$329/339$\\
32& 56, 57&$-1.05_{-0.04}^{+0.04}$&$-9_{-\infty}^{+19}$&$268_{-20}^{+23}$ &$0.29_{-0.02}^{+0.02}$ &$339/339$\\
33& 57, 58&$-1.01_{-0.04}^{+0.04}$&$-9_{-\infty}^{+19}$&$252_{-18}^{+19}$&$0.36_{-0.02}^{+0.02}$ & $378/339$\\
34& 58, 59&$-1.00_{-0.04}^{+0.05}$&$-2.9_{-1.0}^{+0.3}$&$245_{-21}^{+25}$& $0.37_{-0.01}^{+0.03}$ &$390/339$\\
35& 59, 60&$-1.05_{-0.05}^{+0.05}$&$-2.9_{-1.1}^{+0.3}$&$233_{-23}^{+26}$&$0.31_{-0.02}^{+0.03}$  &$383/339$\\
36& 60, 61&$-1.00_{-0.06}^{+0.06}$&$-2.6_{-0.6}^{+0.2}$&$212_{-22}^{+28}$  &$0.37_{-0.03}^{+0.04}$&$369/339$\\
37& 61, 63&$-1.00_{-0.05}^{+0.05}$&$-2.5_{-0.2}^{+0.2}$&$165_{-12}^{+13}$  &$0.27_{-0.02}^{+0.03}$&$421/339$\\
38& 63, 64&$-1.0_{-0.1}^{+0.1}$&$-2.7_{-0.6}^{+0.3}$&$141_{-16}^{+17}$&$0.25_{-0.04}^{+0.05}$  &$305/339$\\
39& 64, 65&$-1.0_{-0.1}^{+0.1}$&$-2.8_{-\infty}^{+0.3}$&$154_{-16}^{+24}$&$0.29_{-0.04}^{+0.04}$ & $316/339$\\
40& 65, 67&$-1.00_{-0.04}^{+0.04}$&$-2.8_{-0.3}^{+0.2}$&$165_{-10}^{+11}$ &$0.39_{-0.03}^{+0.03}$ &$551/339$\\
41& 67, 68&$-1.04_{-0.06}^{+0.06}$&$-3_{-1}^{+0.3}$&$137_{-11}^{+12}$& $0.39_{-0.04}^{+0.05}$& $415/339$\\
42& 68, 69&$-1.03_{-0.05}^{+0.05}$&$-3_{-0.7}^{+0.3}$&$153_{-12}^{+12}$ &$0.49_{-0.04}^{+0.05}$ &$382/339$\\
43& 69, 70&$-1.0_{-0.1}^{+0.2}$&$-2.30_{-0.15}^{+0.13}$&$87_{-16}^{+13}$ &$0.43_{-0.09}^{+0.23}$ &$389/339$\\
44& 70, 71&$-1.08_{-0.06}^{+0.06}$&$-2.80_{-0.64}^{+0.25}$&$140_{-12}^{+15}$  &$0.34_{-0.04}^{+0.04}$&$379/339$\\
45& 71, 72&$-1.00_{-0.02}^{+0.04}$&$-9_{-\infty}^{+19}$&$158_{-4}^{+8}$&$0.46_{-0.03}^{+0.03}$  &$363/339$\\
46& 72, 73&$-1.02_{-0.08}^{+0.09}$&$-2.8_{-0.4}^{+0.3}$&$93_{-8}^{+8}$& $0.39_{-0.06}^{+0.08}$ &$316/339$\\
47& 73, 74&$-0.9_{-0.2}^{+0.8}$&$-2.4_{-0.2}^{+0.2}$&$56_{-19}^{+12}$& $0.45_{-0.18}^{+1.79}$ &$315/339$\\
48& 74, 76&$-1.10_{-0.06}^{+0.07}$&$-2.8_{-0.3}^{+0.2}$&$89_{-6}^{+6}$  &$0.331_{-0.04}^{+0.05}$&$506/339$\\
49& 76, 77&$-1.1_{-0.1}^{+0.1}$&$-2.9_{-0.7}^{+0.4}$&$84_{-8}^{+9}$  &$0.27_{-0.05}^{+0.07}$&$313/339$\\
50& 77, 78&$-0.3_{-0.9}^{+0.5}$&$-2.1_{-0.8}^{+0.1}$&$47_{-8}^{+57}$& $1.68_{-1.68}^{+3.71}$& $318/339$\\
51& 78, 82&$-1.2_{-0.1}^{+0.1}$&$-2.6_{-0.2}^{+0.2}$&$75_{-7}^{+6}$  &$0.15_{-0.02}^{+0.03}$&$500/339$\\
52& 82, 84&$-1.1_{-0.1}^{+0.1}$&$-2.4_{-0.2}^{+0.2}$&$139_{-14}^{+15}$&$0.20_{-0.02}^{+0.03}$ & $441/339$\\
53& 84, 85&$-1.1_{-0.1}^{+0.2}$&$-2.70_{-\infty}^{+0.35}$&$109_{-17}^{+28}$&$0.11_{-0.03}^{+0.04}$  &$252/339$\\
54& 85, 90&$-1.2_{-0.1}^{+0.2}$&$-2.6_{-0.1}^{+0.3}$&$76_{-14}^{+7}$&$0.13_{-0.02}^{+0.08}$ & $560/339$\\
55& 90, 101&$-1.2_{-0.1}^{+0.1}$&$-2.4_{-0.3}^{+0.2}$&$83_{-10}^{+9}$ &$0.06_{-0.01}^{+0.02}$ &$529/339$\\
56& 101, 113&$-1.30_{-0.07}^{+0.08}$&$-3.0_{-1.0}^{+0.4}$&$86_{-8}^{+8}$&$0.04_{-0.01}^{+0.01}$  &$506/339$\\
57& 113, 136&$-1.3_{-0.1}^{+0.1}$&$-3_{-0.8}^{+0.5}$&$66_{-6}^{+4}$&$0.036_{-0.004}^{+0.008}$ & $676/339$\\
58& 136, 137&$-1.3_{-0.1}^{+0.1}$&$-9_{-\infty}^{+19}$&$83_{-10}^{+13}$&$0.08_{-0.02}^{+0.03}$ & $266/339$\\
59& 137, 150&$-1.3_{-0.1}^{+0.1}$&$-3_{-\infty}^{+0.4}$&$66_{-5}^{+5}$  &$0.04_{-0.01}^{+0.01}$&$491/339$\\
60& 150, 171&$-1.1_{-0.3}^{+1.4}$&$-2.3_{-0.2}^{+0.2}$&$33_{-12}^{+8}$  &$0.04_{-0.02}^{+0.26}$&$382/339$\\
\hline
Broken power-law&&& &  &&\\
\hline 
Sr. no. &(t$_1$,t$_2$) & $\alpha_1$      &   $E_1$ (keV)      &   $\alpha_2$          & $E_2$ (keV) &  $\beta$  & $K$ & \sw{pgstat}/dof\\ \hline
1& -1, 7&$[-3.0]$&$12.3_{-1.1}^{+0.2}$&$1.6_{-0.6}^{+0.6}$&$256_{-122}^{+47}$&$7_{-4}^{+\infty}$&$0.0001_{-0.00004}^{+0.002}$& $408/337$\\
2& 7, 9&$[-3.0]$&$12.7_{-2.0}^{+1.8}$&$1.4_{-0.1}^{+0.1}$&$142_{-28}^{+59}$&$2.4_{-0.3}^{+0.6}$&$0.0002_{-0.00007}^{+0.0002}$&$275/338$ \\
3& 9, 10&$0.9_{-0.5}^{+0.7}$&$17_{-17}^{+\infty}$&$1.3_{-0.1}^{+0.1}$&$194_{-51}^{+73}$&$2.5_{-0.4}^{+0.7}$&$5_{-5}^{+29}$&$215/338$ \\
4& 10, 12&$-0.5_{-\infty}^{+0.5}$&$17.0_{-1.5}^{+1.2}$&$1.25_{-0.04}^{+0.05}$&$170_{-23}^{+41}$&$2.1_{-0.1}^{+0.1}$&$0.14_{-0.01}^{+0.02}$&$287/338$ \\
5& 12, 14&$0.2_{-0.7}^{+0.6}$&$18_{-4}^{+5}$&$1.20_{-0.05}^{+0.04}$&$212_{-27}^{+52}$&$2.2_{-0.1}^{+0.1}$&$0.9_{-0.9}^{+2.6}$&$315/337$ \\\hline
\end{tabular}
\end{center}
\end{tiny}
\end{table*}

\addtocounter{table}{-1}

\begin{table*}
\caption{Time-resolved Spectral fitting (continued)}
\label{tab:specfitting}
\begin{tiny}
\begin{center}
\begin{tabular}{c|cccccccD}
\hline
Broken power-law&&& &  &&\\
\hline 
Sr. no. &(t$_1$,t$_2$) & $\alpha_1$      &   $E_1$ (keV)      &   $\alpha_2$          & $E_2$ (keV) &  $\beta$  & $K^b$ & \sw{pgstat}/dof\\ \hline
6& 14, 17&$-1_{-1}^{+1}$&$16_{-2}^{+3}$&$1.20_{-0.04}^{+0.05}$&$185_{-22}^{+50}$&$2.1_{-0.1}^{+0.1}$&$0.023_{-0.023}^{+0.296}$&$373/337$ \\
7& 17, 18&$-0.8_{0.8}^{+0.8}$&$15_{-2}^{+2}$&$1.10_{-0.06}^{+0.07}$&$124_{-13}^{+23}$&$2.2_{-0.1}^{+0.1}$&$0.11_{-0.02}^{+0.03}$&$288/338$ \\
8& 18, 19&$0.4_{-0.4}^{+0.3}$&$19_{-19}^{+\infty}$&$1.00_{-0.06}^{+0.06}$&$138.0_{-15.4}^{+22.3}$&$2.2_{-0.1}^{+0.1}$&$2.98_{-2.98}^{+4.78}$ &$218/338$ \\
9& 19, 20&$-0.2_{-0.9}^{+0.8}$&$17_{-4}^{+4}$&$1.10_{-0.05}^{+0.05}$&$146_{-12}^{+14}$&$2.3_{-0.1}^{+0.1}$&$0.67_{-0.67}^{+4.73}$&$337/337$ \\
10& 20, 24&$-0.2_{-0.7}^{+0.4}$&$17_{-2}^{+2}$&$1.10_{-0.02}^{+0.02}$&$158_{-8}^{+8}$&$2.30_{-0.05}^{+0.05}$&$0.80_{-0.66}^{+1.43}$&$611/337$\\
11& 24, 26&$-0.8_{-0.8}^{+1.2}$&$15_{-2}^{+5}$&$1.40_{-0.04}^{+0.06}$&$118_{-9}^{+10}$&$2.5_{-0.1}^{+0.1}$&$0.25_{-0.25}^{+2.07}$&$386/337$ \\
12& 26, 29&$0.2_{-0.5}^{+0.3}$&$18_{-1}^{+2}$&$1.50_{-0.02}^{+0.02}$&$352_{-26}^{+24}$&$5.3_{-0.6}^{+0.8}$&$0.013_{-0.001}^{+0.001}$&$571/337$ \\
13& 29, 30&$-1.3_{-0.8}^{+1.2}$&$15.5_{-1.5}^{+2.5}$&$1.30_{-0.04}^{+0.04}$&$205_{-28}^{+30}$&$2.6_{-0.2}^{+0.2}$&$0.044_{-0.044}^{+0.490}$&$283/337$ \\
14& 30, 31&$-0.1_{-0.6}^{+0.5}$&$15_{-15}^{+\infty}$&$1.20_{-0.03}^{+0.03}$&$177_{-14}^{+16}$&$2.4_{-0.1}^{+0.1}$&$1.16_{-1.16}^{+3.50}$&$298/338$ \\
15& 31, 32&$-0.6_{-0.5}^{+0.4}$&$16.7_{-16.7}^{+\infty}$&$1.10_{-0.03}^{+0.03}$&$186.0_{-17.3}^{+20.4}$&$2.3_{-0.1}^{+0.1}$&$0.35_{-0.35}^{+0.70}$&$408/338$ \\
16& 32, 34&$-0.2_{-0.8}^{+0.4}$&$18.0_{-2.3}^{+2.7}$&$1.10_{-0.03}^{+0.03}$&$175.0_{-11}^{+13}$&$2.30_{-0.05}^{+0.06}$&$1.09_{-1.09}^{+2.17}$&$485/337$ \\
17& 34, 35&$0.06_{-0.60}^{+0.40}$&$19_{-4}^{+4}$&$1.10_{-0.04}^{+0.04}$&$174_{-13}^{+15}$&$2.3_{-0.1}^{+0.1}$&$2.15_{-2.15}^{+4.08}$&$317/337$ \\
18& 35, 36&$0.1_{-0.7}^{+0.4}$&$18_{-3}^{+3}$&$1.10_{-0.03}^{+0.04}$&$159_{-12}^{+14}$&$2.4_{-0.1}^{+0.1}$&$2.31_{-2.31}^{+4.52}$&$319/337$ \\
19& 36, 38&$-0.3_{-0.6}^{+0.4}$&$17_{-1.5}^{+1.4}$&$1.30_{-0.03}^{+0.03}$&$116_{-6}^{+7}$&$2.50_{-0.07}^{+0.07}$&$1.10_{-1.10}^{+1.69}$&$386/337$ \\
20& 38, 39&$-0.3_{-1.0}^{+0.5}$&$17_{-2}^{+2}$&$1.60_{-0.05}^{+0.05}$&$114_{-13}^{+16}$&$2.7_{-0.2}^{+0.2}$&$0.94_{-0.94}^{+2.66}$&$294/337$ \\
21& 39, 41&$0.3_{-0.3}^{+0.3}$&$19.0_{-1.4}^{+1.5}$&$1.80_{-0.05}^{+0.04}$&$126_{-22}^{+22}$&$3_{-0.3}^{+0.3}$&$5.24_{-3.10}^{+5.23}$&$391/337$ \\
22& 41, 44&$-0.6_{-0.3}^{+0.2}$&$15_{-15}^{+\infty}$&$1.60_{-0.02}^{+0.02}$&$133_{-10}^{+11}$&$2.80_{-0.14}^{+0.14}$&$0.51_{-0.29}^{+0.44}$&$443/338$ \\
23& 44, 45&$[-2.5]$&$12.3_{-0.7}^{+0.6}$&$1.40_{-0.04}^{+0.04}$&$108_{-11}^{+13}$&$2.5_{-0.1}^{+0.2}$&$0.005_{-0.001}^{+0.001}$&$271/338$ \\
24& 45, 46&$[-2.0]$&$14.4_{-0.6}^{+0.6}$&$1.40_{-0.03}^{+0.03}$&$169_{-21}^{+16}$&$2.6_{-0.2}^{+0.2}$&$0.013_{-0.001}^{+0.002}$&$292/338$ \\
25& 46, 47&$[-3.0]$&$13.0_{-0.5}^{+0.5}$&$1.50_{-0.05}^{+0.04}$&$110_{-17}^{+12}$&$2.5_{-0.2}^{+0.2}$&$0.0012_{-0.0002}^{0.0002}$&$313/338$ \\
26& 47, 48&$[-3.0]$&$13.0_{-0.6}^{+0.6}$&$1.60_{-0.05}^{+0.04}$&$121_{-14}^{+17}$&$3.0_{-0.3}^{+0.3}$&$0.00128_{-0.0002}^{+0.0003}$&$290/338$ \\
27& 48, 50&$[-3.0]$&$12.6_{-0.4}^{+0.4}$&$1.80_{-0.05}^{+0.04}$&$104_{-23}^{+13}$&$3.0_{-0.4}^{+0.4}$&$0.00114_{-0.0002}^{+0.0002}$&$288/338$ \\
28& 50, 51&$[-3.0]$&$13.3_{-0.7}^{+0.7}$&$1.60_{-0.06}^{+0.06}$&$109_{-17}^{+20}$&$2.8_{-0.3}^{+0.4}$&$0.00085_{-0.0002}^{+0.0002}$&$255/338$ \\
29& 51, 53&$-0.8_{-1.0}^{+0.8}$&$15.0_{-1.4}^{+1.6}$&$1.60_{-0.04}^{+0.04}$&$113_{-13}^{+15}$&$2.7_{-0.2}^{+0.2}$&$0.24_{-0.24}^{+1.53}$&$335/337$ \\
30& 53, 55&$[-3.0]$&$13_{-0.4}^{+0.4}$&$1.40_{-0.03}^{+0.03}$&$157_{-14}^{+17}$&$2.5_{-0.1}^{+0.1}$&$0.0011_{-0.0001}^{+0.0002}$&$370/338$ \\
31& 55, 56&$-0.2_{-0.7}^{+0.6}$&$17_{-3}^{+3}$&$1.30_{-0.04}^{+0.04}$&$200_{-24}^{+31}$&$2.6_{-0.2}^{+0.2}$&$0.88_{-0.88}^{+3.29}$&$309/337$ \\
32& 56, 57&$-1.2_{-0.9}^{+1.5}$&$14_{-1}^{+4}$&$1.30_{-0.03}^{+0.04}$&$177_{-15}^{+19}$&$2.5_{-0.1}^{+0.1}$&$0.09_{-0.09}^{+0.96}$&$330/337$ \\
33& 57, 58&$-0.6_{-0.8}^{+1}$&$15.0_{-2.3}^{+3.1}$&$1.30_{-0.03}^{+0.03}$&$169_{-13}^{+15}$&$2.50_{-0.12}^{+0.14}$&$0.56_{-0.56}^{+2.12}$&$359/337$ \\
34& 58, 59&$-0.6_{-0.5}^{+0.9}$&$15.0_{-2.3}^{+3.0}$&$1.30_{-0.03}^{+0.03}$&$168_{-12}^{+14}$&$2.4_{-0.1}^{+0.1}$&$0.57_{-0.57}^{+2.99}$&$366/337$ \\
35& 59, 60&$0.4_{-0.4}^{+0.3}$&$21.0_{-2.8}^{+3.4}$&$1.40_{-0.04}^{+0.04}$&$211_{-25}^{+36}$&$2.6_{-0.1}^{+0.2}$&$5.93_{-4.02}^{+6.67}$&$364/337$ \\
36& 60, 61&$0.2_{-0.5}^{+0.4}$&$19.0_{-2.3}^{+3.0}$&$1.40_{-0.03}^{+0.04}$&$166_{-13}^{+14}$&$2.4_{-0.1}^{+0.1}$&$3.40_{-2.58}^{+5.85}$&$330/337$ \\
37& 61, 63&$[-3]$&$13.0_{-0.5}^{+0.5}$&$1.30_{-0.03}^{+0.03}$&$120_{-9}^{+9}$&$2.3_{-0.1}^{+0.1}$&$0.0012_{-0.0002}^{+0.0002}$&$375/338$ \\
38& 63, 64&$-1.4_{-\infty}^{+1.4}$&$14_{-1}^{+1}$&$1.30_{-0.05}^{+0.05}$&$120_{-13}^{+15}$&$2.50_{-0.15}^{+0.16}$&$0.04_{-0.005}^{+0.007}$&$283/338$ \\
39& 64, 65&$0.40_{-0.40}^{+0.35}$&$21_{-3}^{+5}$&$1.50_{-0.05}^{+0.06}$&$164_{-26}^{+23}$&$2.7_{-0.2}^{+0.2}$&$4.51_{-3.04}^{+7.11}$&$289/337$ \\
40& 65, 67&$-0.8_{-1.0}^{+0.7}$&$15.3_{-1.6}^{+1.9}$&$1.00_{-0.03}^{+0.03}$&$132_{-9}^{+10}$&$2.50_{-0.08}^{+0.08}$&$0.32_{-0.32}^{+1.68}$&$461/337$ \\
41& 67, 68&$-0.5_{-1}^{+0.7}$&$16.0_{-1.6}^{+2.1}$&$1.40_{-0.04}^{+0.04}$&$130_{-14}^{+14}$&$2.6_{-0.2}^{+0.2}$&$0.81_{-0.81}^{+3.82}$&$366/337$ \\
42& 68, 69&$0.2_{-0.6}^{+0.4}$&$18_{-4}^{+3}$&$1.4_{-0.1}^{+0.05}$&$146_{-30}^{+18}$&$2.6_{-0.3}^{+0.2}$&$5_{-5}^{+10}$&$340/337$ \\
43& 69, 70&$-0.07_{-0.5}^{+0.4}$&$17.0_{-1.7}^{+1.8}$&$1.60_{-0.05}^{+0.05}$&$109_{-16}^{+18}$&$2.4_{-0.1}^{+0.1}$&$2.15_{-2.15}^{+4.62}$&$334/337$ \\
44& 70, 71&$0.05_{-0.50}^{+0.30}$&$18_{-2}^{+2}$&$1.5_{-0.04}^{+0.04}$&$154_{-18}^{+20}$&$2.7_{-0.2}^{+0.2}$&$2.62_{-2.624}^{+3.64}$& $301/337$\\
45& 71, 72&$-0.2_{-0.5}^{+0.5}$&$17_{-1.6}^{+1.8}$&$1.40_{-0.03}^{+0.03}$&$149_{-10}^{+12}$&$2.8_{-0.1}^{+0.2}$&$1.59_{-1.59}^{+3.97}$&$307/337$ \\
46& 72, 73&$0.3_{-0.5}^{+0.3}$&$18.2_{-2.0}^{+2.2}$&$1.60_{-0.05}^{+0.05}$&$105_{-9}^{+10}$&$2.70_{-0.20}^{+0.14}$&$5.57_{-5.57}^{+8.05}$&$275/337$ \\
47& 73, 74&$0.4_{-0.9}^{+0.3}$&$18_{-5}^{+2}$&$1.8_{-0.2}^{+0.1}$&$110_{-44}^{+31}$&$2.7_{-0.4}^{+0.3}$&$5.60_{-5.60}^{8.12}$&$278/337$ \\
48& 74, 76&$-0.1_{-0.8}^{+0.4}$&$16_{-1.7}^{+1.5}$&$1.6_{-0.05}^{+0.04}$&$108_{-15}^{+12}$&$2.7_{-0.2}^{+0.2}$&$1.86_{-1.86}^{3.62}$&$410/337$ \\
49& 76, 77&$0.4_{-0.5}^{+0.4}$&$19_{-2}^{+3}$&$1.70_{-0.05}^{+0.06}$&$129_{-17}^{+15}$&$3.0_{-0.3}^{+0.4}$&$4.52_{-4.52}^{+8.12}$&$285/337$ \\
50& 77, 78&$-1_{-1}^{+1}$&$15.0_{-1.6}^{+3.3}$&$1.70_{-0.06}^{+0.06}$&$136_{-34}^{+21}$&$3.0_{-0.3}^{+0.4}$&$0.18_{-0.18}^{+3.52}$&$271/337$ \\
51& 78, 82&$0.3_{-0.5}^{+0.3}$&$17.3_{-1.7}^{+1.5}$&$1.80_{-0.04}^{+0.04}$&$108_{-10}^{+12}$&$2.7_{-0.2}^{+0.2}$&$2.91_{-2.15}^{+3.58}$&$381/337$ \\
52& 82, 84&$-1.3_{-0.9}^{+0.9}$&$15.0_{-1.1}^{+1.5}$&$1.50_{-0.04}^{+0.04}$&$128_{-15}^{+20}$&$2.4_{-0.1}^{+0.2}$&$0.06_{-0.06}^{+0.57}$&$352/337$ \\
53& 84, 85&$0.3_{-0.8}^{+0.6}$&$17.0_{-5.3}^{+4.4}$&$1.6_{-0.1}^{+0.1}$&$119_{-24}^{+23}$&$2.6_{-0.3}^{+0.4}$&$2.15_{-2.15}^{+4.85}$&$238/337$ \\
54& 85, 90&$-0.2_{-0.8}^{+0.6}$&$16_{-2}^{+2}$&$1.80_{-0.05}^{+0.04}$&$118_{-21}^{+15}$&$2.8_{-0.2}^{+0.2}$&$0.71_{-0.71}^{2.27}$ &$462/337$\\
55& 90, 101&$[-2]$&$12.7_{-0.4}^{+0.4}$&$1.60_{-0.05}^{+0.03}$&$101_{-23}^{+13}$&$2.5_{-0.3}^{+0.2}$&$0.00456_{-0.00046}^{0.00054}$ &$470/338$\\
56& 101, 113&$[-3]$&$12.8_{-0.4}^{+0.4}$&$1.70_{-0.03}^{+0.03}$&$122_{-12}^{+14}$&$2.9_{-0.3}^{+0.3}$&$0.00027_{-0.00003}^{+0.00004}$& $403/338$\\
57& 113, 136&$[-3.0]$&$12.20_{-0.30}^{+0.04}$&$1.8_{-0.06}^{+0.03}$&$97_{-15}^{+15}$&$2.9_{-0.3}^{+0.4}$&$0.0003_{-0.00003}^{+0.000005}$&$556/337$ \\
58& 136, 137&$[-3.0]$&$12.4_{-1.0}^{+0.9}$&$1.80_{-0.06}^{+0.07}$&$181_{-68}^{+51}$&$5.1_{-2.2}^{+\infty}$& $0.0008_{-0.0002}^{+0.0003}$&$256/338$\\
59& 137, 150&$[-3.0]$&$11.6_{-0.4}^{+0.1}$&$1.70_{-0.04}^{+0.04}$&$94_{-10}^{+12}$&$2.9_{-0.2}^{+0.4}$&$0.0004_{-0.00006}^{+0.00001}$&$ 449/337$\\
60& 150, 171&$0.4_{-0.8}^{+0.8}$&$14.0_{-1.5}^{+4}$&$2.0_{-0.1}^{+0.1}$&$95_{-62}^{+39}$&$2.9_{-0.6}^{+0.8}$&$0.93_{-0.93}^{+5.78}$ &$372/337$\\\hline
\end{tabular}
\\ $b$: $photons~ keV^{-1} ~ cm^{-2} ~s^{-1}$ at 1 keV
\end{center}
\end{tiny}
\end{table*}

\addtocounter{table}{-1}

\begin{table*}
\caption{Time-resolved Spectral fitting (continued)}
\label{tab:specfitting}
\begin{tiny}
\begin{center}
\begin{tabular}{c|ccccccccccD}
\hline
BB+Band&&&   &&\\
\hline 
Sr. no. &(t$_1$,t$_2$) &$\alpha$& $\beta$      & $E_p$ (keV)  &   $K_B$       & $kT_{BB}$ (keV)                   & $K_{BB}$ & \sw{pgstat}/dof\\ \hline
1& -1, 7 &$-0.9_{-0.3}^{+0.5}$&$-9.4_{-\infty}^{+19}$&$153_{-23}^{+34}$&$0.02_{-0.01}^{+0.01}$ & $6.5_{-1.6}^{+2.4}$&$0.50_{-0.34}^{+0.39}$& $414/337$     \\
2&7, 9& $0.01_{-0.90}^{+1.80}$&$-2.4_{-\infty}^{+0.3}$&$149_{-34}^{+68}$&$0.08_{-0.05}^{+0.36}$ & $5.5_{-1.1}^{+1.3}$&$1.36_{-0.81}^{+0.78}$&  $ 276/337$         \\
3&9, 10& $-1.1_{-0.2}^{+0.2}$&$-9_{-\infty}^{+19}$&$299_{-41}^{+191}$&$0.05_{-0.01}^{+0.01}$ &$13_{-8}^{+\infty}$&$0.73_{-0.73}^{+2.00}$& $212/337$ \\
4&10, 12& $-0.4_{-0.4}^{+0.4}$&$-2.1_{-0.2}^{+0.1}$&$215_{-34}^{+81}$ & $0.12_{-0.04}^{+0.06}$ &$7.0_{-1.0}^{+1.5}$& $2.55_{-1.03}^{+0.78}$&  $296   /337$     \\
5&12, 14& $-0.8_{-0.3}^{+0.2}$&$-2.2_{-\infty}^{+0.1}$&$331_{-59}^{+383}$&$0.10_{-0.04}^{+0.02}$ &$7.2_{-3.1}^{+\infty}$&$0.79_{-0.79}^{+1.05}$&  $309   /337$   \\
6&14, 17& $-0.5_{-0.2}^{+0.4}$&$-2.1_{-0.1}^{+0.1}$&$260_{-54}^{+66}$&$0.11_{-0.02}^{+0.05}$& $8.0_{-1.1}^{+1.5}$&$2.09_{-0.88}^{+1.06}$  & $383  /337$    \\
7&17, 18& $-0.5_{-0.2}^{+0.5}$&$-2.3_{-0.3}^{+0.2}$&$185_{-34}^{+51}$&$0.22_{-0.04}^{+0.17}$&$7_{-2}^{+7}$&  $2.13_{-1.90}^{+2.20}$ & $287   /337$    \\
8&18, 19& $-0.7_{-0.4}^{+0.3}$&$-2.4_{-\infty}^{+0.2}$&$219_{-34}^{+219}$&$0.23_{-0.12}^{+0.08}$&$10_{-10}^{+\infty}$&$1.20_{-1.20}^{+2.24}$ & $209   /337$   \\
9&19, 20& $-0.6_{-0.2}^{+0.5}$&$-2.4_{-0.3}^{+0.2}$&$218_{-41}^{+42}$&$0.31_{-0.06}^{+0.17}$&$7.0_{-1.5}^{+3.0}$&$3.16_{-2.21}^{+2.99}$   &$335   /337$     \\
10&20, 24&$-0.5_{-0.1}^{+0.1}$&$-2.4_{-0.1}^{+0.1}$&$212_{-12}^{+14}$&$0.39_{-0.04}^{+0.04}$&$7.3_{-0.6}^{+0.8}$&  $4.26_{-1.14}^{+1.19}$&  $579   /337$    \\
11&24, 26&$-0.1_{-0.3}^{+0.4}$&$-2.5_{-0.1}^{+0.1}$&$122_{-9}^{+10}$&$0.63_{-0.19}^{+0.41}$&$6.0_{-0.4}^{+0.4}$& $7.79_{-1.94}^{+2.14}$ & $390   /337$    \\
12&26, 29&$-0.8_{-0.1}^{+0.1}$&$-10.0_{-\infty}^{+0.0}$&$196_{-7}^{+8}$&$0.25_{-0.02}^{+0.02}$&$6.0_{-0.5}^{+0.5}$& $4.55_{-0.94}^{+0.99}$   &       $458   /337$      \\
13&29, 30&$-0.5_{-0.2}^{+0.1}$&$-2.6_{-0.3}^{+0.2}$&$213_{-21}^{+33}$&$0.30_{-0.06}^{+0.11}$&$7.0_{-0.8}^{+0.9}$&$5.63_{-2.08}^{+2.14}$  & $293   /337$    \\
14&30, 31&$-0.7_{-0.2}^{+0.2}$&$-3.0_{-0.6}^{+0.2}$&$256_{-31}^{+59}$&$0.29_{-0.04}^{+0.06}$&$5.7_{-1.2}^{+1.5}$& $2.74_{-1.37}^{+2.03}$& $289  /337$    \\
15&31, 32&$-0.6_{-0.2}^{+0.2}$&$-2.4_{-0.2}^{+0.1}$&$254_{-30}^{+38}$&$0.40_{-0.06}^{+0.08}$& $7.4_{-1.5}^{+2.3}$&$4.14_{-2.36}^{+2.57}$  &    $409   /337$   \\
16&32, 34&$-0.5_{-0.1}^{+0.1}$&$-2.4_{-0.1}^{+0.1}$&$241_{-18}^{+20}$&$0.52_{-0.05}^{+0.07}$&$8.0_{-0.6}^{+0.7}$&  $7.67_{-2.00}^{+2.13}$& $460   /337$    \\
17&34, 35&$-0.5_{-0.1}^{+0.2}$&$-2.4_{-0.1}^{+0.1}$&$241_{-26}^{+27}$&$0.54_{-0.07}^{+0.11}$&$7.7_{-1.2}^{+1.6}$& $5.62_{-2.78}^{+3.14}$& $302   /337$    \\
18&35, 36&$-0.6_{-0.2}^{+0.3}$&$-2.6_{-0.3}^{+0.2}$&$224_{-30}^{+32}$&$0.46_{-0.07}^{+0.13}$&$6.7_{-1.2}^{+1.9}$&  $4.25_{-2.56}^{+3.03}$&$313  /337$      \\
19&36, 38&$0.1_{-0.3}^{+0.4}$&$-2.5_{-0.1}^{+0.1}$&$122_{-7}^{+8}$&$1.23_{-0.36}^{+0.70}$&$6.0_{-0.3}^{+0.3}$&    $11.35_{-2.35}^{+2.50}$ & $400  /337$     \\
20&38, 39&$0.3_{-0.6}^{+1}$&$-2.7_{-0.2}^{+0.2}$&$103.0_{-9.5}^{+10.5}$&$1.21_{-0.61}^{+3.03}$&$5.9_{-0.4}^{+0.4}$&$11.89_{-3.13}^{+3.33}$ & $305  /337$      \\
21&39, 41&$-0.4_{-0.3}^{+0.5}$&$-2.8_{-0.3}^{+0.2}$&$99_{-9}^{+8}$&$0.50_{-0.16}^{0.45}$&$6.0_{-0.3}^{+0.4}$&$11.20_{-1.93}^{+2.22}$   & $399  /337 $    \\
22&41, 44&$-0.4_{-0.2}^{+0.3}$&$-2.8_{-0.2}^{+0.2}$&$116_{-7}^{+7}$&$0.49_{-0.11}^{+0.20}$&$5.5_{-0.2}^{+0.2}$&  $10.36_{-1.48}^{+1.60}$&$ 456  /337$     \\
23&44, 45&$-0.5_{-0.4}^{+0.5}$&$-2.6_{-0.3}^{+0.2}$&$116_{-12}^{+17}$&$0.52_{-0.19}^{+0.49}$&$5.0_{-0.6}^{+0.8}$&  $6.01_{-2.89}^{+3.02}$&  $276  /337$    \\
24&45, 46&$-0.4_{-0.1}^{+0.2}$&$-2.5_{-0.2}^{+0.1}$&$167_{-9}^{+16}$&$0.53_{-0.10}^{+0.06}$&$6.5_{-0.5}^{+0.4}$&  $10.77_{-1.48}^{+1.56}$ & $317   /337$    \\
25&46, 47&$-0.2_{-0.5}^{+0.8}$&$-2.5_{-0.3}^{+0.2}$&$107_{-13}^{+16}$&$0.70_{-0.31}^{+1.17}$&$6.0_{-0.5}^{+0.7}$& $9.29_{-2.90}^{+3.36}$ & $320  /337 $   \\
26&47, 48&$-0.3_{-0.5}^{+0.7}$&$-3.1_{-\infty}^{+0.4}$&$108_{-10}^{+14}$&$0.45_{-0.19}^{+0.51}$&$5.7_{-0.5}^{+0.6}$& $8.26_{-2.73}^{+2.63}$  &      $293   /337$    \\
27&48, 50&$-0.3_{-0.5}^{+1.1}$&$-2.8_{-0.6}^{+0.3}$&$81_{-6}^{+8}$&$0.42_{-0.21}^{+1.75}$&$4.95_{-0.34}^{+0.40}$&  $5.87_{-1.80}^{+2.05}$ & $303   /337 $   \\
28&50, 51&$0.1_{-0.7}^{+1.3}$&$-2.8_{-0.6}^{+0.3}$&$99_{-12}^{+13}$&$0.67_{-0.38}^{+2.82}$&$5.5_{-0.5}^{+0.5}$&    $6.90_{-2.59}^{+2.59}$ & $ 268   /337$    \\
29&51, 53&$-0.4_{-0.3}^{+0.3}$&$-2.7_{-0.2}^{+0.2}$&$100_{-7}^{+8}$&$0.58_{-0.17}^{+0.29}$&$5.0_{-0.1}^{+\infty}$& $6.53_{-1.77}^{+1.43}$ & $354   /337 $   \\
30&53, 55&$-0.7_{-0.2}^{+0.2}$&$-2.7_{-0.4}^{+0.2}$&$177_{-17}^{+21}$&$0.29_{-0.05}^{+0.07}$&$6.0_{-0.5}^{+0.6}$&  $5.09_{-1.55}^{+1.56}$& $402  /337$    \\
31&55, 56&$-0.7_{-0.1}^{+0.2}$&$-4_{-\infty}^{+1}$&$241_{-26}^{+23}$&$0.28_{-0.03}^{+0.05}$&$7_{-1}^{+1}$&   $5.48_{-1.98}^{+2.13}$  &$307  /337$    \\
32&56, 57&$-0.9_{-0.1}^{+0.1}$&$-9_{-\infty}^{+19}$&$267_{-20}^{+23}$&$0.305_{-0.022}^{+0.065}$&$7.5_{-1.2}^{+1.4}$&$4.87_{-2.12}^{+2.21}$  & $325   /337$    \\
33&57, 58&$-0.8_{-0.1}^{+0.2}$&$-3.1_{-\infty}^{+0.5}$&$234_{-25}^{+25}$&$0.41_{-0.05}^{+0.07}$&$7_{-1}^{+1}$& $6.84_{-2.49}^{+2.57}$  &  $356   /337$    \\
34&58, 59&$-0.7_{-0.2}^{+0.2}$&$-2.6_{-0.4}^{+0.2}$&$219_{-23}^{+29}$&$0.44_{-0.06}^{+0.09}$&$7_{-1}^{+1}$& $7.37_{-2.67}^{+2.83}$ & $367  /337 $   \\
35&59, 60&$-0.8_{-0.1}^{+0.2}$&$-2.8_{-0.6}^{+0.3}$&$225_{-26}^{+31}$&$0.34_{-0.04}^{+0.06}$&$8_{-1}^{+1}$&$6.39_{-2.24}^{+2.38}$ &  $ 359  /337$    \\
36&60, 61&$-0.6_{-0.2}^{+0.3}$&$-2.5_{-0.3}^{+0.14}$&$190_{-20}^{+35}$&$0.48_{-0.10}^{+0.14}$& $7.0_{-0.7}^{+1.0}$&$9.53_{-2.75}^{+3.05}$  &$330   /337$   \\
37&61, 63&$-0.5_{-0.3}^{+0.4}$&$-2.40_{-0.15}^{+0.12}$&$143_{-15}^{+17}$&$0.46_{-0.12}^{+0.24}$&$5.7_{-0.4}^{+0.5}$& $5.83_{-1.88}^{+2.04}$  &  $389   /337$   \\
38&63, 64&$-0.4_{-0.4}^{+0.6}$&$-2.5_{-0.2}^{+0.2}$&$130_{-18}^{+22}$&$0.41_{-0.14}^{+0.40}$&$6.0_{-0.8}^{+1.3}$& $4.68_{-2.28}^{+2.49}$&  $293   /337 $  \\
39&64, 65&$-0.80_{-0.15}^{+0.40}$&$-3.4_{3.4}^{+0.8}$&$182_{-34}^{+25}$&$0.28_{-0.05}^{+0.15}$&$8.4_{-1.5}^{+1.4}$& $7.31_{-2.24}^{+2.35}$ &  $285 /337$   \\
40&65, 67&$-0.3_{-0.2}^{+0.2}$&$-2.6_{-0.1}^{+0.1}$&$146_{-9}^{+10}$&$0.69_{-0.13}^{+0.19}$&$6.3_{-0.2}^{+0.3}$&    $10.53_{-1.98}^{+2.00}$ &   $464  /337$   \\
41&67, 68&$-0.4_{-0.3}^{+0.4}$&$-2.7_{-0.3}^{+0.2}$&$132_{-11}^{+13}$&$0.66_{-0.17}^{+0.34}$&$6.0_{-0.5}^{+0.6}$& $10.57_{-2.93}^{+3.19}$ &   $373   /337 $  \\
42&68, 69&$-0.6_{-0.2}^{+0.3}$&$-2.7_{-0.4}^{+0.2}$&$147_{-13}^{+15}$&$0.69_{-0.14}^{+0.24}$ & $6.4_{-0.5}^{+0.7}$& $11.30_{-3.00}^{+3.22}$&  $ 336   /337$   \\
43&69, 70&$-0.15_{-0.5}^{+1.1}$&$-2.4_{-0.2}^{+0.1}$&$97_{-15}^{+18}$&$0.96_{-0.51}^{+2.77}$&$6.0_{-0.4}^{+0.7}$&  $11.53_{-3.46}^{+4.10}$&  $345   /337$   \\
44&70, 71&$-0.5_{-0.2}^{+0.3}$&$-2.7_{-0.3}^{+0.2}$&$141_{-13}^{+14}$&$0.53_{-0.12}^{+0.23}$&$6.0_{-0.5}^{+0.5}$& $11.14_{-2.53}^{+2.70}$ &$ 318   /337 $  \\
45&71, 72&$-0.5_{-0.2}^{+0.2}$&$-3.2_{-1.0}^{+0.3}$&$153_{-10}^{+10}$&$0.65_{-0.11}^{+0.17}$&$6.0_{-0.5}^{+0.6}$&  $10.67_{-2.84}^{+2.90}$ & $317  /337 $  \\
46&72, 73&$0.1_{-0.6}^{+1.1}$&$-2.7_{-0.3}^{+0.2}$&$97_{-8}^{+10}$&$1.25_{-0.67}^{3.44}$&$6.0_{-0.4}^{+0.4}$&   $11.90_{-3.81}^{+4.22}$   &  $276   /337$   \\
47&73, 74&$0.1_{-0.6}^{+\infty}$&$-2.6_{-0.2}^{+0.2}$&$81_{-9}^{+10}$&$1.04_{-0.03}^{+0.29}$&$5.2_{-0.3}^{+0.4}$&  $9.73_{-0.47}^{+1.28}$ &      $275   /337 $  \\
48&74, 76&$0.1_{-0.4}^{+0.8}$&$-2.7_{-0.2}^{+0.2}$&$95_{-8}^{+7}$&$1.14_{-0.49}^{+2.26}$&$5.5_{-0.3}^{+0.3}$&      $11.56_{-2.38}^{+2.80}$&   $413 /337$   \\
49&76, 77&$-0.5_{-0.4}^{+0.5}$&$-3_{-0.7}^{+0.3}$&$98_{-9}^{+11}$&$0.40_{-0.15}^{+0.36}$&$6.0_{-0.6}^{+0.9}$&$6.66_{-2.30}^{+2.54}$   &   $290   /337$   \\
50&77, 78&$-0.5_{-0.4}^{+0.6}$&$-2.8_{-0.6}^{+0.3}$&$111_{-13}^{+15}$&$0.38_{-0.14}^{+0.42}$&$6.2_{-0.5}^{+0.7}$&  $9.02_{-2.52}^{+2.62}$ &       $275   /337 $  \\
51& 78, 82&$0.0_{-0.4}^{+0.7}$&$-2.7_{-0.2}^{+0.2}$&$91_{-6}^{+6}$&$0.49_{-0.21}^{+0.66}$&$5.0_{-0.2}^{+0.2}$&     $6.57_{-1.20}^{+1.23}$ &  $386   /337$   \\
52&82, 84&$-0.3_{-0.3}^{+0.5}$&$-2_{-0.2}^{+0.1}$&$131_{-15}^{+15}$&$0.38_{-0.11}^{0.28}$&$6.0_{-0.4}^{+0.5}$&$6.61_{-1.55}^{+1.67}$   & $ 380 /337$   \\
53&84, 85&$0.4_{-1.0}^{+4.0}$&$-2.6_{-0.4}^{+0.3}$&$104_{-19}^{+17}$&$0.54_{-0.36}^{+60.34}$&$5.5_{-0.6}^{+0.8}$& $5.13_{-2.06}^{+2.04}$ & $ 235   /337$   \\
54&85, 90&$-0.3_{-0.4}^{+1}$&$-2.6_{-0.2}^{+0.2}$&$91_{-11}^{+8}$&$0.31_{-0.12}^{+0.43}$&$5.3_{-0.3}^{+0.3}$&$4.64_{-0.97}^{+1.39}$     & $ 463   /337 $  \\
55&90, 101&$-0.2_{-0.4}^{+4.5}$&$-2.5_{-0.2}^{+0.2}$&$88_{-20}^{+101}$&$0.17_{-0.08}^{+75.02}$&$5_{-0.3}^{+0.3}$&  $2.02_{-0.53}^{+1.19}$  &  $ 477   /337$   \\
56&101, 113&$-0.3_{-0.4}^{+0.5}$&$-3_{-0.5}^{+0.3}$&$100_{-8}^{+10}$&$0.09_{-0.03}^{+0.08}$&$5_{-0.3}^{+0.3}$&      $1.69_{-0.36}^{+0.36}$&  $441  /337$   \\
57&113, 136&$0.25_{-0.90}^{-0.25}$&$-2.6_{-0.6}^{+0.2}$&$72_{-11}^{+11}$&$0.27_{-0.19}^{+40.62}$&$4.5_{-0.2}^{+0.2}$& $1.65_{-0.46}^{+0.56}$ &   $610   /337  $ \\
58&136, 137&$-0.8_{-0.4}^{+0.6}$&$-9.3_{9.3}^{+19}$&$103_{-15}^{+18}$&$0.11_{-0.04}^{+0.102}$&$5.0_{-0.7}^{+0.8}$& $2.85_{-1.70}^{+1.71}$ & $257   /337  $ \\
59&137, 150&$-0.5_{-0.4}^{+0.7}$&$-3_{-0.6}^{+0.3}$&$79_{-6}^{+6}$&$0.10_{-0.04}^{+0.15}$&$4.0_{-0.3}^{+0.3}$& $1.40_{-0.47}^{+0.49}$    &  $ 463   /337 $  \\
60&150, 171&$-1.1_{-0.4}^{+1.1}$&$-2.8_{-\infty}^{+0.5}$&$63_{-24}^{+16}$&$0.02_{-0.01}^{+0.36}$&$4_{-1}^{+1}$& $0.57_{-0.30}^{+0.84}$   & $ 374  /337$   \\

\hline

\end{tabular}
\end{center}
\end{tiny}
\end{table*}

\addtocounter{table}{-1}

\begin{table*}
\caption{Time-resolved Spectral fitting (continued)}
\label{tab:specfitting}
\begin{tiny}
\begin{center}
\begin{tabular}{c|ccccccccccccD}
\hline
\sw{grbcomp}&&& &  &&\\
\hline 
Sr. no. &(t$_1$,t$_2$) &$kT_{s}$ (keV)      &   $kT_{e}$ (keV)      &   $\tau$          & $\delta$ &  $\alpha_b$  & $R_{ph}$ ($10^{10}$) cm& \sw{pgstat}/dof\\ \hline
1&-1, 7 &$[5.3]$&$90_{-14}^{+4}$&$[3.2]$&$1.50\pm-0.05$&$30_{-24}^{+\infty}$ &$8.7_{-0.3}^{+0.2}$ &$412/339$ \\
2&7, 9 &$[4.1]$&$77_{-32}^{+20}$&$4.0_{-0.9}^{+1.5}$&$1.80\pm0.07$&$2.6_{-1.3}^{+\infty}$ & $17_{-1}^{+1}$ &$281/338$ \\
3&9, 10 &$[4.5]$&$78_{-32}^{+40}$&$4_{-1}^{+1}$&$1.70\pm0.06$&$1.8_{-0.7}^{+1.6}$ & $20_{-1}^{+1}$&$213/338$\\
4&10, 12 &$7.7_{-1.8}^{+1.5}$&$117_{-62}^{+90}$&$3.0_{-0.8}^{+2.1}$&$1.20\pm0.03 $ &$1.4_{-0.3}^{+0.6}$ & $10_{-2}^{+6}$&$287/337$\\
5&12, 14 &$6.5_{-2.6}^{+1.7}$&$86_{-22}^{+30}$&$4_{-1}^{+1}$& $1.60\pm0.05$ &$1.3_{-0.2}^{0.2}$ &$14_{-4}^{+17}$ &$ 305/337$\\
6&14, 17 &$8.1_{-1.7}^{+1.5}$&$91_{-31}^{+45}$&$3.6_{-0.8}^{+1.5}$& $1.50\pm0.05$ &$1.2_{-0.1}^{+0.2}$ &$9_{-2}^{+4}$ & $371/33$7\\
7&17, 18 &$[7]$&$57_{-18}^{+32}$&$5.0_{-1.4}^{+2.7}$&$2.40\pm0.13$&$1.4_{-0.2}^{+0.4}$ & $14.7_{-0.3}^{+0.3}$&$283/338$\\
8&18, 19 &$[6]$&$58_{-12}^{+16}$&$5_{-1}^{+1}$& $2.36\pm0.12$&$1.4_{-0.2}^{+0.3}$ & $21.6_{-0.4}^{+0.4}$&$207/338$ \\
9&19, 20 &$7.4_{-2.3}^{+1.7}$&$62_{-16}^{+19}$&$4.5_{-1.0}^{+2.0}$& $2.20\pm0.10$ &$1.5_{-0.2}^{+0.3}$ &$17_{-4}^{+16}$ &$329/337$\\
10&20, 24 &$7.3_{-0.8}^{+0.7}$&$63_{-6}^{+7}$&$4.8_{-0.4}^{+0.5}$& $2.2\pm0.1$ &$1.5_{-0.1}^{+0.1}$ &$19_{-2}^{+4}$ &$551/337$\\
11&24, 26 &$7.6_{-0.7}^{+0.7}$&$35.5_{-9.3}^{+11.5}$&$7_{-2}^{+15}$& $3.8\pm0.3$ &$1.6_{-0.2}^{+0.3}$ &$17_{-2}^{+3}$ &$386/337$\\
12&26, 29 &$6.8_{-0.5}^{+0.5}$&$53.0_{-7.8}^{+9.8}$&$5.0_{-0.7}^{+1.0}$& $2.6\pm0.14$ &$1.70_{-0.15}^{+0.2}$ &$20_{-2}^{+3}$ &$462/337 $\\
13&29, 30 &$7.7_{-1.2}^{+1.1}$&$69_{-17}^{+22}$&$4.5_{-0.9}^{+1.5}$& $2.0 \pm 0.1$ &$1.7_{-0.2}^{+0.4}$ & $16_{-3}^{+6}$& $286/337$\\
14&30, 31& $6.3_{-1.5}^{+1.1}$&$67.2_{-10.1}^{+13.7}$&$4.9_{-0.7}^{+0.9}$& $2.0 \pm 0.1$&$1.6_{-0.2}^{+0.3}$ & $23_{-5}^{+16}$&$285/337$\\
15&31, 32& $7.1_{-1.5}^{+1.3}$&$74.3_{-12.6}^{+15.3}$&$4.5_{-0.6}^{+0.8}$&$1.83\pm0.07$&$1.5_{-0.1}^{+0.2}$ &$21_{-4}^{+11}$ &$397/337$ \\
16&32, 34& $8.3_{-0.9}^{+0.8}$&$71_{-9}^{+10}$&$4.6_{-0.5}^{+0.6}$& $1.9\pm0.1$&$1.5_{-0.1}^{+0.1}$ & $19_{-2}^{+4}$&$437/337$\\
17&34, 35& $7.6_{-1.6}^{+1.3}$&$68_{-11}^{+12}$&$4.8_{-0.6}^{+1.0}$& $2 \pm 0.1$&$1.5_{-0.1}^{+0.2}$ & $22_{-4}^{+11}$&$291/337$\\
18&35, 36& $7.0_{-1.5}^{+1.2}$&$64_{-11}^{+12}$&$5.0_{-0.7}^{+1.0}$& $2\pm0.1$ &$1.6_{-0.2}^{+0.2}$ & $24_{-5}^{+14}$&$304/337$\\
19&36, 38& $8.0_{-0.6}^{+0.5}$&$30.0_{-5.2}^{+6.4}$&$12.6_{-5.3}^{+\infty}$& $4.5\pm0.5$ &$1.6_{-0.1}^{+0.1}$ & $18_{-8}^{+2}$&$395/337$\\
20&38, 39&$7.5_{-0.5}^{+0.5}$&$28.5_{-2.5}^{+2.6}$&$[10]$& $4.7\pm0.5$ &$1.8_{-0.2}^{+0.2}$ & $18_{-2}^{+2}$&$302/338$\\
21&39, 41& $7.0_{-0.5}^{+0.5}$&$50_{-20}^{+25}$&$3.8_{-0.9}^{+2.5}$& $2.7\pm0.2$ &$2.4_{-0.6}^{+1.7}$ & $20_{-2}^{+3}$&$388/337$\\
22&41, 44& $7.0_{-0.4}^{+0.4}$&$40_{-8.4}^{+10}$&$5.8_{-1.2}^{+2.7}$&$3.4\pm0.3$&$2.0_{-0.3}^{+0.3}$ & $21_{-2}^{+2}$&$448/337$\\
23&44, 45& $6.0_{-1.3}^{+0.9}$&$42_{-15}^{+18}$&$5.0_{-1.3}^{+5.1}$&$3.2\pm0.2$ &$1.8_{-0.3}^{+0.6}$ & $25_{-5}^{+14}$&$271/337$\\
24&45, 46& $7.8_{-0.8}^{+0.8}$&$54_{-13}^{+16}$&$5.0_{-1.1}^{+2.3}$& $2.5\pm0.1$ &$1.7_{-0.2}^{+0.4}$ & $19_{-3}^{+4}$&$310/337$ \\
25&46, 47& $7_{-1}^{+1}$&$47_{-23}^{+32}$&$4.4_{-1.4}^{+11.6}$& $2.9\pm0.2$ &$1.9_{-0.4}^{+1.1}$ &$20_{-3}^{+7}$ &$314/337$\\
26&47, 48& $7_{-1}^{+1}$&$47_{-18}^{+25}$&$5.0_{-1.4}^{+6.2}$& $2.9\pm0.2$ &$2.8_{-0.9}^{+\infty}$ &$19_{-3}^{+6}$ &$290/337$\\
27&48, 50& $6.1_{-0.8}^{+0.7}$&$33_{-17}^{+24}$&$5_{-2}^{+\infty}$& $4.1\pm0.4$ &$2.2_{-0.6}^{+2.1}$ &$20_{-3}^{+6}$ &$302/337$\\
28&50, 51& $7_{-1}^{+1}$&$27_{-8}^{+24}$&$11_{-7}^{+\infty}$& $5.1\pm0.6$ &$2.0_{-0.4}^{+1.3}$ & $16_{-2}^{+6}$&$267/337$\\
29&51, 53& $6.8_{-0.7}^{+0.6}$&$43_{-13}^{+16}$&$4.8_{-1.1}^{+3}$&$ 3.1\pm0.2$&$2.2_{-0.4}^{+0.8}$ &$20_{-3}^{+5}$ &$335/337$\\
30&53, 55& $6.4_{-0.7}^{+0.6}$&$64_{-12}^{+14}$&$4.3_{-0.6}^{+0.9}$&$ 2.1\pm0.1 $&$1.9_{-0.3}^{+0.4}$ &$23_{-3}^{+5}$ &$387/337$\\
31&55, 56& $7.2_{-1.5}^{+1.1}$&$85_{-18}^{+20}$&$3.9_{-0.5}^{+0.8}$&$ 1.60\pm0.06 $&$2.3_{-0.5}^{+1.2}$ & $19_{-3}^{+8}$&$301/337$\\
32&56, 57& $6.7_{-1.3}^{+1.1}$&$83_{-17}^{+20}$&$3.8_{-0.5}^{+0.7}$&$ 1.60\pm0.06 $&$2.0_{-0.3}^{+0.6}$ & $24_{-4}^{+11}$&$317/337$\\
33&57, 58& $6.9_{-1.3}^{+1.0}$&$78_{-14}^{+16}$&$4.0_{-0.5}^{+0.7}$&$1.75\pm0.07 $&$2.1_{-0.4}^{+0.6}$ &$25_{-4}^{+11}$ &$349/337$ \\
34&58, 59& $7.0_{-1.1}^{+1.0}$&$72_{-13}^{+15}$&$4.2_{-0.6}^{+0.8}$&$1.9\pm0.1 $&$1.8_{-0.2}^{+0.3}$ &$24_{-4}^{+9}$ &$359/337$ \\
35&59, 60& $6.9_{-1.2}^{+1.0}$&$86_{-18}^{+20}$&$3.5_{-0.5}^{+0.6}$&$1.60\pm0.06$&$2.0_{-0.3}^{+0.5}$ & $23_{-4}^{+101}$&$345/337$\\
36&60, 61& $7.5_{-1.2}^{+1.0}$&$81_{-21}^{+25}$&$3.5_{-0.6}^{+1.0}$&$1.67\pm0.06 $&$1.8_{-0.3}^{+0.5}$ &$21_{-3}^{+7}$ &$315/337$\\
37&61, 63& $6.7_{-0.7}^{+0.7}$&$44_{-10}^{+12}$&$5.6_{-1.1}^{+2.5}$&$3.1\pm0.2$&$1.5_{-0.2}^{+0.2}$ &$21_{-3}^{+5}$ &$378/337$\\
38&63, 64& $6.6_{-1.4}^{+1.1}$&$45_{-16}^{+20}$&$5_{-1}^{+5}$&$3.0\pm0.2$&$1.7_{-0.3}^{+0.6}$ &$19_{-4}^{+11}$ &$288/337$\\
39&64, 65& $7.3_{-1.8}^{+1.2}$&$82.7_{-25.2}^{+30}$&$3.3_{-0.6}^{+1}$&$1.60\pm0.06$&$2.5_{-0.7}^{+2.5}$ & $19_{-4}^{+7}$&$277/337$\\
40&65, 67& $7.5_{-0.6}^{+0.5}$&$47_{-9}^{+10}$&$5.5_{-1}^{+1.8}$&$2.9\pm0.2$&$1.7_{-0.2}^{+0.2}$ & $21_{-2}^{+3}$&$447/337$\\
41&67, 68& $7.1_{-0.8}^{+0.7}$&$49_{-13}^{+16}$&$5_{-1.1}^{+2.4}$&$2.8\pm0.2$&$2.0_{-0.3}^{+0.6}$ &$23_{-3}^{+6}$ &$366/337$\\
42&68, 69& $7.0_{-0.8}^{+0.7}$&$57_{-14}^{+16}$&$4.3_{-0.7}^{+1.3}$&$2.4\pm0.1$&$2.0_{-0.3}^{+0.6}$ &$26_{-3}^{+6}$ &$323/337$ \\
43&69, 70& $6.6_{-1}^{+0.7}$&$75_{-47}^{+42}$&$2.9_{-0.7}^{+4.3}$&$1.8\pm0.1$&$2.0_{-0.6}^{+1.3}$ &$24_{-3}^{+8}$ &$335/337$\\
44&70, 71& $7.4_{-0.8}^{+0.7}$&$56_{-16}^{+21}$&$4.2_{-0.9}^{+2.0}$&$2.4\pm0.1$&$2.0_{-0.3}^{+0.7}$ &$21_{-3}^{+4}$ &$308/337$\\
45&71, 72& $7.2_{-0.7}^{+0.7}$&$56_{-9.5}^{+10}$&$4.8_{-0.7}^{+1}$&$2.4\pm0.1$&$2.4_{-0.4}^{+0.7}$ &$24_{-3}^{+5}$ &$306/337$\\
46&72, 73& $7.4_{-0.9}^{+0.7}$&$26_{-6.2}^{+18}$&$11.8_{-7}^{-12}$&$5.2\pm0.6$&$1.9_{-0.2}^{+0.5}$ & $20_{-2}^{+5}$&$274/337$ \\
47&73, 74&$7.0_{-0.6}^{+0.5}$&$23.0_{-3.2}^{+3.3}$&$[10]$&$6.0\pm0.8$&$1.7_{-0.2}^{+0.3}$&$20_{-2}^{+3}$ &$275/338$ \\
48&74, 76& $7.2_{-0.5}^{+0.5}$&$25.0_{-6.4}^{+11.7}$&$12.8_{-7.2}^{+\infty}$&$5.5\pm0.6$&$1.8_{-0.2}^{+0.2}$ &$-20_{-2}^{+3}$ &$406/337$\\
49&76, 77& $6.7_{-1.3}^{+0.9}$&$45_{-21}^{+28}$&$4.3_{-1.25}^{+6.4}$&$3.0\pm0.2 $&$2.4_{-0.7}^{+3.7}$ &$19_{-3}^{+10}$ &$284/337$ \\
50&77, 78& $7.2_{-1}^{+0.9}$&$52_{-27}^{+34}$&$4.0_{-1.2}^{+8.4}$&$2.6\pm0.2$&$2.3_{-0.7}^{+4}$ &$18_{-3}^{+5}$ &$270/337$\\
51&78, 82& $[7]$&$24_{-5.5}^{+8.2}$&$11_{-5}^{+\infty}$&$5.7\pm0.7 $&$2_{-0.2}^{+0.2}$ & $15.3_{-0.2}^{+0.2}$&$384/338$\\
52&82, 84& $7.4_{-0.7}^{+0.7}$&$38.4_{-12.3}^{+17.1}$&$6_{-1.9}^{+8.6}$&$3.5\pm0.3 $&$1.4_{-0.2}^{+0.3}$ & $16_{-2}^{+3}$&$375/337$ \\
53&84, 85&$7_{-1}^{+1}$&$38_{-6}^{+6}$&$[6]$&$3.6\pm0.3$&$2.0_{-0.5}^{+0.8}$& $14_{-2}^{+4}$&$237/338$\\
54&85, 90& $6.6_{-0.5}^{+0.5}$&$32_{-15}^{+19}$&$5.8_{-2}^{+\infty}$&$4.2\pm0.4$&$2.0_{-0.3}^{+0.6}$ & $15_{-2}^{+2}$&$462/337$ \\
55&90, 101&$6.1_{-0.4}^{+0.4}$&$30_{-3.4}^{+3.4}$&$[6]$&$4.5\pm0.4$&$1.7_{-0.2}^{+0.3} $&$12_{-1}^{+2}$ &$473/338$\\
56&101, 113& $6.5_{-0.6}^{+0.6}$&$39_{-15}^{+20}$&$5.0_{-1.6}^{+8.3}$&$3.5\pm0.3$&$2.2_{-0.5}^{+1.5}$ & $9_{-1}^{+2}$& $437/337$\\
57&113, 136& $5.8_{-0.5}^{+0.5}$&$25_{-10}^{+17}$&$7.8_{-3.4}^{+\infty}$&$2.0\pm0.1$&$1.9_{-0.4}^{+1.0}$ & $11_{-1}^{+2}$&$603/337$\\
58&136, 137& $5.8_{-2}^{+1}$&$67_{-26}^{+26}$&$3.5_{-0.8}^{+1.9}$&$2.0\pm0.1 $&$30_{-27}^{+\infty}$ & $17_{-4}^{+18}$&$256/337$ \\
59&137, 150& $5.6_{-0.7}^{+0.6}$&$28_{-12}^{+16}$&$6.6_{-2.3}^{+160}$&$4.8\pm0.5 $&$2.2_{-0.5}^{+1.4}$ &$12_{-2}^{+4}$ &$462/337 $\\
60&150, 171& $4.5_{-1.7}^{+0.9}$&$61_{-50}^{+146}$&$3.0_{-1.3}^{+2.1}$&$ 2.2\pm 0.1$&$3.4_{-2.0}^{+\infty}$ & $13_{-3}^{+19}$&$372/337$ \\\hline
\end{tabular}
\end{center}
\end{tiny}
\end{table*}

\addtocounter{table}{-1}
\begin{table*}
\caption{Time-resolved Spectral fitting (continued)}
\label{tab:specfitting}
\begin{tiny}
\begin{center}
\begin{tabular}{c|cccccccccD}
\hline
\sw{Synchrotron model}&&& &  &&\\
\hline 
Sr. no. &(t$_1$,t$_2$) &$\gamma_{m}/\gamma_{c}$  &   $E_{c}$ (keV)      &   p    & \sw{pgstat}/dof\\ \hline
\hline
1 & -1, 7 & $3.6_{-1.4}^{+0.5}$ & $18.2_{-5.7}^{+3.4}$ & $>3.04$ & $420/339$ \\ 
2 & 7, 9 & $3.7_{-2.1}^{+3.5}$ & $29.6_{-12.1}^{+22.0}$ & $>2.33$ & $286/339$ \\ 
3 & 9, 10 & $3.5_{-2.1}^{+0.8}$ & $41.2_{-13.2}^{+15.0}$ & $>2.43$ & $214/339$ \\ 
4 & 10, 12 & $5.0_{-1.6}^{+1.4}$ & $45.2_{-6.5}^{+8.7}$ & $>2.98$ & $296/339$ \\ 
5 & 12, 14 & $4.1_{-1.3}^{+0.7}$ & $69.8_{-8.9}^{+12.4}$ & $>3.18$ & $302/339$ \\ 
6 & 14, 17 & $4.8_{-1.8}^{+1.3}$ & $60.5_{-7.4}^{+11.3}$ & $>2.69$ & $384/339$ \\ 
7 & 17, 18 & $2.6_{-2.1}^{+0.9}$ & $62.4_{-11.3}^{+125.8}$ & $>2.86$ & $285/339$ \\ 
8 & 18, 19 & $1.8_{-0.8}^{+0.8}$ & $81.9_{-19.2}^{+150.7}$ & $3.93_{-0.87}^{+0.89}$ & $207/339$ \\ 
9 & 19, 20 & $1.9_{-0.9}^{+0.8}$ & $84.8_{-17.9}^{+146.3}$ & $>3.41$ & $334/339$ \\ 
10 & 20, 24 & $1.6_{-0.1}^{+0.1}$ & $89.5_{-10.5}^{+133.7}$ & $3.88_{-0.37}^{+0.54}$ & $602/339$ \\ 
11 & 24, 26 & $2.3_{-0.7}^{+0.3}$ & $35.4_{-2.3}^{+7.3}$ & $>3.28$ & $434/339$ \\ 
12 & 26, 29 & $3.3_{-0.3}^{+0.1}$ & $32.8_{-2.6}^{+0.9}$ & $>4.24$ & $558/339$ \\ 
13 & 29, 30 & $3.1_{-0.7}^{+0.6}$ & $51.7_{-6.9}^{+10.7}$ & $>3.81$ & $311/339$ \\ 
14 & 30, 31 & $2.3_{-0.4}^{+0.5}$ & $77.6_{-12.1}^{+11.9}$ & $>4.13$ & $298/339$ \\ 
15 & 31, 32 & $2.3_{-0.5}^{+0.4}$ & $90.9_{-6.4}^{+20.5}$ & $>4.09$ & $406/339$ \\ 
16 & 32, 34 & $1.8_{-0.1}^{+0.3}$ & $99.4_{-10.4}^{+19.2}$ & $>3.62$ & $485/339$ \\ 
17 & 34, 35 & $1.5_{-0.2}^{+0.6}$ & $113.7_{-19.9}^{+200.9}$ & $>3.37$ & $304/339$ \\ 
18 & 35, 36 & $2.1_{-0.8}^{+0.2}$ & $81.4_{-5.3}^{+85.8}$ & $>4.33$ & $314/339$ \\ 
19 & 36, 38 & $1.7_{-0.7}^{+0.0}$ & $62.9_{-12.6}^{+1.2}$ & $3.60_{-0.27}^{+0.34}$ & $478/339$ \\ 
20 & 38, 39 & $2.3_{-0.7}^{+0.5}$ & $26.1_{-4.3}^{+6.0}$ & $>3.56$ & $347/339$ \\ 
21 & 39, 41 & $2.7_{-0.6}^{+0.1}$ & $15.5_{-1.4}^{+2.5}$ & $>3.67$ & $470/339$ \\ 
22 & 41, 44 & $3.5_{-0.3}^{+0.2}$ & $14.9_{-0.9}^{+1.8}$ & $>4.55$ & $572/339$ \\ 
23 & 44, 45 & $2.4_{-0.6}^{+0.1}$ & $30.2_{-3.3}^{+6.5}$ & $>3.77$ & $285/339$ \\ 
24 & 45, 46 & $2.9_{-0.4}^{+0.1}$ & $37.6_{-5.2}^{+2.4}$ & $>3.89$ & $364/339$ \\ 
25 & 46, 47 & $2.8_{-0.8}^{+0.5}$ & $22.9_{-3.4}^{+4.9}$ & $>3.41$ & $339/339$ \\ 
26 & 47, 48 & $2.5_{-0.5}^{+0.1}$ & $19.7_{-3.4}^{+1.4}$ & $>3.99$ & $321/339$ \\ 
27 & 48, 50 & $2.7_{-0.8}^{+0.1}$ & $13.2_{-1.7}^{+3.2}$ & $>3.45$ & $326/339$ \\ 
28 & 50, 51 & $2.0_{-0.6}^{+0.2}$ & $26.3_{-6.5}^{+6.0}$ & $>3.58$ & $287/339$ \\ 
29 & 51, 53 & $2.5_{-0.5}^{+0.1}$ & $23.0_{-1.2}^{+3.9}$ & $>3.99$ & $378/339$ \\ 
30 & 53, 55 & $3.5_{-0.5}^{+0.4}$ & $30.1_{-2.8}^{+4.0}$ & $>4.28$ & $446/339$ \\ 
31 & 55, 56 & $2.6_{-0.4}^{+0.1}$ & $53.3_{-8.5}^{+6.2}$ & $>4.04$ & $345/339$ \\ 
32 & 56, 57 & $3.0_{-0.5}^{+0.5}$ & $47.3_{-5.3}^{+7.3}$ & $>4.20$ & $351/339$ \\ 
33 & 57, 58 & $2.8_{-0.5}^{+0.4}$ & $50.8_{-5.4}^{+7.8}$ & $>4.10$ & $392/339$ \\ 
34 & 58, 59 & $2.8_{-0.4}^{+0.4}$ & $50.7_{-5.7}^{+6.4}$ & $>4.05$ & $397/339$ \\ 
35 & 59, 60 & $3.3_{-0.4}^{+0.6}$ & $40.5_{-4.8}^{+4.8}$ & $>4.13$ & $374/339$ \\ 
36 & 60, 61 & $3.2_{-0.5}^{+0.4}$ & $40.7_{-4.1}^{+4.9}$ & $>4.05$ & $346/339$ \\ 
37 & 61, 63 & $2.9_{-0.6}^{+0.1}$ & $36.0_{-1.1}^{+4.2}$ & $>3.66$ & $414/339$ \\ 
38 & 63, 64 & $2.2_{-1.5}^{+0.3}$ & $39.5_{-6.6}^{+69.8}$ & $>3.02$ & $302/339$ \\ 
39 & 64, 65 & $2.7_{-0.6}^{+0.5}$ & $34.0_{-4.7}^{+6.4}$ & $>3.90$ & $300/339$ \\ 
40 & 65, 67 & $2.5_{-0.3}^{+0.5}$ & $40.0_{-3.1}^{+1.8}$ & $>4.39$ & $531/339$ \\ 
41 & 67, 68 & $2.5_{-0.4}^{+0.1}$ & $31.4_{-4.3}^{+4.0}$ & $>3.98$ & $406/339$ \\ 
42 & 68, 69 & $2.7_{-0.4}^{+0.3}$ & $33.3_{-3.4}^{+4.5}$ & $>4.01$ & $367/339$ \\ 
43 & 69, 70 & $2.9_{-0.8}^{+0.1}$ & $19.5_{-2.2}^{+4.1}$ & $>3.42$ & $361/339$ \\ 
44 & 70, 71 & $3.1_{-0.5}^{+0.4}$ & $25.4_{-3.0}^{+4.0}$ & $>3.92$ & $358/339$ \\ 
45 & 71, 72 & $2.0_{-0.3}^{+0.1}$ & $44.7_{-5.6}^{+6.3}$ & $>4.54$ & $372/339$ \\ 
46 & 72, 73 & $2.2_{-0.6}^{+0.4}$ & $25.0_{-3.7}^{+5.3}$ & $>3.77$ & $305/339$ \\ 
47 & 73, 74 & $2.5_{-0.9}^{+0.7}$ & $14.9_{-3.3}^{+5.8}$ & $>2.97$ & $303/339$ \\ 
48 & 74, 76 & $2.5_{-0.4}^{+0.1}$ & $20.5_{-1.6}^{+2.9}$ & $>3.96$ & $481/339$ \\ 
49 & 76, 77 & $2.3_{-0.8}^{+0.5}$ & $21.4_{-4.1}^{+6.2}$ & $>3.40$ & $303/339$ \\ 
50 & 77, 78 & $2.8_{-0.8}^{+0.1}$ & $17.7_{-2.3}^{+4.2}$ & $>3.36$ & $300/339$ \\ 
51 & 78, 82 & $2.9_{-0.7}^{+0.6}$ & $13.6_{-1.7}^{+2.5}$ & $>3.31$ & $465/339$ \\ 
52 & 82, 84 & $3.2_{-1.0}^{+1.0}$ & $24.0_{-3.4}^{+4.4}$ & $>2.68$ & $427/339$ \\ 
53 & 84, 85 & $3.1_{-1.3}^{+0.4}$ & $19.2_{-3.8}^{+8.2}$ & $>2.98$ & $250/339$ \\ 
54 & 85, 90 & $2.8_{-0.6}^{+0.1}$ & $15.0_{-1.3}^{+2.5}$ & $>3.60$ & $524/339$ \\ 
55 & 90, 101 & $3.2_{-1.1}^{+0.2}$ & $14.2_{-1.3}^{+3.2}$ & $>2.95$ & $511/339$ \\ 
56 & 101, 113 & $3.3_{-0.7}^{+0.2}$ & $12.8_{-1.2}^{+2.3}$ & $>3.52$ & $485/339$ \\ 
57 & 113, 136 & $2.9_{-0.6}^{+0.1}$ & $11.3_{-1.7}^{+1.9}$ & $>3.72$ & $654/339$ \\ 
58 & 136, 137 & $3.0_{-1.0}^{+1.2}$ & $13.0_{-4.6}^{+1.8}$ & $>3.39$ & $265/339$ \\ 
59 & 137, 150 & $3.0_{-0.6}^{+0.8}$ & $11.2_{-1.2}^{+2.4}$ & $>3.70$ & $485/339$ \\ 
60 & 150, 171 & $2.1_{-1.7}^{+1.6}$ & $7.6_{-3.3}^{+15.5}$ & $>2.56$ & $378/339$ \\ \hline
\end{tabular}
\end{center}
\end{tiny}
\end{table*}

\section*{Acknowledgements}
This research has
made  use  of  data  obtained  through  the  HEASARC  Online  Service,  provided  by  the  NASA-GSFC,  in  support  of  NASA  High
Energy  Astrophysics  Programs. This publication also uses the data from the AstroSat mission 
of  the Indian Space Research Organisation (ISRO), archived at the Indian Space Science Data Centre 
(ISSDC). CZT-Imager is built by a consortium of Institutes across India including Tata Institute of 
Fundamental Research, Mumbai, Vikram Sarabhai Space Centre, Thiruvananthapuram, ISRO Satellite Centre, 
Bengaluru, Inter University Centre for Astronomy and Astrophysics, Pune, Physical Research 
Laboratory, Ahmedabad, Space Application Centre, Ahmedabad: contributions from the vast technical team 
from all these institutes are gratefully acknowledged. 
The polarimetric computations were performed on the HPC resources at the Physical
Research Laboratory (PRL). We thank Lev Titarchuk for discussions on the $grbcomp$ model.

\bibliography{GRB171010A_ref} 
\bibliographystyle{aasjournal}
\end{document}